\documentclass[journal]{IEEEtran}

\ifCLASSINFOpdf
\else
\fi

\usepackage{setspace}
\usepackage{bbm}
\usepackage{ragged2e}
\usepackage{array}
\usepackage[cmex10]{amsmath}
\usepackage{amssymb}
\usepackage{cite}
\usepackage{graphicx}
\usepackage{array,color}
\usepackage{amsmath}
\allowdisplaybreaks
\usepackage{stfloats}
\usepackage{graphicx}
\usepackage{epstopdf}
\usepackage{tabularx}
\usepackage{epsfig,epsf,color,balance,cite}
\usepackage{algorithmic}
\usepackage{algorithm}
\usepackage{url}
\usepackage{bm}
\usepackage{multirow}
\usepackage{hyperref}
\usepackage{amsthm}
\usepackage{gensymb}
\usepackage[caption=false,font=footnotesize,labelfont=rm,textfont=rm]{subfig}

\ifodd 1
\usepackage{soul}
\usepackage{color}
\setstcolor{red}

\newcommand{\del}[1]{\st{#1}} 

\newcommand{\com}[1]{\textbf{\color{red} (COMMENT: #1)}} 
\newcommand{\response}[1]{\textbf{\color{green} (RESPONSE: #1)}} 
\else

\newcommand{\del}[1]{}

\newcommand{\com}[1]{}
\newcommand{\comg}[1]{}
\newcommand{\response}[1]{}
\fi

\title{\LARGE {Rotatable Antenna-Empowered Wireless Networks: A Tutorial}}
\author{ 
	Beixiong Zheng,~\IEEEmembership{Senior Member,~IEEE},
	Qingjie Wu,  
	Xue Xiong,
    Yanhua Tan,
    Tiantian Ma,
    Qi Dai,
    Weihua Zhu,
    Changsheng You,~\IEEEmembership{Member,~IEEE},
    Xiaodan Shao,~\IEEEmembership{Member,~IEEE},
    Lipeng Zhu,~\IEEEmembership{Senior Member,~IEEE},\\
    Weidong Mei,~\IEEEmembership{Member,~IEEE},
    Qingqing Wu,~\IEEEmembership{Senior Member,~IEEE},
    Jie Tang,~\IEEEmembership{Senior Member,~IEEE},\\
	Robert Schober,~\IEEEmembership{Fellow,~IEEE}, 
    Kai-Kit Wong,~\IEEEmembership{Fellow,~IEEE},
	and Rui Zhang,~\IEEEmembership{Fellow,~IEEE}

\thanks{
	Beixiong Zheng, Qingjie Wu, Xue Xiong, Yanhua Tan, Tiantian Ma, Qi Dai, and Weihua Zhu are with the School of Microelectronics, South China University of Technology, Guangzhou 511442, China (e-mail: bxzheng@scut.edu.cn; miqjwu@mail.scut.edu.cn;~ftxuexiong@mail.scut.edu.cn;~bctanyanhua06@mail.scut.edu.cn;~mitiantianma@mail.scut.edu.cn;~7qidai@gmail.com;~miweihuazhu@mail.scut.edu.cn).

    Changsheng You is with the Department of Electronic and Electrical Engineering, Southern University of Science and Technology (SUSTech), Shenzhen 518055, China (e-mail: youcs@sustech.edu.cn).

    Xiaodan Shao is with the Department of Electrical and Computer Engineering, University of Waterloo, Waterloo, ON N2L 3G1, Canada (e-mail: x6shao@uwaterloo.ca).
    
    Lipeng Zhu is with the State Key Laboratory of CNS/ATM and the School of Interdisciplinary Science, Beijing Institute of Technology, Beijing 100081, China (e-mail: zhulp@bit.edu.cn).

    Weidong Mei is with the National Key Laboratory of Wireless Communications, University of Electronic Science and Technology of China, Chengdu 611731, China (e-mail: wmei@uestc.edu.cn).

    Qingqing Wu is with the Department of Electronic Engineering, Shanghai Jiao Tong University, Shanghai 200240, China (e-mail: qingqingwu@sjtu.edu.cn).

   Jie Tang is with the School of Electronic and Information Engineering, South China University of Technology, Guangzhou 510640, China (e-mail: eejtang@scut.edu.cn). 

    Robert Schober is with the Institute for Digital Communications, Friedrich-Alexander-University Erlangen-N$\ddot{\mathrm{u}}$rnberg (FAU), 91054 Erlangen, Germany (e-mail: robert.schober@fau.de).

    Kai-Kit Wong is affiliated with the Department of Electronic and Electrical Engineering, University College London, Torrington Place, WC1E 7JE, United Kingdom, and he is also affiliated with the Department of Electronic Engineering, Kyung Hee University, Yongin-si, Gyeonggi-do 17104, Korea (e-mail: kai-kit.wong@ucl.ac.uk).

    Rui Zhang is with the Department of Electrical and Computer Engineering, National University of Singapore, Singapore 117583 (e-mail: elezhang@nus.edu.sg).
}
}

\begin{document}
\maketitle

\begin{abstract}
Rotatable antenna (RA) has emerged as a promising technology for enhancing wireless communication and sensing performance through flexible antenna orientation/boresight rotation in three-dimensional (3D) space. By enabling mechanical or electronic boresight adjustment without altering physical antenna positions, RA introduces additional spatial degrees of freedom (DoFs) beyond conventional beamforming based on fixed antennas, offering a lightweight and hardware-efficient paradigm for antenna architecture design.
In this paper, we provide a comprehensive tutorial on the fundamentals, architectures, and applications of RA-empowered wireless networks. Specifically, we begin by reviewing the historical evolution of RA-related technologies and {\color{black}clarifying the distinctive role of RA among other flexible antenna architectures.} Then, we establish a unified mathematical framework for RA-enabled systems, including general antenna/array rotation models that capture boresight-dependent directional gain, as well as channel models that cover near- and far-field propagation characteristics, wideband frequency selectivity, and polarization effects.
Building upon this foundation, we investigate antenna/array rotation optimization in representative communication and sensing scenarios for different performance objectives. Furthermore, we examine RA channel estimation/acquisition strategies encompassing orientation scheduling mechanisms and signal processing methods that exploit multi-view channel observations.
Beyond theoretical modeling and algorithmic design, we discuss practical RA configurations and deployment strategies, highlighting key design trade-offs for hardware implementation and system architectures. We also present recent RA prototypes and experimental results that validate the practical performance gains enabled by antenna rotation. Finally, we highlight promising extensions of RA to emerging wireless paradigms and outline open challenges to inspire future research.
\end{abstract}
\IEEEpeerreviewmaketitle

\vspace{-0.3cm}
\section{Introduction}\label{sec:introduction}
\subsection{Background and Motivation}
Over the past few decades, wireless communication technologies have advanced at an unprecedented pace, profoundly reshaping human life and modern society~\cite{Chowdhury20206G}. Beyond merely facilitating interpersonal connectivity, wireless networks have evolved into critical infrastructure that underpins economic development, supports public services, and safeguards national security. As the fifth-generation (5G) network reaches global maturity, the research community is actively shaping the sixth-generation (6G) era~\cite{Akyildiz20206G}. {\color{black}Future 6G networks are envisioned to support a set of representative usage scenarios identified in the International
Mobile Telecommunications (IMT)-2030 framework, including immersive communication, hyper-reliable and low-latency communication, massive communication, artificial intelligence (AI) and communication, ubiquitous connectivity, and integrated sensing and communication (ISAC)~\cite{International2023Framework}.} These emerging scenarios impose stringent performance requirements, including ultra-high data rates, robust reliability, ubiquitous coverage, ultra-low latency, as well as integrated intelligence and sensing capabilities, which pose significant challenges to current wireless systems~\cite{Wang2023On,Wang2023Realizing,You2025Next,Tariq2020ASpeculative,Saad2019AVision}.

To accommodate these increasingly ambitious demands, the evolution of wireless networks from generation to generation has primarily followed two technical pathways: expanding system bandwidth in the frequency domain and increasing the number of antennas to exploit additional degrees of freedom (DoFs) in the spatial domain. However, spectrum resources are becoming increasingly scarce, particularly in the commercial low-frequency bands. Due to the limited availability of fragmented frequency spectrum, further expanding bandwidth has become challenging. Although shifting to higher-frequency bands (e.g., millimeter-wave (mmWave) and terahertz (THz)) offers wider bandwidths, this is accompanied by much higher path loss, increased susceptibility to blockage, and more stringent hardware design requirements. Consequently, deploying a larger number of antennas at the base station (BS) is being pursued to enhance spectral efficiency in cellular networks~\cite{Paulraj2004AnOverview}. This evolution has driven multiple-input multiple-output (MIMO) technology toward more advanced architectures, such as massive MIMO~\cite{Lu2014AnOverview,Larsson2014Massive} and extremely large-scale MIMO (XL-MIMO)~\cite{Lu2024ATutorial,Wang2024ATutorial}. By deploying hundreds or even thousands of antennas at the BS, these architectures can achieve significant spatial multiplexing and beamforming gains, thereby compensating for the limitations in spectral resources.

Despite their potential, these antenna-scaling approaches are increasingly limited by hardware complexity and energy consumption. The realization of massive MIMO and XL-MIMO systems entails substantial challenges, including high radio-frequency (RF) chain cost, increased circuit power consumption, and considerable signal processing complexity. Furthermore, the benefits of simply increasing the number of antennas diminish rapidly due to the law of diminishing returns, where the marginal performance gain fails to offset the substantial increase in hardware complexity and cost~\cite{Hoydis2011Massive}. Although various more cost-effective alternatives such as sparse arrays~\cite{Wang2023Can}, lens antenna arrays~\cite{Zeng2016Millimeter}, and intelligent reflecting surfaces (IRSs)~\cite{Zheng2022ASurvey,Wu2021Intelligent,Zheng2020Intelligent,Huang2019Reconfigurable,Zhou2025RotatableIRS,Wu2024Intelligent} have been investigated to mitigate these issues, most of them still rely on the conventional fixed-antenna architecture. In fixed-antenna systems, both the position and orientation of the antennas remain
static post-deployment, resulting in an inherent mismatch between the physical transceiver and the dynamic wireless propagation environment. This lack of antenna adaptability prevents existing systems from fully exploiting the rich spatial DoFs available in dynamic wireless environments, often leading to degraded coverage and limited capacity in complex scenarios. To overcome these intrinsic limitations, there is a growing need to move beyond fixed-antenna designs toward flexible architectures that can actively adapt to the propagation environment and make better use of antenna resources.

\begin{figure}
    \centering
    \includegraphics[width=\linewidth]{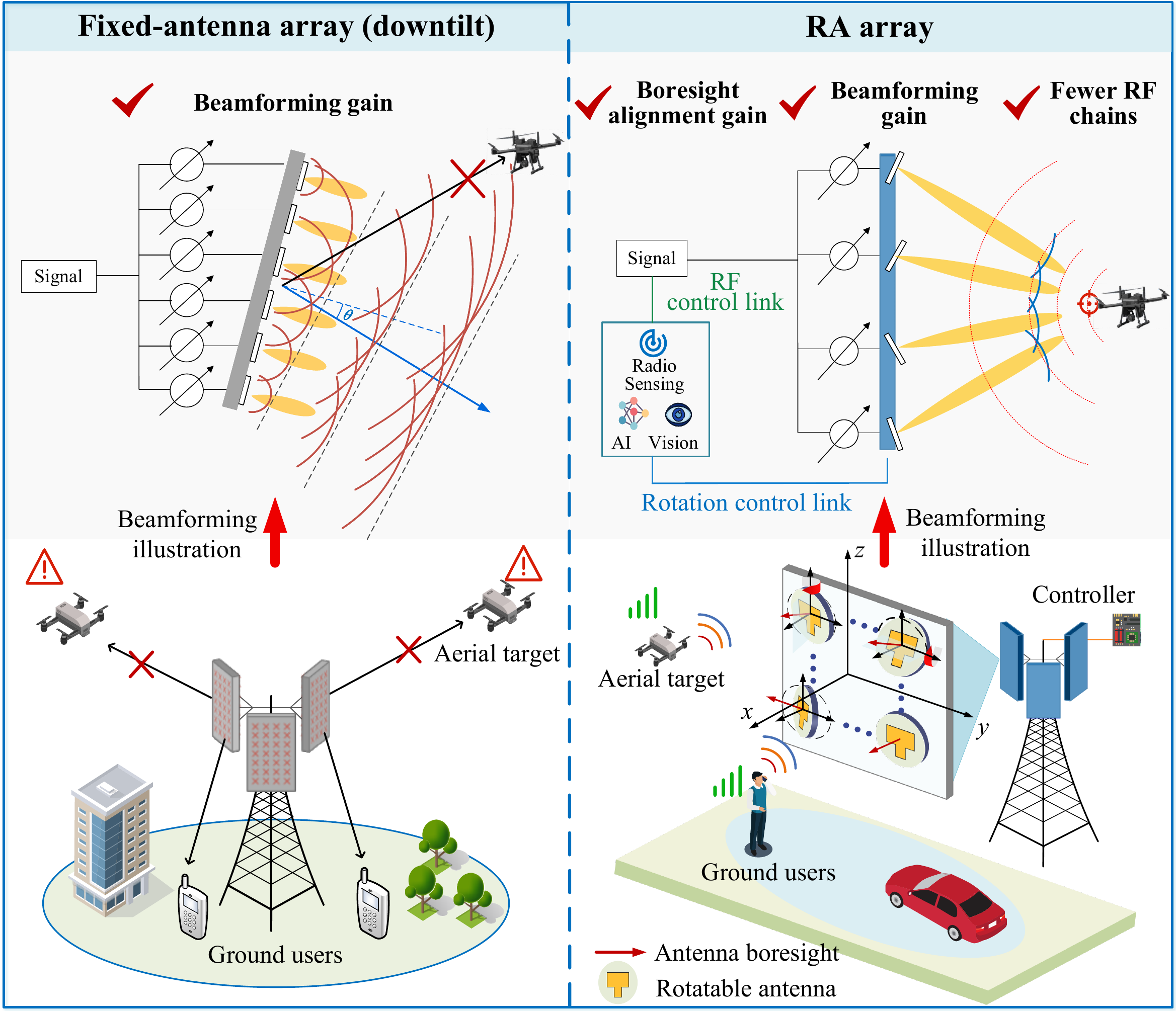}
    \vspace{-0.35cm}
   {\color{black}\caption{Architectural comparison between conventional fixed-antenna array (downtilt) and RA array.}
    \label{fig:RA_archtec}}
    \vspace{-0.3cm}
\end{figure}

\subsection{What Is RA and How It Works}
Motivated by the above, rotatable antenna (RA) has recently emerged as an efficient technology for enhancing wireless communication and sensing performance through adaptive control of antenna orientation/boresight~\cite{Zheng2026Rotatable,Wu2025Modeling,Zheng2025Rotatable,Xiong2026Intelligent}. 
{\color{black}In particular, by allowing each antenna to independently rotate its orientation/boresight toward a desired direction in three-dimensional (3D) space, RA introduces additional spatial DoFs that complement conventional beamforming over fixed antennas.}
This capability allows the radiated energy to be not only confined within a narrow angular region but also physically steered toward a specific spatial point, forming a directional ``spotlight'' effect that enhances the effective array gain toward intended users or targets, as illustrated in Fig.~\ref{fig:RA_archtec}.
Additionally, the rotational capability also enables RA to scan the 3D space in an eye-like manner, thereby achieving broader spatial coverage for both communication and sensing.
Moreover, through fine-grained orientation/boresight control, RA enhances the effective directional gain toward desired users or targets while suppressing radiation in undesired directions, thereby improving power efficiency and system performance with significantly fewer RF chains than traditional fixed-antenna systems.

\begin{figure}
    \centering
    \includegraphics[width=2.4in]{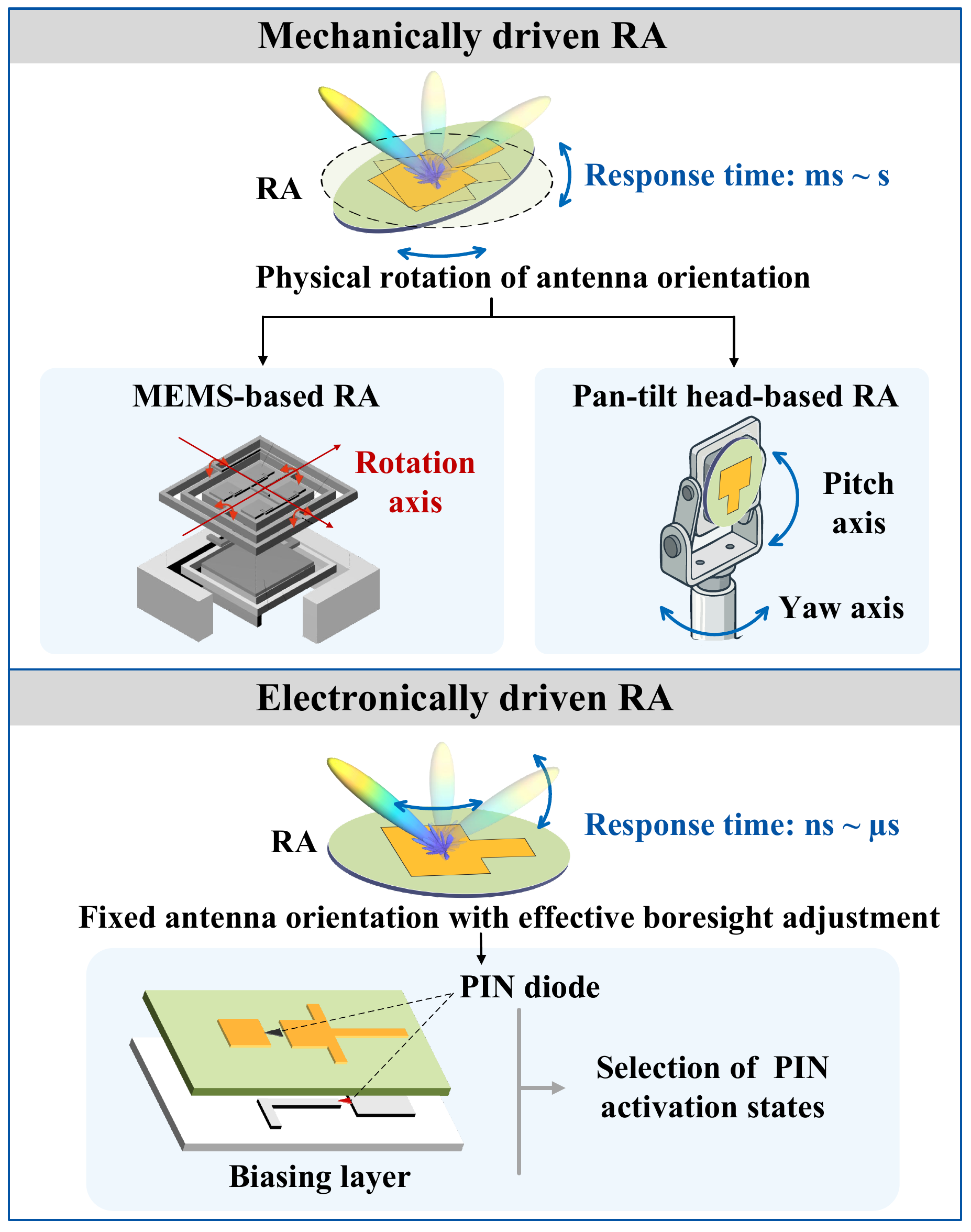}
    {\color{black}
    \caption{Hardware implementations of RA: Mechanically and electronically driven.}
    \label{fig:RA_implement}}
    \vspace{-0.3cm}
\end{figure}


In practical implementations, RA designs can be achieved through mechanical or electronic control, as illustrated in Fig.~\ref{fig:RA_implement}. Specifically, mechanical control approaches typically utilize servo motors or micro-electromechanical systems (MEMS) to physically rotate the orientation of directional antennas, offering a wide angular adjustment range with milliwatt-level power consumption and millisecond (ms)-scale response times~\cite{Baek2003AVband,Yang2025ALow}.
In contrast, electronic control methods retain a fixed antenna orientation while enabling rapid boresight adjustment through electronic techniques such as Positive-Intrinsic-Negative (PIN) diode switching or reconfigurable parasitic radiator element loading, thus achieving much faster response times at the microsecond (µs)/nanosecond (ns) scale and better compatibility with existing wireless systems~\cite{Zhang2022Highly,Towfiq2018AReconfigurable,Lotfi2017Printed}.
To harness the benefits of both approaches, co-designed RA architectures that integrate both driving mechanisms can also be adopted to enable wide-angle and rapid rotation of antenna orientation/boresight. 

\begin{figure}
    \centering
    \includegraphics[width=3in]{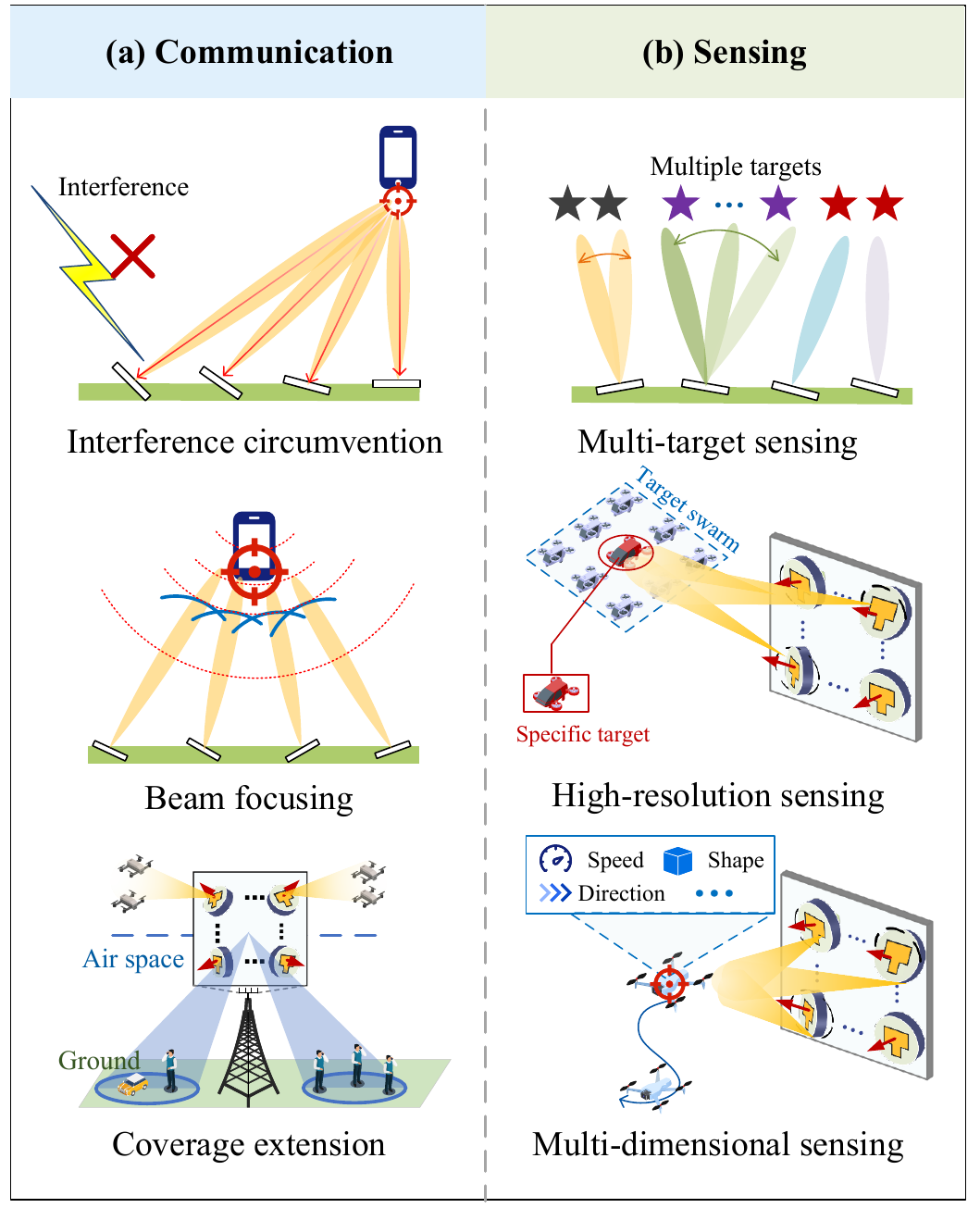}
    \caption{Functional advantages of RA for communication and sensing.}
    \label{fig_intro_RAfunction}
\end{figure}

Through antenna rotation capability, RA achieves notable performance gains in wireless communication systems, particularly in adaptive interference circumvention, beam focusing, and coverage extension, as illustrated in Fig.~\ref{fig_intro_RAfunction}.
{\color{black}For instance, since interference typically originates from specific directions, RAs can reorient their antenna boresights away from those directions, thereby reducing reliance on sophisticated signal processing for interference suppression.
Conventionally, beamforming is achieved by applying complex-valued weights to identical transmitted symbols across multiple antennas in the array, thus enabling directional transmission through coherent superposition and cancellation (see Fig.~\ref{fig:RA_archtec}). 
Building upon conventional beamforming, RA arrays can further reorient each individual antenna’s radiation pattern to achieve much sharper beam focusing.}
Furthermore, RA arrays can dynamically adjust antenna orientations to flexibly extend effective communication coverage across a wide 3D space without increasing transmit power or requiring denser infrastructure.
This capability supports users and targets distributed across different altitudes and spatial regions, which is particularly desirable in low-altitude ISAC and space-air-ground integrated networks (SAGINs).
Moreover, in hybrid configurations where RAs are co-deployed with fixed-sector antenna arrays, RAs’ reorientation capability can complement static coverage by mitigating blind spots and enhancing link reliability in distributed or irregular deployments.

In parallel, RA-enabled sensing systems demonstrate significant advantages over conventional fixed-antenna systems due to their unique boresight tracking and scanning capabilities, as illustrated in Fig.~\ref{fig_intro_RAfunction}.
Even with a very limited number of antennas, RA arrays can support efficient multi-target sensing by exploiting antenna rotation and spatial scanning to observe targets from diverse directions, allowing the number of detectable targets to significantly exceed the number of antennas.
Beyond wide-area scanning for detecting multiple targets, RA arrays can further enable high-resolution sensing by coherently rotating the boresight directions of multiple RAs toward a common spatial location.
Building on high-resolution sensing, flexible boresight control of RAs further facilitates multi-dimensional sensing by enabling more effective extraction of rich target features. By adaptively tracking targets from different orientations, RA-enabled sensing systems can capture additional information such as target velocity, movement direction, shape, and other fine-grained characteristics with higher reliability. This capability is particularly beneficial in dynamic sensing scenarios involving mobile or clustered targets, where conventional fixed-antenna systems often suffer from limited angular diversity and incomplete feature observation.

{\color{black}\subsection{Historical Development}}
From early concepts of mechanically rotating antennas used for direction finding to the modern realization of flexible antenna architectures, improving the efficiency and controllability of wireless transmission has remained a central theme~\cite{Ma2026ASurvey}.
A foundational milestone in this evolution occurred in 1901, when Guglielmo Marconi’s transatlantic radio experiment marked the beginning of long-distance wireless communication and demonstrated the importance of spatial radiation control~\cite{Holzmann1995Early}.
Subsequently, various directional antennas, including loop antennas, Yagi-Uda antennas, and parabolic reflectors, were developed and deployed across communication, radar, and radio astronomy applications~\cite{Balanis1996Antenna}.
These antennas were commonly installed on mechanically rotatable platforms that allowed dynamic adjustment of their radiation directions, representing an early precursor to the modern RA architectures.
{\color{black}Building on this design paradigm, practical implementations of mechanical antenna rotation were widely utilized during World War II~\cite{USNavy1944YE1,Getting1990SCR}.
Further miniaturization was achieved through advances in MEMS toward the end of the twentieth century, yielding more compact and efficient antenna rotation mechanisms with fast, low-power, and flexible mechanical control~\cite{Chauvel1997Micro,Chiao1999MEMS,Baek2003AVband}.}

Despite its effectiveness, mechanical antenna rotation became increasingly inadequate as target speeds increased, motivating the development of electronic radiation beam control.
Initially, phased-array antennas were introduced to direct beams by imposing controlled phase differences across antennas~\cite{Balanis1996Antenna}.
{\color{black}However, such systems require many phase shifters and suffer from degraded radiation efficiency at large steering angles.
To reduce the hardware burden of conventional phased arrays, reactively controlled directive arrays were later investigated, where the main beam is electronically steered by selecting appropriate reactive loads~\cite{Harrington1978Reactively,Thiel2002Switched}.}
{\color{black}Around the early 2000s, semiconductor devices such as PIN diodes and varactor diodes further enabled practical electronic beam control.}
By manipulating current flows through RF switching diodes, the antenna radiation pattern can be dynamically modified~\cite{Nikolaou2006Pattern}.
Moreover, advances in pixel antenna technology further accelerated the development of electronic beam control.
{\color{black}By controlling the PIN diodes connecting the pixels, pixel antenna systems offer large beam rotation coverage, high scanning resolution, and highly reconfigurable radiation patterns, closely aligning with the electronically controlled RA concept~\cite{Lotfi2017Printed,Towfiq2018AReconfigurable,Zhang2022Highly}.}

While flexible antenna architectures remained largely application-specific in the past, the increasing demand for higher spatial resolution and more adaptive coverage in advanced wireless systems has renewed attention on flexible antenna technologies recently~\cite{Zhu2024Historical}. In this context, the notion of ``movable antenna (MA)'' was formally mentioned in 2007~\cite{Balanis2007Modern} and applied to wireless communications in 2009~\cite{Zhao2009Single}. In parallel, while the term ``fluid antenna'' was first mentioned in~\cite{Tam2008Patent} and traditionally corresponds to liquid-based radiators~\cite{Kosta1989ALiquid,Kosta2004Liquid}, Wong {\em et al.}~redefined the concept of fluid antenna system (FAS) in 2020~\cite{Wong2020Performance,Wong2020Fluid,Wong2021Fluid} to advocate software-controlled position- and shape-flexible antennas capable of exploiting spatial diversity. More precisely, FAS is not an antenna technology per se, but a hardware-agnostic system concept that treats the antenna as a reconfigurable physical layer resource to broaden system design, encompassing software-controllable fluidic, conductive, and dielectric structures capable of reconfiguring fundamental antenna characteristics such as gain, radiation pattern, and operating frequency~\cite{New2025ATutorial,Lu2025Fluid,New2025Fluid,Hong2025AContemporary,Wu2025Fluid}.

Both MA and FAS share the same concept of antenna positional reconfigurability to enrich the effective spatial diversity of wireless channels, while differing in their underlying hardware. Building upon these advances, a joint transmit-receive MA-aided wireless communication system with two-dimensional (2D) antenna movement was proposed in 2022~\cite{Zhu2022Modeling,Ma2024MIMO}, followed by its extension to 3D movement in~\cite{Zhu2024Movable}. To extend spatial reconfigurability beyond existing MA designs, the six-dimensional movable antenna (6DMA) framework was proposed in 2024~\cite{Shao20256D,Shao20256DMovable,Shao20256DMA}, incorporating both 3D positional and rotational control, thereby achieving enhanced flexibility without increasing the number of antennas. Moreover, pinching antenna, first mentioned in 2022~\cite{Fukuda2022Pinching}, can be conceptually traced back to earlier visions of programmable surface-wave pathways with controllable leaky radiation~\cite{Wong2020AVision}, an application of position‑reconfigurable FAS on a large scale. 
It is worth noting that the practical pinching antenna developed by NTT DOCOMO is not reconfigurable; the position‑reconfigurable variants studied in recent theoretical works are conceptual extensions rather than existing hardware implementations~\cite{Ding2025Flexible,Liu2026Pinching,Liu2026PASS,Xu2026Generalized}.
This technology provides another path to large-scale spatial adaptation by radiating from arbitrary points along dielectric waveguides.

\begin{table*}[]
	\centering	
	\caption{Comparison of Different Flexible Antenna Architectures}
	\vspace{-0.2cm}
	\label{tab:sum}
	\renewcommand{\arraystretch}{1.3}
    \begin{small}
	\begin{tabular}{|c|c|c|c|c|}
		\hline
		\textbf{\begin{tabular}[c]{@{}c@{}}Antenna \\ Architecture\end{tabular}} &
		\textbf{Hardware  Implementation} &
		\textbf{\begin{tabular}[c]{@{}c@{}} Reconfigurable Parameter\end{tabular}} &
		\textbf{\begin{tabular}[c]{@{}c@{}}Deployment \\ Complexity\end{tabular}} &
		\textbf{\begin{tabular}[c]{@{}c@{}}System\\ Overhead\end{tabular}} \\ \hline\hline
		RA \cite{Zheng2026Rotatable,Wu2025Modeling,Zheng2025Rotatable} &
		Mechanical or electronic means &
		Antenna 3D boresight direction &
		Low &
		Low \\ \hline
		FAS \cite{Wong2020Performance,Wong2020Fluid,Wong2021Fluid} &
		Hardware agnostic &
		Antenna shape, position, etc. &
		Moderate &
		Moderate \\ \hline
		MA \cite{Zhu2022Modeling,Ma2024MIMO,Zhu2024Movable} &
		Motors &
		Antenna 3D position &
		Moderate &
		Moderate \\ \hline
		6DMA \cite{Shao20256D,Shao20256DMovable,Shao20256DMA} &
		Motors and flexible cables &
		Antenna 3D position and 3D rotation &
		Moderate to high &
		Moderate to high \\ \hline
        \begin{tabular}[c]{@{}c@{}}PASS~\cite{Fukuda2022Pinching,Wong2020AVision,Ding2025Flexible,Liu2026Pinching,Liu2026PASS,Xu2026Generalized}\end{tabular} &
        \begin{tabular}[c]{@{}c@{}}Waveguides and distributed \\ pinching antennas\end{tabular} &
        Large-scale antenna position &
        Moderate to high &
        Moderate to high \\ \hline
	\end{tabular}
    \end{small}
\end{table*}

\begin{figure}
    \centering
    \includegraphics[width=3in]{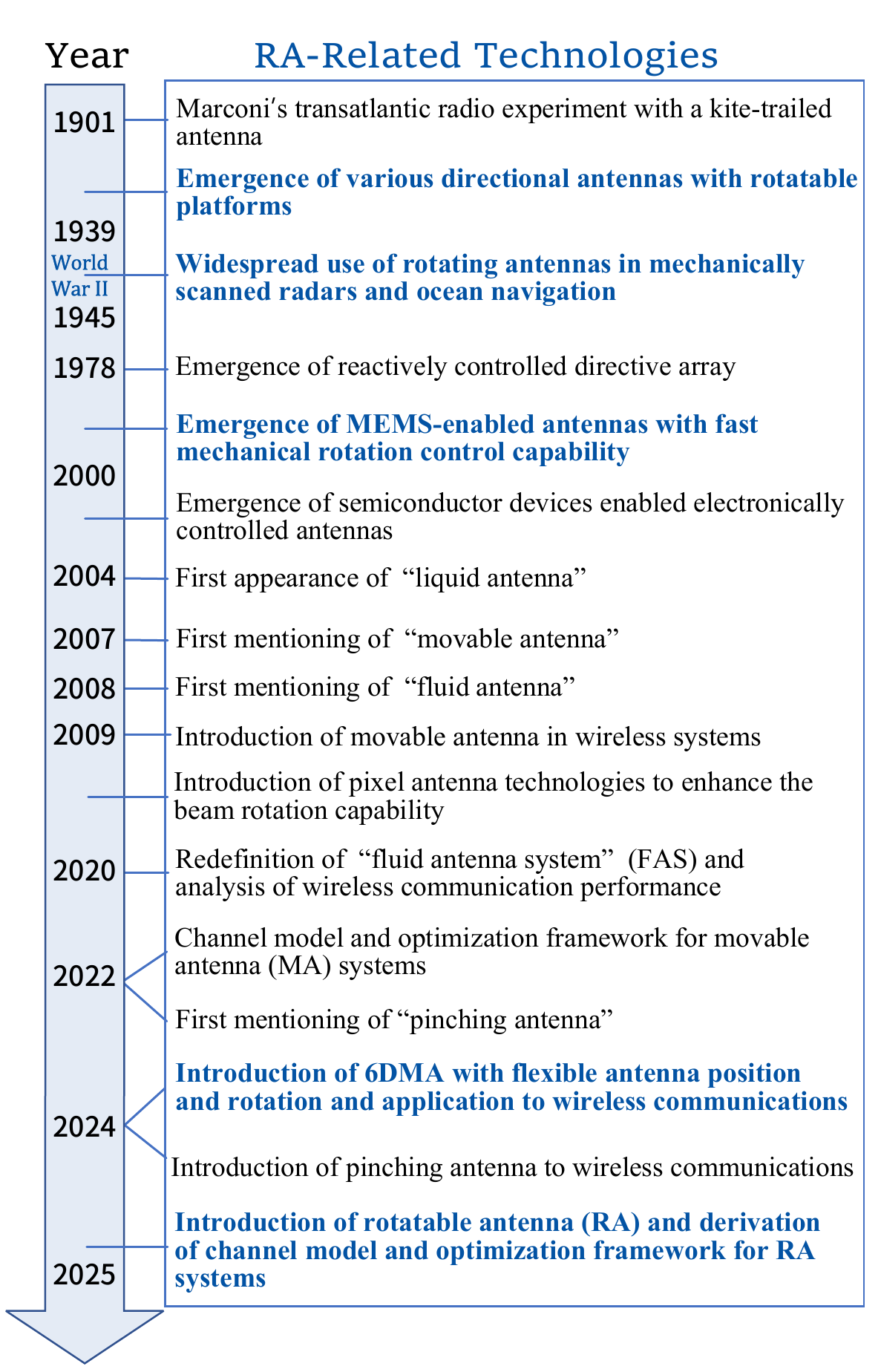}
    \caption{Illustration of the historical development of RA and other flexible antenna architectures.}
    \label{fig:history}
\end{figure}

Driven by the growing interest in spatially adaptive systems, RA technology emerged in late 2024-2025 as a rotation-centric flexible antenna architecture that enables spatial adaptability through controllable 3D antenna orientation/boresight rotation~\cite{Wu2025Modeling,Zheng2026Rotatable,Zheng2025Rotatable}. In particular, inspired by the physical antenna motion in MA/6DMA, mechanically driven RA was proposed to achieve spatial adaptability by physically rotating the antenna orientation. 
Meanwhile, motivated by the internal reconfiguration mechanism of FAS based on electronic switching elements, electronically driven RA was developed to reshape the radiation pattern by switching the radiation mainlobe direction (without physically changing antenna orientation). 

{\color{black}To clearly highlight the key characteristics of different flexible antenna architectures, Table~\ref{tab:sum} provides a comparative summary of typical designs, including FAS, MA, 6DMA, pinching-antenna systems (PASS), and RA. In some sense, RA can also be viewed as a low-complexity yet promising member of the FAS/MA/6DMA family that retains only the antenna rotation capability (without changing antenna position), thereby offering a cost-effective and compact solution. 
Compared with the translational motion in MA and 6DMA, which requires additional physical space (e.g., sliding tracks or movable regions), RA achieves reconfigurability through localized rotation, allowing more efficient integrated mechanical or electronic implementations. Thus, MA and 6DMA are more suitable when sufficient movement space is available and position-dependent channel variations can be exploited, while PASS is more attractive for large-scale waveguide-supported deployments that require flexible antenna placement over extended regions. By contrast, RA confines reconfiguration to antenna orientation/boresight control, striking a favorable balance between flexibility and complexity.
This makes RA especially suitable for deployment scenarios constrained by physical space, payload, hardware cost, power budget, or control overhead, such as micro BS/access point (AP) deployments, unmanned aerial vehicle (UAV)- or satellite-mounted nodes, and dense indoor hotspots.} 
Key milestones in the historical development of RA-related technologies are summarized in Fig.~\ref{fig:history}.

\begin{figure*}
\center
    \includegraphics[width=0.93\textwidth]{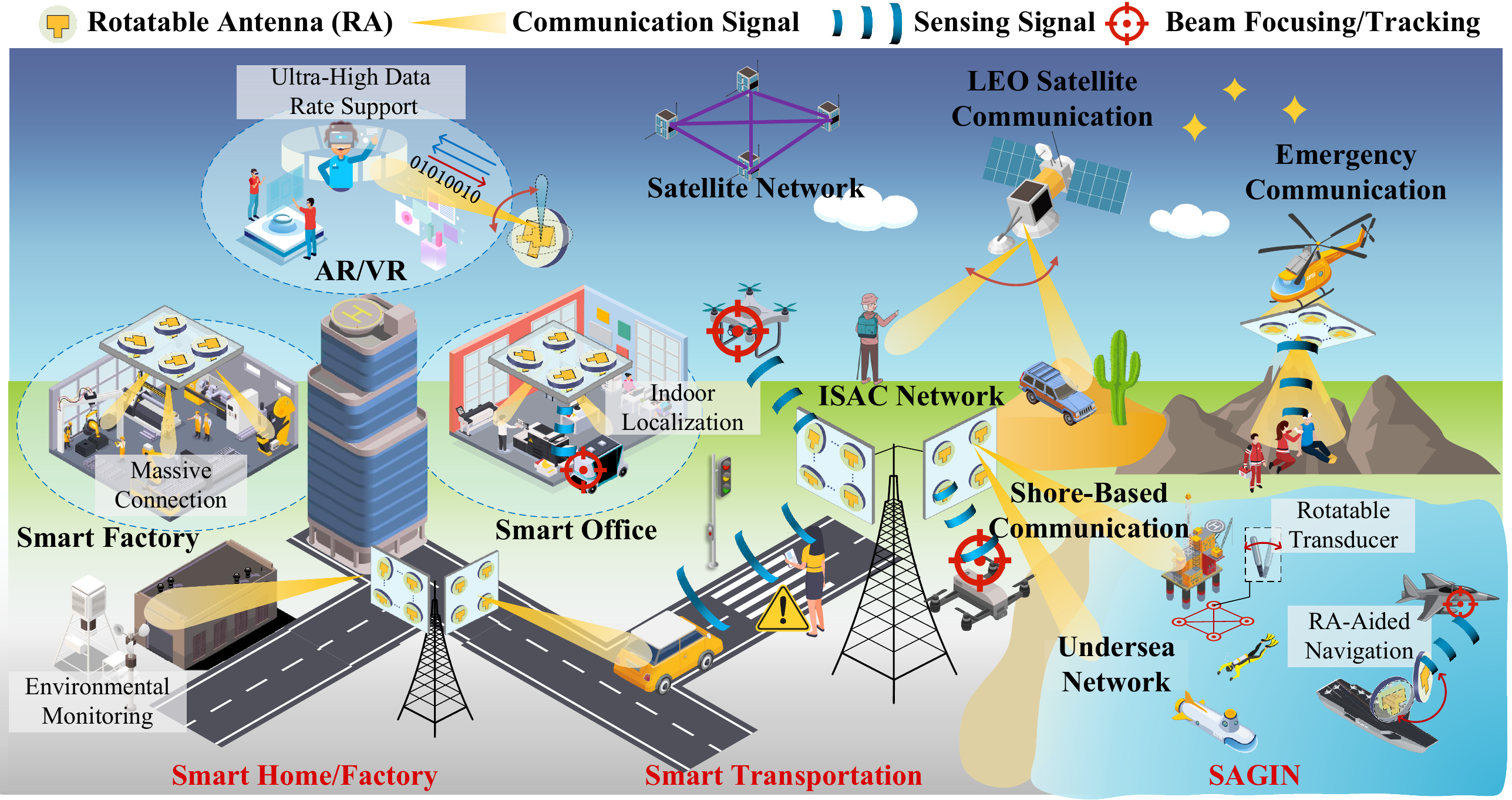}
    \caption{Application scenarios of RA-enabled wireless networks.}
    \label{fig:intro_scenario}
\end{figure*}

\vspace{0.1cm}
{\color{black}\subsection{Potential of RA Deployment in Future Wireless Networks}}
Looking ahead, RA technology is expected to play an increasingly important role in compact and energy-efficient wireless communication and sensing systems. {\color{black}In particular, as shown in Fig.~\ref{fig:RA_archtec}, the conventional 2D ground coverage enabled by fixed-sector antennas (downtilt)} is no longer sufficient to accommodate the growing demands of emerging 3D connectivity and perception scenarios involving aerial and non-terrestrial nodes, such as in the low-altitude economy (LAE) and SAGINs~\cite{Li2025Rotatable,Ma2026Rotatable}. With low-complexity boresight control and fine-grained directional adaptability, RA enables a paradigm shift from static sector-based 2D coverage to flexible 3D spatial transmission across diverse deployment environments (see Fig.~\ref{fig:RA_archtec}). 
As illustrated in Fig.~\ref{fig:intro_scenario}, RA arrays can be deployed across space, air, and ground segments, allowing wireless infrastructures to dynamically adapt their radiation directions in response to spatially distributed and time-varying service demands.
Beyond enhancing conventional urban and suburban outdoor coverage, RA is particularly attractive in scenarios where dense BS deployments are infeasible or inefficient. For example, in deserts, mountainous regions, and ocean environments, RA-equipped aerial platforms or satellites can dynamically rotate their antenna boresights to provide scalable coverage and reliable backhaul links over sparse, mobile, or infrastructure-limited networks. In more challenging environments such as underwater and deep-ocean operations (where terrestrial connectivity cannot be established), RA-enabled maritime or submarine platforms offer a promising means to maintain directional, robust, and task-oriented communication and sensing links.

In addition to large-scale outdoor deployments, RA also exhibits significant potential in indoor and short-range scenarios that demand high spatial selectivity and flexible coverage adaptation. In indoor environments such as smart factories, smart offices, and residential spaces, RAs can dynamically adjust their boresight directions to align with the locations of active users, devices, or machines, thereby improving link reliability, mitigating interference, and enhancing energy efficiency. Moreover, in emerging immersive applications such as virtual reality (VR), augmented reality (AR), and indoor digital twin systems, RA enables more precise spatial focusing of signal energy, supporting high-data-rate, low-latency communications and high-accuracy sensing within confined 3D spaces.
Beyond traditional static deployment scenarios, RA can further benefit emerging applications involving mobile platforms and dynamic environments (such as intelligent transportation systems, high-speed vehicles, and industrial robotics), where rapid changes in relative geometry pose significant challenges to fixed-antenna solutions. By continuously adapting antenna orientation/boresight in response to environmental and mobility variations, RA offers a practical means to sustain directional links and stable performance under highly dynamic conditions.
Meanwhile, such directional agility also supports accurate sensing, localization, and navigation by adaptively aligning the radiation mainlobe toward mobile targets or dynamic areas of interest.
These advantages highlight the practical scalability of RA across heterogeneous platforms and evolving network topologies, laying a strong foundation for its broader integration into future wireless systems.

\begin{table*}[t]
{\color{black}
\caption{Comparison of This Tutorial With Representative Overview/Tutorial Articles on Flexible Antennas}
\label{tab:survey_comparison}
\centering
\footnotesize
\setlength{\tabcolsep}{4pt}
\renewcommand{\arraystretch}{1.18}
\begin{tabular}{|
>{\centering\arraybackslash}m{1.15cm}|
>{\centering\arraybackslash}m{1.35cm}|
>{\justifying\arraybackslash}m{14.2cm}|}
\hline
\textbf{Reference} & \textbf{Type} & \multicolumn{1}{c|}{\textbf{Main Contributions}} \\
\hline
\cite{Zhu2024MovableAntennas} & Magazine & An overview of MA-aided wireless communications, covering application scenarios, hardware architectures, channel characterization, performance advantages, and key design challenges based on antenna position reconfiguration. \\
\hline
\cite{Shao20256DMA} & Magazine & An overview of 6DMA-enhanced wireless networks, including flexible antenna position/orientation control, system and channel modeling, performance enhancement, implementation challenges, and potential solutions. \\
\hline
\cite{Zheng2025Rotatable} & Magazine & An overview of RA-enabled wireless communication and sensing, discussing basic principles, potential applications, key challenges, and future opportunities of RA systems. \\
\hline
\cite{Xiong2026Intelligent} & Magazine & An overview of intelligent RA for integrated sensing, communication, and computation, highlighting the integration of antenna rotation with sensing, communication, computation, and intelligent control. \\
\hline
\cite{Li2025Rotatable} & Magazine & An overview of RA-empowered low-altitude economy, focusing on deployment opportunities, key challenges, and potential RA-enabled services in low-altitude scenarios. \\
\hline
\cite{Zhu2026ATutorial} & Tutorial & A comprehensive tutorial on MA-enabled wireless networks, covering channel modeling, performance analysis, antenna movement/position optimization, channel acquisition, hardware implementation, and representative wireless applications. \\
\hline
\cite{New2025ATutorial} & Tutorial & A tutorial on FAS-enabled 6G networks, covering channel modeling, channel estimation, performance analysis, and multiple access, mainly based on spatial-correlation channel models for rich-scattering environments. \\
\hline
\cite{Shao2025ATutorial} & Tutorial & A tutorial on 6DMA for 6G networks, covering position and rotation modeling, channel estimation, optimization methods, and representative application scenarios. \\
\hline
\cite{Liu2026PASS} & Tutorial & A tutorial on PASS, covering fundamental signal, hardware, and power radiation models, performance analysis, pinching beamforming, wideband transmission, channel acquisition, and emerging wireless applications. \\
\hline
\cite{Xu2026Generalized} & Tutorial & A tutorial on generalized PASS, covering representative physical realizations and channel models, system architectures, advanced design strategies, integration with emerging wireless technologies, and practical challenges and future directions. \\
\hline
This paper & Tutorial & A comprehensive tutorial on RA-empowered wireless networks, covering: the motivation and historical development of RA-related technologies; unified antenna/array rotation and channel modeling incorporating near/far-field, multipath, wideband, and polarization effects; antenna/array rotation optimization for communication and ISAC systems; RA channel estimation/acquisition; practical configurations, deployment strategies, prototypes, and commercial products; and extensions to emerging technologies and future research directions. \\
\hline
\end{tabular}}
\end{table*}

\vspace{0.1cm}
\subsection{Objective, Contribution, and Organization}
Given the significant potential of RA for unlocking new DoFs with low hardware complexity, this paper provides a comprehensive tutorial on the fundamentals, implementation, and recent advances of RA‑empowered wireless networks. Specifically, it covers the mathematical framework for antenna/array rotation and channel modeling, the main design challenges for RA‑enabled systems, hardware implementation architectures, proof‑of‑concept prototypes, and emerging application scenarios. Our objective is to establish a solid theoretical and practical foundation for researchers in this field and to inspire future investigations.

Although flexible antenna architectures have recently attracted extensive attention, existing works typically focus on some specific reconfiguration dimensions and their associated modeling assumptions, which are not directly transferable to RA systems. {\color{black}To better illustrate the research gap, a comparison between this paper and representative overview/tutorial articles on flexible antenna systems is provided in Table~\ref{tab:survey_comparison}.} In particular, research on MA/FAS has mainly investigated position reconfiguration within a prescribed region, where performance improvements arise from translational movement and the resulting position‑dependent channel variation~\cite{Zhu2024MovableAntennas,Zhu2026ATutorial,New2025ATutorial,Zhu2023Movable,Ma2023Compressed}.
Meanwhile, most existing 6DMA‑related works rely on simplified signal models that overlook practical aspects such as wideband frequency selectivity and polarization effects~\cite{Shao20256DMA,Shao2025ATutorial}. More importantly, existing RA‑related studies are largely limited to conceptual discussions or specific optimization problems, and thus do not yet provide a comprehensive introduction that bridges the gap between theoretical modeling and practical implementation. In addition, although several magazine papers~\cite{Zheng2025Rotatable,Xiong2026Intelligent,Li2025Rotatable} have discussed the basic principles, key challenges, and opportunities of RA systems, they generally lack comprehensiveness in comparing different channel modeling methods, channel estimation strategies, detailed optimization approaches, and state‑of‑the‑art prototypes. Compared with these existing works, the main contributions of this tutorial are summarized as follows:
\begin{itemize}
    \item We provide the motivation and historical development of RA‑empowered wireless networks and establish a unified fundamental framework for RA‑enabled systems, including a general antenna/array rotation model and representative channel models that capture boresight‑dependent directional gain and polarization effects.

    \item We investigate antenna/array rotation optimization in representative RA‑aided systems, covering RA for communication scenarios (e.g.,  single-input single-output (SISO)/multiple-input single-output (MISO)/single-input multiple-output (SIMO)/MIMO, single-user/multi-user, and wideband systems) and RA for ISAC scenarios under different performance metrics.

    \item We present RA channel estimation/acquisition strategies, including fixed‑orientation and dynamic‑orientation estimation, and examine representative signal processing methods that exploit multi‑view observations enabled by RA orientation control.

    \item We elaborate practical RA configurations spanning from concept to implementation, compare representative configuration options (e.g., sparse/non‑sparse array structures, continuous/discrete rotation, and distributed/centralized deployment), and provide guidelines for selecting suitable RA configurations under hardware and control constraints.

    \item We provide an overview of RA prototypes and related products, summarizing representative implementation strategies and experimental results that validate the practical performance gains achievable through antenna rotation. Moreover, we discuss open challenges and future research directions to broaden the application scope of RA‑enabled wireless systems.
\end{itemize}

\begin{figure}
    \centering
    \includegraphics[width=\linewidth]{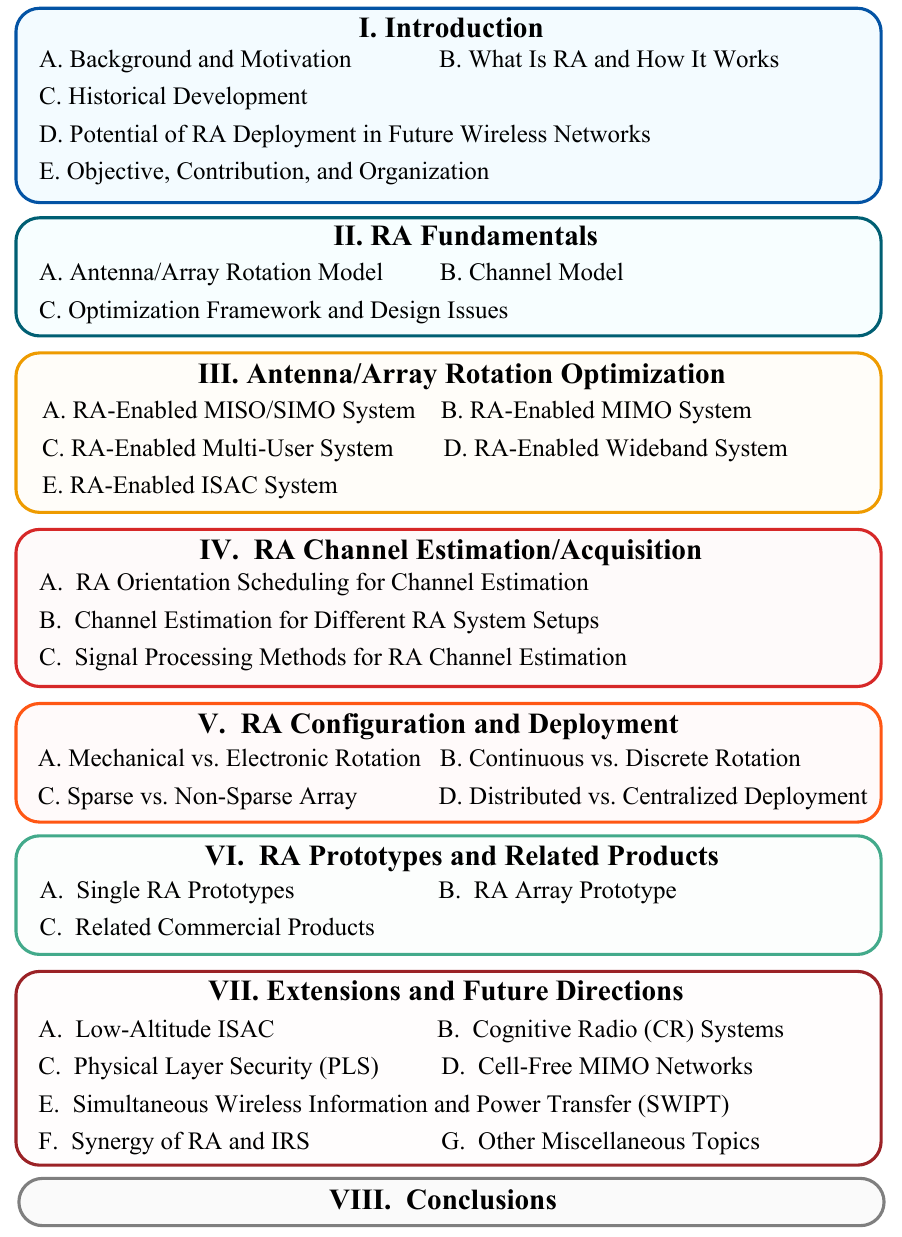}
    \vspace{-0.3cm}
{\color{black}\caption{Organization of this paper.}
    \label{fig_intro_RAorganization}}
\end{figure}

The organization of this paper is illustrated in Fig.~\ref{fig_intro_RAorganization}. Section~\ref{sec:introduction} introduces the motivation and background of RA‑empowered wireless networks and reviews the historical development and future trends of RA‑related technologies. Section~\ref{sec:fundamentals} presents the fundamentals of RA systems, including the antenna/array rotation model and representative channel modeling methods. Section~\ref{sec:optimization} investigates antenna/array rotation optimization in representative RA‑aided communication and ISAC systems. Section~\ref{sec:estimation} reviews RA channel estimation/acquisition strategies. 
Section~\ref{sec:configuration} discusses RA configurations and deployments.
Section~\ref{sec:prototypes} presents proof‑of‑concept prototypes and related commercial products. Section~\ref{sec:extensions} provides extensions and future directions. Finally, this paper is concluded in Section~\ref{sec:conclusions}.

\textit{Notation:} Upper‑case and lower‑case boldface letters denote matrices and column vectors, respectively. $(\cdot)^T$, $(\cdot)^{\ast}$, $(\cdot)^H$, and $(\cdot)^{-1}$ stand for the transpose, conjugate, Hermitian transpose, and matrix inversion operations, respectively. The sets of $a\times b$ dimensional complex and real matrices are denoted by $\mathbb{C}^{a\times b}$ and $\mathbb{R}^{a\times b}$, respectively. $\lfloor \cdot \rfloor$ is the floor function, $\otimes$ denotes the Kronecker product, and $\mathbb{E}\{\cdot\}$ denotes expectation. For a vector $\bm{x}$, $\|\bm{x}\|$ denotes its $\ell_2$‑norm, $\mathrm{diag}(\bm{x})$ returns a diagonal matrix with the elements of $\bm{x}$ on its main diagonal, $\Re \{\bm{x}\}$ and $\Im \{\bm{x}\}$ denote its real and imaginary parts, respectively, and $[\bm{x}]_i$ denotes its $i$‑th entry. For a matrix $\mathbf{X}$, $\mathrm{Tr}(\mathbf{X})$ and $\det(\mathbf{X})$ denote its trace and determinant, respectively; $\mathrm{vec}(\mathbf{X})$ is the vectorization operator; and $\mathbf{X}\succeq 0$ implies that $\mathbf{X}$ is positive semidefinite. $\mathbf{I}$ and $\mathbf{0}$ denote an identity matrix and an all‑zero matrix, respectively, with appropriate dimensions. The distribution of a circularly symmetric complex Gaussian (CSCG) random vector with zero mean and covariance matrix $\bm{\Sigma}$ is denoted by $\mathcal{N}_c(\bm{0},\bm{\Sigma})$; and $\sim$ stands for ``distributed as''.


\section{RA Fundamentals}\label{sec:fundamentals}
This section introduces the fundamentals of RA-enabled wireless systems. We first present the antenna and array rotation models, which capture the flexible orientation/boresight control of individual antennas~\cite{Wu2025Modeling,Zheng2026Rotatable} as well as the coordinated rotation of antenna arrays. Next, we establish a near-field channel model tailored for RA systems, which characterizes how antenna rotation modifies the directional gain pattern observed by wireless channels. Building on this foundation, the RA framework is further extended to cover far-field propagation, multipath environments, wideband scenarios, and polarization effects.
Finally, we develop a unified optimization framework for designing antenna orientations/boresights to enhance system performance.
Unless otherwise specified, we consider a general system model comprising $K$ single-antenna users and a BS equipped with an RA array consisting of $N$ directional antennas for ease of exposition. For notational convenience, we use subscripts ``B'', ``U'', and ``C'' to indicate the BS, user, and scatterer cluster, respectively.

\subsection{Antenna/Array Rotation Model}\label{Rotation_Model}
\begin{figure*}
\center
    \includegraphics[width=0.9\textwidth]{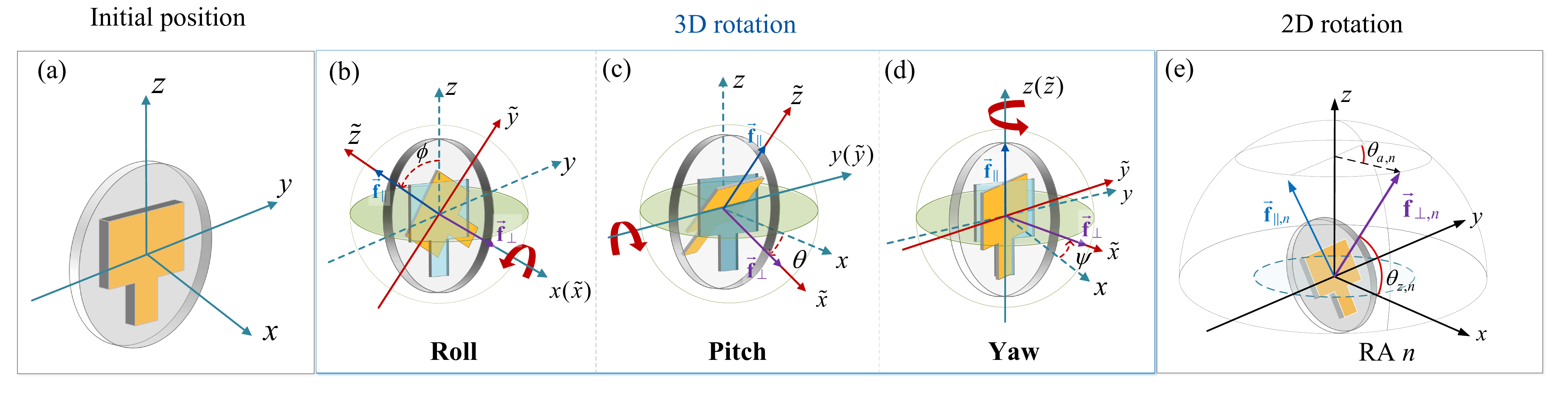}
    \caption{Illustration of 3D and 2D rotations of an individual antenna.}
    \label{fig_rotation_1}
\end{figure*}

For directional antennas/arrays, the channel conditions depend strongly on the relative geometric relationships between transceivers. In this regard, antenna/array rotation alters both the directional gain toward a spatial point and the polarization characteristics of the transmitted/received signal.
According to Euler’s rotation theorem, any 3D rotation around a fixed point can be represented as a composition of three elementary rotations~\cite{Goldstein2001Classical}.
As illustrated in Fig.~\ref{fig_rotation_1}, the orientation of an antenna/array can be described by three rotation angles: a roll angle $\phi \in [0,2\pi)$ around the $x$-axis, a pitch angle $\theta \in [0,2\pi)$ around the $y$-axis, and a yaw angle $\psi \in [0,2\pi)$ around the $z$-axis.
{\color{black}In this paper, we adopt an active extrinsic $x$-$y$-$z$ rotation convention to describe the physical orientation change of the antenna/array. Under this convention, the antenna/array is successively rotated about the fixed global $x$-, $y$-, and $z$-axes according to the right-hand rule.
Let $\bm{\theta}\triangleq[\phi,\theta,\psi]^T$ denote the rotation angle vector, and let $\mathbf{R}_x(\phi)$, $\mathbf{R}_y(\theta)$, and $\mathbf{R}_z(\psi)$ denote the elementary rotation matrices induced by rotations around the $x$-, $y$-, and $z$-axes, respectively.
Based on the active extrinsic $x$-$y$-$z$ rotation sequence, the rightmost matrix $\mathbf{R}_x(\phi)$ is applied first, followed by $\mathbf{R}_y(\theta)$ and $\mathbf{R}_z(\psi)$. The composite rotation matrix $\mathbf{R}(\bm{\theta})$ is therefore given by~\cite{Diebel2006Representing,3GPP20205G,Shao2025Distributed}
\begin{align}
	\label{deqn_ex1a}
    \mathbf{R}(\bm{\theta}) &= \mathbf{R}_z(\psi) \mathbf{R}_y(\theta) \mathbf{R}_x(\phi)\nonumber\\
    &= \begin{bmatrix}
         c_{\psi} & -s_{\psi} & 0\\
         s_{\psi} & c_{\psi} & 0\\
         0 & 0 & 1
       \end{bmatrix}
       \begin{bmatrix}
         c_{\theta} & 0 & s_{\theta}\\
         0 & 1 & 0\\
         -s_{\theta} & 0 & c_{\theta}
       \end{bmatrix}
       \begin{bmatrix}
         1 & 0 & 0\\
         0 & c_{\phi} & -s_{\phi}\\
         0 & s_{\phi} & c_{\phi}
       \end{bmatrix}\nonumber\\
    &=\begin{bmatrix}
         c_{\theta} c_{\psi} & s_{\phi} s_{\theta} c_{\psi} - c_{\phi} s_{\psi} & c_{\phi} s_{\theta} c_{\psi} + s_{\phi} s_{\psi}\\
         c_{\theta} s_{\psi} & s_{\phi} s_{\theta} s_{\psi} + c_{\phi} c_{\psi} & c_{\phi} s_{\theta} s_{\psi} - s_{\phi} c_{\psi}\\
         -s_{\theta} & s_{\phi} c_{\theta} & c_{\phi} c_{\theta}
    \end{bmatrix},
\end{align}
where we use $c_x \triangleq \cos(x)$ and $s_x \triangleq \sin(x)$ for notational simplicity.
Note that the composite rotation matrix $\mathbf{R}(\bm{\theta})$ maps a vector from the antenna’s/array’s local coordinate system to the global coordinate system, i.e., $\mathbf{v}_{\mathrm{global}}=\mathbf{R}(\bm{\theta})\mathbf{v}_{\mathrm{local}}$. Since $\mathbf{R}(\bm{\theta})$ is orthogonal, the inverse transformation is given by $\mathbf{v}_{\mathrm{local}}=\mathbf{R}^T(\bm{\theta})\mathbf{v}_{\mathrm{global}}$.}

\subsubsection{Antenna Rotation}
For an RA, the antenna can independently rotate in 3D space while keeping its position fixed.
The 3D spatial orientation of RA~$n$, $n\in \mathcal{N}\triangleq \{1,2,\dots,N\}$, can be uniquely determined by two non-parallel body-fixed vectors~\cite{Markley2014Fundamentals}:  
\begin{itemize}
	\item{Pointing vector $\vec{\mathbf{f}}_{\perp,n}\in \mathbb{R}^{3\times 1}$, denoting the antenna boresight direction (typically normal to the antenna line/plane);}
	\item{Reference vector $\vec{\mathbf{f}}_{\parallel,n}\in \mathbb{R}^{3\times 1}$, lying in the antenna line/plane and pointing to a fixed azimuth reference.} 
\end{itemize}
{\color{black}The pointing and reference vectors play distinct roles in the RA channel model. The pointing vector $\vec{\mathbf{f}}_{\perp,n}$ determines the boresight-dependent directional gain, while the reference vector $\vec{\mathbf{f}}_{\parallel,n}$ specifies the in-plane reference direction and is used to characterize the antenna’s polarization direction. Thus, the rotation model provides the geometric information required for both the gain-only and polarization-aware channel models provided in Section~\ref{Channel_Model}.}

{\bf 3D Antenna Rotation:}
In the 3D Cartesian coordinate system shown in Fig.~\ref{fig_rotation_1}(a), we initialize the pointing and reference vectors of each antenna to be parallel to the positive $x$- and $z$-axes, respectively, i.e., $\vec{\mathbf{f}}_{\perp,n}^{(0)} = \mathbf{e}_1$ and $\vec{\mathbf{f}}_{\parallel,n}^{(0)} = \mathbf{e}_3$, where $\mathbf{e}_1\triangleq [1,0,0]^T$ and $\mathbf{e}_3\triangleq [0,0,1]^T$.
Given the rotation angle vector $\bm{\theta}_n\triangleq [\phi_n,\theta_n,\psi_n]^T$, the pointing and reference vectors become
\begin{align}
	&\vec{\mathbf{f}}_{\perp,n} = \mathbf{R}(\bm{\theta}_n)\vec{\mathbf{f}}_{\perp,n}^{(0)}= \begin{bmatrix}
c_{\theta_n}c_{\psi_n}\\
c_{\theta_n}s_{\psi_n}\\
-s_{\theta_n}
\end{bmatrix},\label{deqn_ex2a}\\
    &\vec{\mathbf{f}}_{\parallel,n} = \mathbf{R}(\bm{\theta}_n)\vec{\mathbf{f}}_{\parallel,n}^{(0)}= \begin{bmatrix}
c_{\phi_n}s_{\theta_n}c_{\psi_n} + s_{\phi_n}s_{\psi_n}\\
c_{\phi_n}s_{\theta_n}s_{\psi_n} - s_{\phi_n}c_{\psi_n}\\
c_{\phi_n}c_{\theta_n}
\end{bmatrix},\label{deqn_ex3a}
\end{align}
respectively, where we have $\|\vec{\mathbf{f}}_{\perp,n}\| = 1$ and $\|\vec{\mathbf{f}}_{\parallel,n}\| = 1$ due to normalization, and $\vec{\mathbf{f}}_{\perp,n}$ is perpendicular to $\vec{\mathbf{f}}_{\parallel,n}$, i.e., $\vec{\mathbf{f}}_{\perp,n}^T \vec{\mathbf{f}}_{\parallel,n} = 0$.

{\bf 2D Antenna Rotation:} When the antenna directional gain pattern is rotationally symmetric, rotation around the boresight axis does not affect the gain pattern.
{\color{black}If we only focus on the antenna directional gain in RA channel modeling, the above 3D antenna rotation model can be simplified by omitting self-rotation around the boresight axis, i.e., $\phi_n=0, \forall n\in \mathcal{N}$.}
As a result, the pointing and reference vectors for the 2D antenna rotation are given by
\begin{align}
	\vec{\mathbf{f}}_{\perp,n} =
    \begin{bmatrix}
c_{\theta_n}c_{\psi_n}\\
c_{\theta_n}s_{\psi_n}\\
-s_{\theta_n}
\end{bmatrix},\quad \vec{\mathbf{f}}_{\parallel,n} = \begin{bmatrix}
s_{\theta_n}c_{\psi_n}\\
s_{\theta_n}s_{\psi_n}\\
c_{\theta_n}
\end{bmatrix},\label{deqn_ex4a}
\end{align}
respectively. Alternatively, for ease of representing the pointing and reference vectors in the global coordinate system, they can also be parameterized by a zenith angle $\theta_{\mathrm{z},n}$ and an azimuth angle $\theta_{\mathrm{a},n}$,  as in~\cite{Wu2025Modeling,Zheng2026Rotatable}:
\begin{align}
	\vec{\mathbf{f}}_{\perp,n} =
    \begin{bmatrix}
c_{\theta_{\mathrm{z},n}}\\
s_{\theta_{\mathrm{z},n}} c_{\theta_{\mathrm{a},n}}\\
s_{\theta_{\mathrm{z},n}} s_{\theta_{\mathrm{a},n}}
\end{bmatrix},\quad \vec{\mathbf{f}}_{\parallel,n} = \begin{bmatrix}
-s_{\theta_{\mathrm{z},n}}\\
c_{\theta_{\mathrm{z},n}} c_{\theta_{\mathrm{a},n}}\\
c_{\theta_{\mathrm{z},n}} s_{\theta_{\mathrm{a},n}}
\end{bmatrix}.\label{deqn_ex10a}
\end{align}
As shown in Fig.~\ref{fig_rotation_1}(e), the zenith angle $\theta_{\mathrm{z},n}$ denotes the angle between the pointing vector and the $x$-axis, and the azimuth angle $\theta_{\mathrm{a},n}$ denotes the angle between the projection of the pointing vector onto the $y$-$z$ plane and the $y$-axis.
{\color{black}In such 2D antenna rotation models, if polarization mismatch is ignored, compensated, or approximated as a constant factor, only the pointing vector $\vec{\mathbf{f}}_{\perp,n}$ is needed, since the directional gain depends only on the angular separation between the signal propagation direction and the antenna boresight (i.e., the maximum-gain direction of the mainlobe). This directional gain-only simplification is widely used for tractable RA channel modeling. However, when polarization effects are considered, the reference vector $\vec{\mathbf{f}}_{\parallel,n}$ must be retained to evaluate the polarization direction and the resulting polarization matching factor.}

{\bf One-Dimensional (1D) Antenna Rotation:} 
In special cases where the antenna performs only 1D rotation, the pointing and reference vectors are simplified to the following forms~\cite{Zhang2026RotatableAntenna,Zhang2025Rotatable}.  
 \begin{itemize}
	\item Rotation around the $x$-axis shown in Fig.~\ref{fig_rotation_1}(b): $\theta_n=\psi_n=0$,
    which gives the pointing vector $\vec{\mathbf{f}}_{\perp,n}=[1,0,0]^T$ and the reference vector $\vec{\mathbf{f}}_{\parallel,n}=[0,-s_{\phi_n},c_{\phi_n}]^T$.
	\item Rotation around the $y$-axis shown in Fig.~\ref{fig_rotation_1}(c): $\phi_n=\psi_n=0$,
    which gives the pointing vector $\vec{\mathbf{f}}_{\perp,n}=[c_{\theta_n},0,-s_{\theta_n}]^T$ and the reference vector $\vec{\mathbf{f}}_{\parallel,n}=[s_{\theta_n},0,c_{\theta_n}]^T$.  
    \item Rotation around the $z$-axis shown in Fig.~\ref{fig_rotation_1}(d): $\phi_n=\theta_n=0$, which gives the pointing vector
    $\vec{\mathbf{f}}_{\perp,n}=[c_{\psi_n},s_{\psi_n},0]^T$ and the reference vector $\vec{\mathbf{f}}_{\parallel,n}=[0,0,1]^T$.
\end{itemize}


\begin{figure}[!t]\centering
	\includegraphics[width=2.8in]{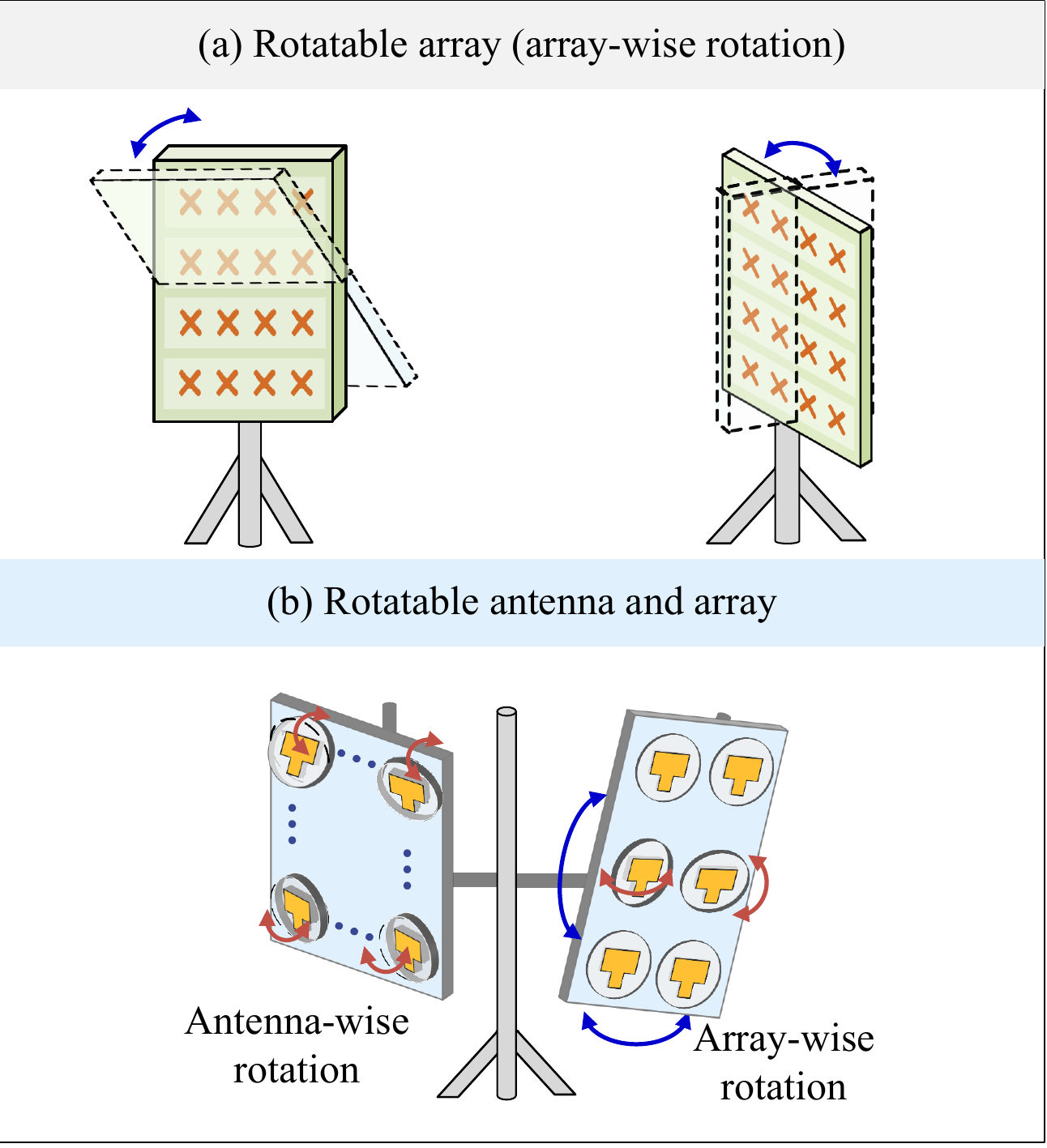}
	\caption{Illustration of two types of rotations in RA arrays. (a) Array-wise rotation. (b) Joint antenna-wise and array-wise rotation.}
	\label{RA_array_rotation}
\end{figure}

{\color{black}
{\bf Energy Consumption Model of Antenna Rotation:} 
For each RA, antenna rotation inevitably incurs mechanical actuation energy. To characterize this energy cost, we consider the physical actuation required to steer the RA orientation through an angular distance $\Theta_n$ within a rotation duration $t_0$. For RA~$n$, the angular distance between its initial orientation $\vec{\mathbf{f}}_{\perp,n}^{(0)}$ and its target orientation $\vec{\mathbf{f}}_{\perp,n}$ is geometrically given by $\Theta_n = \arccos\big(\vec{\mathbf{f}}_{\perp,n}^T \vec{\mathbf{f}}_{\perp,n}^{(0)}\big)$. 
To model the actuation dynamics, we assume that each RA is driven by an armature-controlled direct current (DC) servo motor operating under a time-optimal constant-acceleration/constant-deceleration profile. From rotational kinematics and Newton’s second law, covering the angular distance $\Theta_n$ within $t_0$ seconds requires a constant motor torque $\tau_n = J \frac{4\Theta_n}{t_0^2}$, where $J$ is the structural moment of inertia\cite{Krause2013Analysis}. To generate this torque using a DC motor with torque constant $K_t$, the required armature current magnitude is $|i_n(t)| = \frac{\tau_n}{K_t} = \frac{4 J \Theta_n}{K_t t_0^2}$ \cite{Fitzgerald2003Electric}. Since the dominant electrical energy consumption arises from thermal dissipation across the armature resistance $R_0$, the total rotational energy of RA~$n$ follows directly from Joule’s law:
\begin{align}
\hspace{-0.3cm}	E_{\text{rot}, n}(\Theta_n, t_0) = \int_{0}^{t_0} i_n^2(t) R_0 \, dt = \frac{16 J^2 R_0}{K_t^2} \frac{\Theta_n^2}{t_0^3} = \kappa \frac{\Theta_n^2}{t_0^3},\label{Energy_RA}
\end{align}
where $\kappa \triangleq \frac{16 J^2 R_0}{K_t^2}$ is a consolidated hardware-dependent coefficient. This expression shows that the required rotational energy grows quadratically with the angular distance and decreases cubically with the rotation duration. Intuitively, large-angle steering and fast boresight updates may incur non-negligible energy consumption. It should be noted that energy modeling for RA systems remains an open research topic. The proposed model provides an initial characterization of the mechanical actuation energy required for physically rotating RAs. In practice, the actual energy consumption may also depend on the update frequency, tracking dynamics, platform type, control strategy, and hardware implementation. Moreover, electronically driven RAs may exhibit different energy-consumption characteristics compared to mechanically driven RAs. Therefore, the proposed model is intended as an initial step toward incorporating actuation energy into RA system design.}

\subsubsection{Array Rotation} 
As illustrated in Fig.~\ref{RA_array_rotation}, rotating
the entire array not only changes the orientation of the antennas but also alters their spatial positions relative to the array center. 
Let $\bm{\theta}_{\mathrm{array}}\triangleq [\phi_{\mathrm{array}},\theta_{\mathrm{array}},\psi_{\mathrm{array}}]^T$ denote the rotation angle vector of the array at the BS and let $\mathbf{q}_{\mathrm{B},n}\in \mathbb{R}^{3\times 1}$ denote the initial position of RA~$n$. Accordingly, the position of RA~$n$ after array rotation is given by
\begin{align}
	\label{deqn_ex5a}
	\tilde{\mathbf{q}}_{\mathrm{B},n} = \mathbf{R}(\bm{\theta}_{\mathrm{array}})\big(\mathbf{q}_{\mathrm{B},n} - \mathbf{q}_{\mathrm{B},0}\big) + \mathbf{q}_{\mathrm{B},0}, \quad \forall n\in \mathcal{N},
\end{align}
where $\mathbf{q}_{\mathrm{B},0} \in \mathbb{R}^{3\times 1}$ denotes the position of the array rotation center at the BS.

If all antennas share identical initial orientations and no independent antenna rotation is applied during array rotation as shown in Fig.~\ref{RA_array_rotation}(a), then we have $\bm{\theta}_n=\mathbf{0},\ \forall n\in\mathcal{N}$. In this case, all antennas still have the same pointing and reference vectors after array rotation, which can be respectively expressed as
\begin{align}
	\vec{\mathbf{f}}_{\perp,n} &= \mathbf{R}(\bm{\theta}_{\mathrm{array}}) \vec{\mathbf{f}}_{\perp,n}^{(0)}=\mathbf{R}(\bm{\theta}_{\mathrm{array}}) \mathbf{e}_1,\; \forall n\in \mathcal{N},\label{deqn_ex6a}\\
    \vec{\mathbf{f}}_{\parallel,n} &= \mathbf{R}(\bm{\theta}_{\mathrm{array}}) \vec{\mathbf{f}}_{\parallel,n}^{(0)}=\mathbf{R}(\bm{\theta}_{\mathrm{array}}) \mathbf{e}_3,\; \forall n\in \mathcal{N}.\label{deqn_ex7a}
\end{align}
However, if independent antenna rotation is performed simultaneously during array rotation as shown in Fig.~\ref{RA_array_rotation}(b), the rotation of both the array and the individual antennas needs to be considered~\cite{Wang2026Flexible}, and the pointing and reference vectors of RA~$n$ can be expressed as
\begin{align}
	\vec{\mathbf{f}}_{\perp,n} &= \mathbf{R}(\bm{\theta}_{\mathrm{array}}) \mathbf{R}(\bm{\theta}_n) \vec{\mathbf{f}}_{\perp,n}^{(0)}=\mathbf{R}(\bm{\theta}_{\mathrm{array}}) \mathbf{R}(\bm{\theta}_n) \mathbf{e}_1,\label{deqn_ex8a}\\
    \vec{\mathbf{f}}_{\parallel,n} &= \mathbf{R}(\bm{\theta}_{\mathrm{array}}) \mathbf{R}(\bm{\theta}_n)\vec{\mathbf{f}}_{\parallel,n}^{(0)}=\mathbf{R}(\bm{\theta}_{\mathrm{array}}) \mathbf{R}(\bm{\theta}_n)\mathbf{e}_3,\label{deqn_ex9a}
\end{align}
respectively. Based on the general 3D array rotation defined in \eqref{deqn_ex6a}--\eqref{deqn_ex9a}, the pointing and reference vectors for specific 2D and 1D array rotation special cases can also be derived in a similar way as discussed for individual antenna rotation.

{\color{black}
Regarding the mechanical energy consumption of array rotation, it follows the same physical principle as individual RA rotation and can be evaluated using the general model in \eqref{Energy_RA}, by substituting the array’s equivalent angular displacement and its macroscopic hardware coefficient. For joint rotation (see Fig.~\ref{RA_array_rotation}(b)), the total mechanical energy consumption is the sum of the array-level actuation energy and the individual antenna-level rotation energies.}


\subsection{Channel Model}\label{Channel_Model}
Based on the antenna rotation model in Section~\ref{Rotation_Model}, we present the channel model for RA systems in this subsection. We first introduce two commonly used antenna directional gain patterns that characterize the radiation energy distribution. We then construct the general near-field line-of-sight (LoS) channel model for narrowband systems by incorporating the RA directional gain, and provide its far-field approximation. Finally, we extend the model to multipath propagation, wideband transmission, and polarization effects, followed by a brief discussion of potential extensions to other channel models. 

\begin{figure}[!t]
	\centering
	\hspace{-0.25cm}\subfloat[Incident/departure angles for directional gain pattern]{
		\hspace{-0.4cm}\includegraphics[width=1.8in]{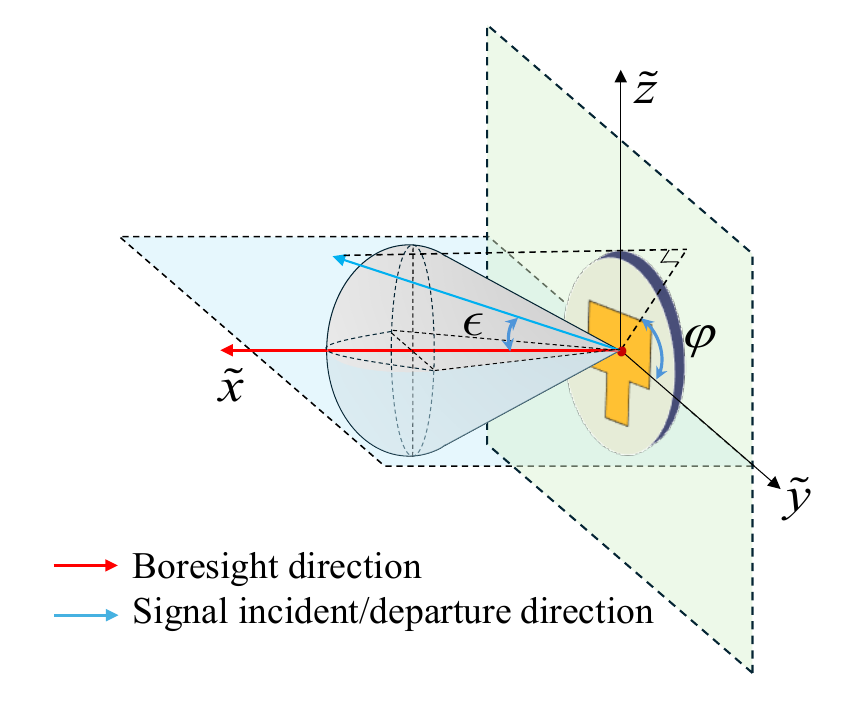}}\hspace{+0.01cm}
	\subfloat[Directional gain patterns for different values of $\rho$]{
		\includegraphics[width=1.8in]{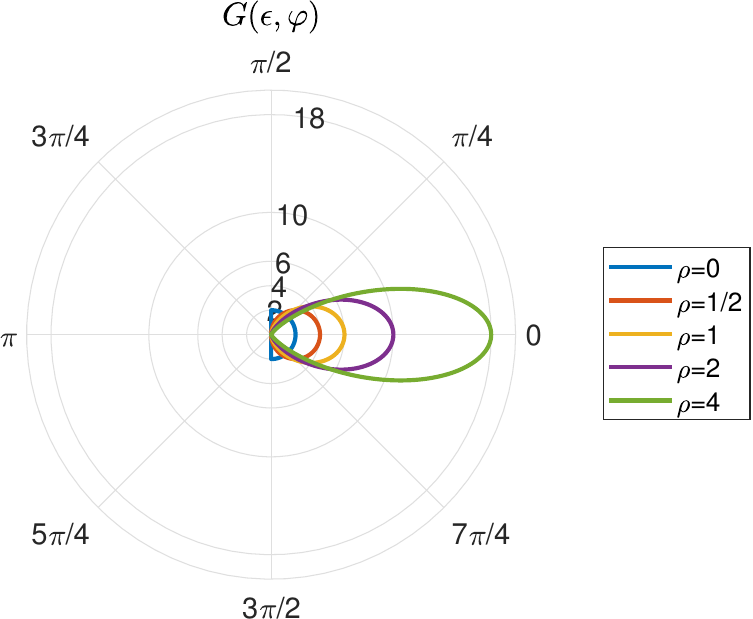}}\hspace{-0.4cm}
	\caption{Illustration of antenna directional gain pattern.}
	\label{fig_pattern}
\end{figure}

\subsubsection{Antenna Directional Gain Pattern}
Given the orientation of each RA, the effective antenna gain depends on both the signal arrival/departure direction and the underlying directional gain pattern.
We consider two commonly used models:
\begin{itemize}
	\item{\bf{Cosine pattern model}}: For an antenna with a single narrow mainlobe and negligible sidelobes, the directional gain pattern can be modeled as~\cite{Balanis1996Antenna}
\begin{align}
	\label{deqn_ex1b}
	\hspace{-0.3cm}{G(\epsilon,\varphi)} =
	\begin{cases}
		{G_{\mathrm{\max}}^{\mathrm{cos}} \cos^{2\rho} (\epsilon),}&{\epsilon \in [0,\frac{\pi}{2}), \varphi \in [0,2\pi)} \\
		{0,}&{\text{otherwise},}
	\end{cases}\hspace{-0.2cm}
\end{align}
where $(\epsilon,\varphi)$ is a pair of incident/departure angles of the signal with respect to the antenna's current boresight direction as illustrated in Fig.~\ref{fig_pattern}(a), the parameter $\rho \geq 0$ reflects the antenna directivity and mainlobe beamwidth, and is determined by the radiation characteristic of the adopted antenna via fitting to measured or full-wave simulated data. In addition, $G_{\mathrm{\max}}^{\mathrm{cos}}$ is the maximum gain in the antenna boresight direction, and is given by $G_{\mathrm{\max}}^{\mathrm{cos}}=2(2\rho +1)$ to meet the law of power conservation. Moreover, as shown in Fig.~\ref{fig_pattern}(b), a larger $\rho$ corresponds to stronger directivity, yielding higher antenna boresight gain and a narrower mainlobe.
	\item{\bf{3GPP element model}:} Another practical and widely adopted model is the 3GPP directional gain pattern~\cite{3GPP20205G,Rebato2019Stochastic}, given by
\begin{align}
	\label{deqn_ex2b}
	G(\epsilon,\varphi) = G_{\max}^{\mathrm{3GPP}}\hspace{-0.1cm} -\hspace{-0.1cm} \min \{-\hspace{-0.1cm}\left(G_{e,H}(\varphi) + G_{e,V}(\epsilon)\right),A_{\max}\},
\end{align}
where $G_{\max}^{\mathrm{3GPP}}$ is the maximum gain and $A_{\max}$ is the front-to-back attenuation limit. The horizontal and vertical components are respectively given by
\begin{subequations}\label{deqn_ex3b}
    \begin{align}
	    G_{e,H}(\varphi) &= - \min \left\{12\left(\frac{\varphi}{\varphi_{\mathrm{3dB}}}\right)^2,A_{\max}\right\},\label{deqn_ex3b1}\\
        G_{e,V}(\epsilon) &= - \min \left\{12\left(\frac{\epsilon}{\epsilon_{\mathrm{3dB}}}\right)^2,A_{\mathrm{side}}\right\},\label{deqn_ex3b2}
    \end{align}
\end{subequations}
where $\varphi_{\mathrm{3dB}}$ and $\epsilon_{\mathrm{3dB}}$ denote the $3$-dB beamwidths (typically 65°), and $A_{\mathrm{side}}$ specifies the vertical sidelobe suppression level.
Note that all gain and attenuation terms in \eqref{deqn_ex2b} and \eqref{deqn_ex3b} are expressed in dB.
\end{itemize}
For the cosine pattern model in \eqref{deqn_ex1b}, which is rotationally symmetric, the directional gain depends only on the angular offset $\epsilon$ between the signal direction and the antenna boresight.
By contrast, for the 3GPP element model in \eqref{deqn_ex2b}, since the pattern is generally irregular, the directional gain depends on the specific signal direction relative to the antenna boresight, and thus both $\epsilon$ and $\varphi$ need to be taken into account.
Nevertheless, both directional gain patterns above characterize the direction-dependent magnitude variation of the electric field, while the phase in instantaneous channel modeling can be absorbed into the propagation channel coefficient.

{\color{black}It is worth noting that the above cosine and 3GPP pattern models only serve as representative examples of directional gain patterns. In practice, the directional gain pattern of an antenna is determined by its physical structure, material properties, feeding mechanism, and operating frequency, and can be expressed as a function of the local incident/departure angles $(\epsilon,\varphi)$. Thus, analytic radiation patterns of common antennas (such as dipoles, monopoles, horn antennas, and microstrip patches) can be incorporated into the proposed modeling framework by substituting $G(\epsilon,\varphi)$ with the corresponding analytic, measured, or full-wave simulated pattern. For a rotated RA, the global signal direction is first transformed into the antenna’s local coordinate system using the rotation model in Section~\ref{Rotation_Model}, yielding the local angles $(\epsilon,\varphi)$, which are then used to evaluate the radiation pattern. Consequently, the proposed RA channel modeling framework is applicable to both rotationally symmetric and asymmetric practical antenna patterns.}


\subsubsection{Near-Field LoS Channel Model}\label{Near_field_model} According to the directional gain patterns in \eqref{deqn_ex1b} and \eqref{deqn_ex2b}, the effective antenna gain at a spatial point depends on the incident/departure angle pair $(\epsilon,\varphi)$ defined with respect to the antenna’s current orientation.
Let $\mathbf{q}_{\mathrm{U},k}\in \mathbb{R}^{3\times 1}$ denote the position of user~$k$ and  $\vec{\mathbf{q}}_{\mathrm{U},n,k}\triangleq \frac{\mathbf{q}_{\mathrm{U},k} - \mathbf{q}_{\mathrm{B},n}}{\|\mathbf{q}_{\mathrm{U},k} - \mathbf{q}_{\mathrm{B},n}\|}$ denote the direction vector from RA~$n$ to user~$k$ with $k\in \mathcal{K}\triangleq \{1,2,\dots,K\}$. Moreover, we define the orientation vector of RA~$n$ as $\vec{\mathbf{f}}_n \triangleq \left[\vec{\mathbf{f}}_{\perp,n}^T,\vec{\mathbf{f}}_{\parallel,n}^T\right]^T\in \mathbb{R}^{6\times 1}$ by concatenating the pointing and reference vectors, which can uniquely determine the antenna orientation. 
Accordingly, the incident/departure angles of user~$k$'s propagation direction with respect to RA~$n$ are derived as
\begin{subequations}\label{deqn_ex4b}
	\begin{align}
		\epsilon(\vec{\mathbf{f}}_n,\vec{\mathbf{q}}_{\mathrm{U},n,k}) & = \arccos\left(\vec{\mathbf{q}}_{\mathrm{U},n,k}^T \vec{\mathbf{f}}_{\perp,n}\right), \label{deqn_ex4b1}\\
		\varphi(\vec{\mathbf{f}}_n,\vec{\mathbf{q}}_{\mathrm{U},n,k}) & = 	\operatorname{arctan2}\left(\vec{\mathbf{q}}_{\mathrm{U},n,k}^T \vec{\mathbf{f}}_{\parallel,n}, \vec{\mathbf{q}}_{\mathrm{U},n,k}^T \tilde{\mathbf{e}}_2\right), \label{deqn_ex4b2}
	\end{align}
\end{subequations}
where $\tilde{\mathbf{e}}_2 \triangleq \mathbf{R}(\bm{\theta}_n)\mathbf{e}_2$ denotes the rotated $\tilde{y}$-axis with $\mathbf{e}_2\triangleq [0,1,0]^T$.
Therefore, the directional gain from RA~$n$ to user~$k$ can be modeled as
\begin{align}
	\label{deqn_ex5b}
    g_k\left(\vec{\mathbf{f}}_n\right) = G\left(\epsilon(\vec{\mathbf{f}}_n,\vec{\mathbf{q}}_{\mathrm{U},n,k}),\varphi(\vec{\mathbf{f}}_n,\vec{\mathbf{q}}_{\mathrm{U},n,k})\right).
\end{align}
This expression highlights that the antenna rotation (i.e., by changing $\vec{\mathbf{f}}_{\perp,n}$ and/or $\vec{\mathbf{f}}_{\parallel,n}$) directly controls the effective directional gain.
For analytical convenience, the cosine pattern model in \eqref{deqn_ex1b}, which is rotationally symmetric around the boresight, is widely adopted~\cite{Wu2025Modeling,Zheng2026Rotatable}. 
Under this practically useful model, \eqref{deqn_ex5b} simplifies to
\begin{align}
	\label{deqn_ex6b}
    g_k\left(\vec{\mathbf{f}}_n\right) = G_{\max}^{\mathrm{cos}}\left[\vec{\mathbf{q}}_{\mathrm{U},n,k}^T \vec{\mathbf{f}}_{\perp,n}\right]^{2\rho}_{+},
\end{align}
where $[x]_{+} \triangleq \max\{x,0\}$. It can be inferred that the directional gain in \eqref{deqn_ex6b} increases as the angle between the pointing vector $\vec{\mathbf{f}}_{\perp,n}$ and the user direction $\vec{\mathbf{q}}_{\mathrm{U},n,k}$ decreases. Intuitively, the maximum directional gain $G_{\max}^{\mathrm{cos}}$ can be achieved when the RA boresight direction is aligned with the user direction, i.e., $\vec{\mathbf{f}}_{\perp,n} = \vec{\mathbf{q}}_{\mathrm{U},n,k}$.
{\color{black} It is important to note that \eqref{deqn_ex5b} and \eqref{deqn_ex6b} describe only the directional gain variation.  This gain-only model is accurate when the transmit and receive polarizations are well aligned or when polarization compensation is applied. When polarization mismatch is non-negligible, the reference vector $\vec{\mathbf{f}}_{\parallel,n}$ must also be used to evaluate the polarization matching factor, as will be discussed in Section~\ref{Polar_Model}.}

Accordingly, the near-field LoS channel between RA~$n$ and user~$k$ is expressed as
\begin{align}
	\label{deqn_ex7b}
    h_{\mathrm{LoS},k}\left(\vec{\mathbf{f}}_n\right) = \frac{\sqrt{\beta_0}} {d_{n,k}} \sqrt{g_k\left(\vec{\mathbf{f}}_n\right)}\; e^{-j\frac{2\pi}{\lambda}d_{n,k}},
\end{align}
where $\beta_0 \triangleq \left(\frac{\lambda}{4\pi}\right)^2$ denotes the channel power gain at a reference distance of $d_0 = 1$~meter~(m) with $\lambda$ being the carrier wavelength, $d_{n,k}\triangleq \|\mathbf{q}_{\mathrm{U},k} - \mathbf{q}_{\mathrm{B},n}\|$ is the distance between RA~$n$ and user~$k$, and  $\beta_{n,k}\triangleq \frac{\sqrt{\beta_0}} {d_{n,k}}$ represents the propagation coefficient.
By stacking the channel coefficients of all RAs in the array, the near-field LoS channel vector between the BS and user~$k$, denoted by $\mathbf{h}_{\mathrm{LoS},k}\left(\mathbf{F}\right)\in \mathbb{C}^{N\times 1}$, is given by
\begin{align}
	\label{deqn_ex8b}
    \hspace{-0.25cm}\mathbf{h}_{\mathrm{LoS},k}\left(\mathbf{F}\right) = \left[h_{\mathrm{LoS},k}(\vec{\mathbf{f}}_1),h_{\mathrm{LoS},k}(\vec{\mathbf{f}}_2),\dots,h_{\mathrm{LoS},k}(\vec{\mathbf{f}}_N)\right]^T\hspace{-0.1cm},\hspace{-0.15cm}
\end{align}
where $\mathbf{F}\triangleq \left[\vec{\mathbf{f}}_1,\vec{\mathbf{f}}_2,\dots,\vec{\mathbf{f}}_N\right]\in \mathbb{R}^{6\times N}$ denotes the stacked orientation matrix for all RAs.

\subsubsection{Far-Field LoS Channel Model}
In many practical scenarios, since the array aperture is much smaller than the link distance, the impinging wavefronts can be approximated as uniform plane waves.
Under the far-field condition, the direction vectors and propagation coefficients across the RAs satisfy $\vec{\mathbf{q}}_{\mathrm{U},1,k}\approx \vec{\mathbf{q}}_{\mathrm{U},2,k}\approx \dots \approx \vec{\mathbf{q}}_{\mathrm{U},N,k} \triangleq \vec{\mathbf{q}}_{\mathrm{U},k}$ and $\beta_{1,k}\approx \beta_{2,k}\approx \dots \approx \beta_{N,k} \triangleq \beta_{k}$.
Taking RA~1 as the reference, the array response vector is thus given by 
\begin{align}
	\label{deqn_ex9b}
    \hspace{-0.3cm}&\mathbf{a}_k\left(N,\vec{\mathbf{q}}_{\mathrm{U},k}\right) \nonumber\\
    \hspace{-0.3cm}=& \left[1,e^{j\frac{2\pi}{\lambda}(\mathbf{q}_{\mathrm{B},2} - \mathbf{q}_{\mathrm{B},1})^T \vec{\mathbf{q}}_{\mathrm{U},k}},\dots,e^{j\frac{2\pi}{\lambda}(\mathbf{q}_{\mathrm{B},N} - \mathbf{q}_{\mathrm{B},1})^T \vec{\mathbf{q}}_{\mathrm{U},k}}\right]^T\hspace{-0.1cm}.\hspace{-0.25cm}
\end{align}
For a uniform planar array (UPA), the above response vector can be transformed into
\begin{align}
	\label{deqn_ex10b}
    \mathbf{a}_k\left(N,\vec{\mathbf{q}}_{\mathrm{U},k}\right) = \mathbf{a}_{y,k}\left(N_y,\vec{\mathbf{q}}_{\mathrm{U},k}\right) \otimes \mathbf{a}_{z,k}\left(N_z,\vec{\mathbf{q}}_{\mathrm{U},k}\right),
\end{align}
where $N_y$ and $N_z$ are the numbers of RAs along the $y$- and $z$-axes, respectively, and
\begin{subequations}\label{deqn_ex11b}
\begin{align}
    \mathbf{a}_{y,k}\left(N_y,\vec{\mathbf{q}}_{\mathrm{U},k}\right) &\hspace{-0.05cm}=\hspace{-0.05cm} \left[1,e^{j\frac{2\pi \Delta_d}{\lambda} \vec{\mathbf{q}}_{\mathrm{U},k}^T \mathbf{e}_2},\dots,e^{j\frac{2\pi \Delta_d}{\lambda} (N_y - 1)\vec{\mathbf{q}}_{\mathrm{U},k}^T \mathbf{e}_2}\right]^T,\label{deqn_ex11b1}\\
    \mathbf{a}_{z,k}\left(N_z,\vec{\mathbf{q}}_{\mathrm{U},k}\right) &\hspace{-0.05cm}=\hspace{-0.05cm} \left[1,e^{j\frac{2\pi \Delta_d}{\lambda} \vec{\mathbf{q}}_{\mathrm{U},k}^T \mathbf{e}_3},\dots,e^{j\frac{2\pi \Delta_d}{\lambda} (N_z - 1)\vec{\mathbf{q}}_{\mathrm{U},k}^T \mathbf{e}_3}\right]^T,\label{deqn_ex11b2}
\end{align}
\end{subequations}
are the 1D steering vector functions for uniform linear arrays (ULAs) along the $y$- and $z$-axes, respectively, with $\Delta_d$ denoting the antenna spacing.

Under the far-field condition, the corresponding directional gain vector is
\begin{align}
	\label{deqn_ex12b}
    \hspace{-0.25cm}\mathbf{g}_k(\mathbf{F},\vec{\mathbf{q}}_{\mathrm{U},k}) =\hspace{-0.1cm} \begin{bmatrix}
    G\left(\epsilon(\vec{\mathbf{f}}_1,\vec{\mathbf{q}}_{\mathrm{U},k}),\varphi(\vec{\mathbf{f}}_1,\vec{\mathbf{q}}_{\mathrm{U},k})\right)\\
    G\left(\epsilon(\vec{\mathbf{f}}_2,\vec{\mathbf{q}}_{\mathrm{U},k}),\varphi(\vec{\mathbf{f}}_2,\vec{\mathbf{q}}_{\mathrm{U},k})\right)\\
    \vdots\\
    G\left(\epsilon(\vec{\mathbf{f}}_N,\vec{\mathbf{q}}_{\mathrm{U},k}),\varphi(\vec{\mathbf{f}}_N,\vec{\mathbf{q}}_{\mathrm{U},k})\right)
\end{bmatrix}\hspace{-0.1cm}\in \mathbb{R}^{N\times 1}.\hspace{-0.1cm}
\end{align}
In particular, the directional gain vector under the cosine pattern model in \eqref{deqn_ex1b} is given by
\begin{align}
	\label{deqn_ex13b}
    \mathbf{g}_k (\mathbf{F},\vec{\mathbf{q}}_{\mathrm{U},k}) = G_{\max}^{\mathrm{cos}} \begin{bmatrix}
    \left[\vec{\mathbf{q}}_{\mathrm{U},k}^T \vec{\mathbf{f}}_{\perp,1}\right]^{2\rho}_{+}\\
    \left[\vec{\mathbf{q}}_{\mathrm{U},k}^T \vec{\mathbf{f}}_{\perp,2}\right]^{2\rho}_{+}\\
    \vdots\\
    \left[\vec{\mathbf{q}}_{\mathrm{U},k}^T \vec{\mathbf{f}}_{\perp,N}\right]^{2\rho}_{+}
\end{bmatrix}\in \mathbb{R}^{N\times 1}.
\end{align}

Accordingly, the far-field LoS channel vector between the BS and user~$k$ can be expressed as
\begin{align} 
	\label{deqn_ex14b}  \hspace{-0.5cm}\mathbf{h}_k\left(\mathbf{F}\right)\hspace{-0.05cm}=\hspace{-0.05cm} \beta_k e^{-j\frac{2\pi}{\lambda}d_{1,k}}\hspace{-0.1cm} \left(\operatorname{diag}(\mathbf{g}_k(\mathbf{F},\vec{\mathbf{q}}_{\mathrm{U},k}))\right)^{\frac{1}{2}} \hspace{-0.1cm}\mathbf{a}_k\left(N,\vec{\mathbf{q}}_{\mathrm{U},k}\right).\hspace{-0.3cm}
\end{align}
It can be observed that the user direction affects both the directional gain and array response vectors.
{\color{black}The above far-field model can be viewed as an approximation of the general near-field model in Section~\ref{Near_field_model}.  
It is accurate when the link distance is sufficiently larger than the array aperture, so that the propagation distances, directions, and path losses across the RAs can be treated as being identical, respectively. In contrast, for large-aperture or high-frequency RA arrays operating within the Rayleigh distance, the near-field model needs to be used, since different RAs may experience different distances, phases, directions, and directional gains.}

\subsubsection{Multipath Channel Model}
For scattering environments, we adopt a geometric propagation model to characterize the multipath channel.
We assume that $Q$ scatterer clusters are distributed in the propagation environment, 
where the position of cluster $q$ is represented by $\mathbf{q}_{\mathrm{C},q}\in \mathbb{R}^{3\times 1}$ with $q\in \mathcal{Q}\triangleq \{1,2,\dots,Q\}$.
Similar to \eqref{deqn_ex5b}, the directional gain from RA~$n$ to cluster $q$ is expressed as
\begin{align}
	\label{deqn_ex15b}
    \tilde{g}_q\left(\vec{\mathbf{f}}_n\right) = G\left(\epsilon(\vec{\mathbf{f}}_n,\vec{\mathbf{q}}_{\mathrm{C},n,q}),\varphi(\vec{\mathbf{f}}_n,\vec{\mathbf{q}}_{\mathrm{C},n,q})\right),
\end{align}
where $\vec{\mathbf{q}}_{\mathrm{C},n,q}\triangleq \frac{\mathbf{q}_{\mathrm{C},q} - \mathbf{q}_{\mathrm{B},n}}{\|\mathbf{q}_{\mathrm{C},q} - \mathbf{q}_{\mathrm{B},n}\|}$ is the direction vector from RA~$n$ to scatterer cluster $q$.
Under the cosine pattern model in \eqref{deqn_ex1b}, the directional gain from RA~$n$ to cluster $q$ reduces to
\begin{align}
	\label{deqn_ex16b}
    \tilde{g}_q\left(\vec{\mathbf{f}}_n\right) = G_{\max}^{\mathrm{cos}}\left[\vec{\mathbf{q}}_{\mathrm{C},n,q}^T \vec{\mathbf{f}}_{\perp,n}\right]^{2\rho}_{+}.
\end{align}
Then, the non-line-of-sight (NLoS) channel between RA~$n$ and user~$k$ is modeled as
\begin{align}
	\label{deqn_ex17b}
    h_{\mathrm{NLoS},k}\left(\vec{\mathbf{f}}_n\right) = \sum_{q=1}^{Q}{\sigma_q\frac{\beta_0\sqrt{\tilde{g}_q(\vec{\mathbf{f}}_n)}}{\tilde{d}_{n,q} \bar{d}_{k,q} }\;e^{-j\frac{2\pi}{\lambda}(\tilde{d}_{n,q}+\bar{d}_{k,q})}},
\end{align}
where $\sigma_q$ denotes the radar cross section (RCS) of cluster $q$, and $\tilde{d}_{n,q}\triangleq \|\mathbf{q}_{\mathrm{C},q} - \mathbf{q}_{\mathrm{B},n}\|$ and $\bar{d}_{k,q}\triangleq \|\mathbf{q}_{\mathrm{C},q} - \mathbf{q}_{\mathrm{U},k}\|$ denote the distances from RA~$n$ to scatterer cluster~$q$ and from scatterer cluster~$q$ to user~$k$, respectively.
Thus, by superimposing the LoS and NLoS components, the overall multipath channel between the BS and user~$k$ is given by
\begin{align}
	\label{deqn_ex18b}
    \mathbf{h}_{\mathrm{MP},k}\left(\mathbf{F}\right) = \mathbf{h}_{\mathrm{LoS},k}\left(\mathbf{F}\right) + \mathbf{h}_{\mathrm{NLoS},k}\left(\mathbf{F}\right)\in \mathbb{C}^{N\times 1}
\end{align}
with
\begin{align}
	\label{deqn_ex25b}
    \mathbf{h}_{\mathrm{NLoS},k}\left(\mathbf{F}\right)=\left[h_{\mathrm{NLoS},k}(\vec{\mathbf{f}}_1),h_{\mathrm{NLoS},k}(\vec{\mathbf{f}}_2),\dots,h_{\mathrm{NLoS},k}(\vec{\mathbf{f}}_N)\right]^T
\end{align}
being the NLoS channel vector.

\subsubsection{Wideband Channel Model}\label{Wideband_channel}
For RA-enabled wideband communications, to effectively exploit the large bandwidth resources and overcome frequency-selective fading, we consider an orthogonal frequency-division multiplexing (OFDM) system with bandwidth $B$ and $L$ subcarriers. 
Let $\tau _{n,k,0} = \frac{d_{n,k}}{c}$ and $\tau_{n,k,q} = \frac{\tilde{d}_{n,q} + \bar{d}_{k,q}}{c}$ denote the propagation delay of the ``RA~$n$-user~$k$'' and ``RA~$n$-cluster~$q$-user~$k$'' links, respectively, with $c$ being the speed of light.
Accordingly, the space-time baseband equivalent channel impulse response between RA~$n$ and user~$k$ is given by
\begin{align}
	\label{deqn_ex19b}
	\hspace{-0.25cm}h_{\mathrm{WB},k}(\vec{\mathbf{f}}_n,t) = \sum_{q = 0}^{Q}{\Gamma_{k,q}(\vec{\mathbf{f}}_n)\; e^{-j2\pi{f_c}\tau_{n,k,q}} \delta \left(t - \tau_{n,k,q}\right)},\hspace{-0.2cm}
\end{align}
where $\Gamma_{k,0}(\vec{\mathbf{f}}_n) = \beta_{n,k}\sqrt{g_k(\vec{\mathbf{f}}_n)}$ is the channel gain of the ``RA~$n$-user~$k$'' link, $\Gamma_{k,q}(\vec{\mathbf{f}}_n) = \sigma_q \beta_0\left(\tilde{d}_{n,q} \bar{d}_{k,q}\right)^{-1} \sqrt{\tilde{g}_q(\vec{\mathbf{f}}_n)}$ with $q = 1,2,\dots,Q$ is the channel gain of the ``RA~$n$-cluster~$q$-user~$k$'' link, $f_c$ is the carrier frequency, and $\delta(t)$ denotes the Dirac delta function. The continuous-time Fourier transform (CTFT) of $h_{\mathrm{WB},k}(\vec{\mathbf{f}}_n,t)$ is then obtained as
\begin{align}
	\label{deqn_ex20b}
	H_{\mathrm{WB},k}(\vec{\mathbf{f}}_n,f) = \sum_{q = 0}^{Q}{\Gamma_{k,q}(\vec{\mathbf{f}}_n)\; e^{- j2\pi{f_c}\left(1 + \frac{f}{f_c}\right)\tau_{n,k,q}}},
\end{align}
which represents the space-frequency channel response between RA~$n$ and user~$k$. 

Let $\Delta_f\triangleq \frac{B}{L}$ denote the OFDM subcarrier spacing. Then, the frequency of the $l$-th subcarrier is given by ${f_l} = (l - 1)\Delta_f$ with $l\in \mathcal{L}\triangleq \{1,2,\ldots,L\}$. Therefore, the space-frequency channel response between the BS and user~$k$ at subcarrier~$l$ is given by
\begin{align}
	&\mathbf{h}_{\mathrm{WB},k,l}(\mathbf{F})\nonumber\\
    =&\left[H_{\mathrm{WB},k}(\vec{\mathbf{f}}_{1}, \hspace{-0.05cm}f_{l} ), H_{\mathrm{WB},k}(\vec{\mathbf{f}}_{2},\hspace{-0.05cm} f_{l} ), \dots, H_{\mathrm{WB},k}(\vec{\mathbf{f}}_{N},\hspace{-0.05cm} f_{l} ) \right]^{T}\hspace{-0.1cm}.
    \label{deqn_ex21b}
\end{align}

\subsubsection{Extension to Incorporate Polarization Effects}\label{Polar_Model}
Polarization is a fundamental property of electromagnetic (EM) waves that describes the orientation of the electric field vector during propagation.
In wireless communication systems, proper polarization alignment between the transmit and receive antennas is essential for efficient signal transmission and reception.
{\color{black}In RA systems, antenna rotation changes not only the boresight direction but also the polarization direction, which may either enhance or degrade the alignment between the transmit and receive polarizations, thereby affecting the effective channel gain~\cite{Zheng2026Rotatable,Zhang2025Polarization,Zhang2026Polarization,Wei2026Polarization}.}

\begin{figure}[!t]\centering
	\includegraphics[width=3in]{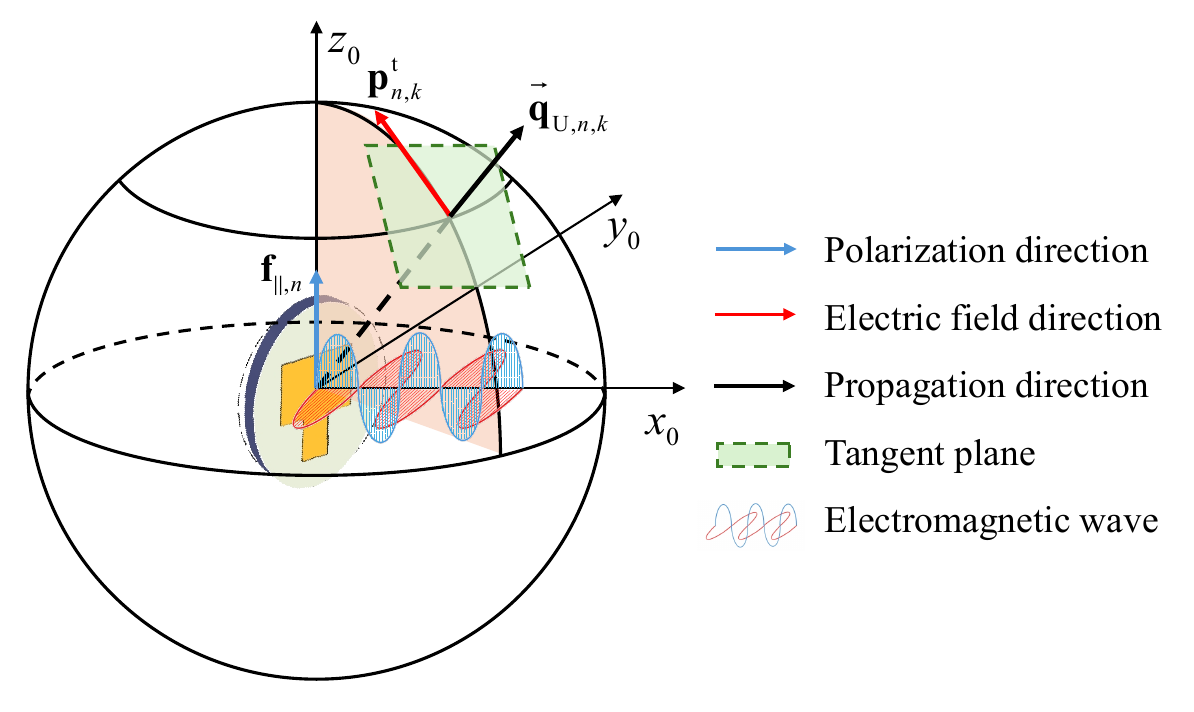}\vspace{-0.25cm}
	\caption{Illustration of polarization directions of antenna and electric field.}
	\label{fig_polarization}
\end{figure}

To incorporate polarization effects into the RA channel model, we assume that both the BS and users employ linearly polarized antennas. {\color{black}This subsection explicitly links the reference vector in the antenna/array rotation model of Section~\ref{Rotation_Model} to the polarization-aware channel model. In particular, the main polarization direction of each RA is assumed to be aligned with its reference vector $\vec{\mathbf{f}}_{\parallel,n}$, which serves as the antenna’s in-plane polarization axis. This establishes a direct geometric relationship between antenna rotation and polarization orientation.}
As illustrated in Fig.~\ref{fig_polarization}, the electric field at an observation point is always orthogonal to the propagation direction and lies in the plane spanned by the transmit polarization direction and the propagation direction.
Thus, the effective transmit electric-field vector, obtained by projecting $\vec{\mathbf{f}}_{\parallel,n}$ onto the tangent plane orthogonal to the propagation direction $\vec{\mathbf{q}}_{\mathrm{U},n,k}$, is given by~\cite{Zheng2026Rotatable}
\begin{align}
	\label{deqn_ex22b}
	\mathbf{p}_{n,k}^{\mathrm{t}}=\vec{\mathbf{f}}_{\parallel,n} - (\vec{\mathbf{f}}_{\parallel,n}^T \vec{\mathbf{q}}_{\mathrm{U},n,k})\vec{\mathbf{q}}_{\mathrm{U},n,k}.
\end{align}
This expression shows that the effective electric-field direction depends jointly on the antenna polarization direction and the user direction.

\begin{figure}[!t]\centering
	\includegraphics[width=3.2in]{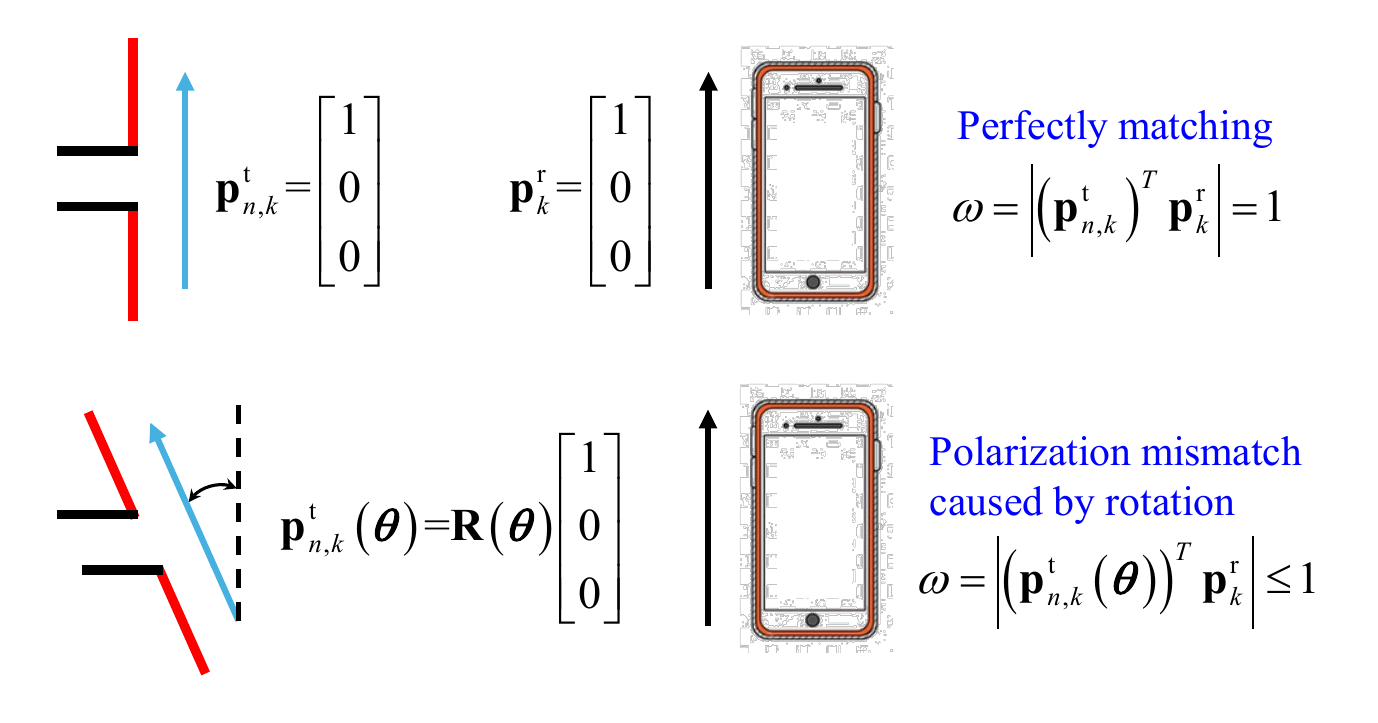}
	\caption{Illustration of rotation-induced polarization mismatch.}
	\label{fig_mismatch}
\end{figure}
As illustrated in Fig.~\ref{fig_mismatch}, perfect polarization matching occurs when the electric-field direction of the incoming wave aligns with the receive polarization direction.
Otherwise, polarization mismatch reduces the effective received power. Let $\mathbf{p}_{k}^{\mathrm{r}}$ denote the receive polarization direction at user~$k$. The polarization matching efficiency is characterized by the inner product between $\mathbf{p}_{n,k}^{\mathrm{t}}$ and $\mathbf{p}_{k}^{\mathrm{r}}$, yielding the polarization matching gain~\cite{Zheng2026Rotatable}
\begin{align}
	\label{deqn_ex23b}
	\omega \left(\vec{\mathbf{f}}_n,\vec{\mathbf{q}}_{\mathrm{U},n,k}\right) = \left(\mathbf{p}_{n,k}^{\mathrm{t}}\right)^T \mathbf{p}_{k}^{\mathrm{r}}.
\end{align}
Then, by incorporating the polarization matching gain into the channel model, the LoS channel coefficient between RA~$n$ and user~$k$ is given by
\begin{align}
	\label{deqn_ex24b}
	h_{\mathrm{PE},k}\left(\vec{\mathbf{f}}_n\right) = \omega \left(\vec{\mathbf{f}}_n,\vec{\mathbf{q}}_{\mathrm{U},n,k}\right) h_{\mathrm{LoS},k}\left(\vec{\mathbf{f}}_n\right).
\end{align}
This polarization-aware model shows that antenna rotation affects the channel amplitude through both directional gain and polarization matching.
Therefore, proper antenna orientation control can balance these two effects to improve channel quality.

The above polarization-aware channel model applies to linearly polarized antennas with arbitrary slant angles, including purely vertical or horizontal polarization. 
It can also be extended to cross-polarized (dual-polarized) antenna pairs, such as $\pm45^\circ$ polarized elements. In such cases, each element can be treated as a linearly polarized antenna sharing the same RF chain, and its channel can be modeled using \eqref{deqn_ex24b}.
By introducing polarforming technology that jointly controls the two elements, the combined transmit/receive polarization can be adaptively adjusted through phase and amplitude control~\cite{Ding2026Polarforming}.
Furthermore, although \eqref{deqn_ex24b} is derived for narrowband LoS channels, it can be extended to multipath and wideband scenarios by incorporating the polarization effects of scatterer clusters into the multipath model in \eqref{deqn_ex18b} and the wideband model in \eqref{deqn_ex19b}.

\subsubsection{Extensions to Other Channel Models}
The RA channel models developed above follow a geometry-/field-response formulation, where the channel explicitly depends on the antenna orientations in a deterministic propagation environment. This enables direct optimization of RA orientations to enhance communication and sensing performance. The baseline model assumes an RA array deployed at the BS and a single fixed antenna at each user, but it can be readily generalized to MIMO links where users are also equipped with RA arrays. In such cases, the MIMO channel is constructed by stacking the SISO channels corresponding to each transmit-receive RA pair, and the directional gain must account for rotations at both ends.

Beyond deterministic environments, the RA channel model can be extended to stochastic channel models by treating user locations, scatterer positions, and small-scale fading coefficients as random variables. 
For example, under a Rician fading model, the channel can be decomposed into a deterministic LoS component modeled by \eqref{deqn_ex7b} and a stochastic NLoS component capturing spatially correlated multipath.
These statistics are often assumed stationary or quasi-static, but in practice, mobility of the transceivers or surrounding objects induces spatio-temporal variations and Doppler shifts. {\color{black}Such dynamics can be incorporated by modeling the time-varying geometry and scheduling RA orientations to track channel evolution, which naturally aligns with the orientation-dependent structure of RA channels.}
The RA modeling framework also extends to distributed antenna deployments, such as cell-free MIMO systems~\cite{Pan2025Rotatable,Peng2026Cell}. In this case, the overall channel is formed by aggregating orientation-dependent channel responses across geographically distributed APs, where coordinated orientation control influences spatial correlation, macro-diversity, and coverage performance. {\color{black}This provides a flexible foundation for analyzing RA-enabled distributed systems and evaluating the potential gains they yield in large-scale networks.}

{\color{black}It should also be noted that the above channel models are based on ideal or well-calibrated RA responses. However, in practical compact RA arrays, antenna rotation may change the relative electromagnetic interaction among closely spaced antennas and surrounding stationary structures, leading to orientation-dependent mutual coupling, radiation pattern distortion, phase-center shifts, impedance mismatch, and polarization variation. As a result, the actual RA channel may deviate from the ideal model.
Therefore, in-situ calibration is necessary in practical RA deployment, where each RA is rotated over a predefined orientation grid after being integrated with its mechanical platform, and the corresponding radiation patterns are measured or full-wave simulated for the final assembled configuration~\cite{Lebron2020Validation}.
Such effects can be incorporated by replacing the ideal directional gain pattern with these calibrated orientation-dependent radiation patterns, or by introducing an orientation-dependent coupling/calibration matrix into the effective channel model~\cite{Su2021On,Wallace2004Mutual}. This provides a basis for more practical RA channel modeling and calibration-aware channel acquisition.}

\subsection{Optimization Framework and Design Issues}
To fully exploit the additional spatial DoFs introduced by antenna rotation, we consider a generic optimization framework for RA-enabled wireless systems:
\begin{subequations}\label{eq:framework}
	\begin{alignat}{2}
		\max_{\bm{\Theta},\mathcal{S}} \quad & \mathbb{U}\left(\bm{\Theta},\mathcal{S}\right) & \label{eq:framework-A}\\
		\mbox{s.t.} \quad
		& \mathbb{F}_i\left(\bm{\Theta}\right) \geq 0,\; 1\leq i\leq I_{\mathbb{F}}, \label{eq:framework-B}\\
		& \mathbb{G}_i\left(\mathcal{S}\right) \geq 0,\; 1\leq i\leq I_{\mathbb{G}}, \label{eq:framework-C}\\
		& \mathbb{Q}_i\left(\bm{\Theta},\mathcal{S}\right) \geq 0,\; 1\leq i\leq I_{\mathbb{Q}}, \label{eq:framework-D}
	\end{alignat}
\end{subequations}
where $\bm{\Theta}\triangleq \left[\bm{\theta}_1,\bm{\theta}_2,\dots,\bm{\theta}_N,\bm{\theta}_{\mathrm{array}}\right]\in {{\mathbb{R}}^{3\times (N+1)}}$ collects the rotation angles of all RAs and the array platform, and $\mathcal{S}$ denotes the set of system resources (e.g., transmit power, bandwidth, beamforming vectors, user association). The number of constraints $I_{\mathbb{X}}$ with $\mathbb{X}\in\{\mathbb{F},\mathbb{G},\mathbb{Q}\}$ depends on the system configuration, including the numbers of RAs and users.
The utility function $\mathbb{U}(\cdot)$ quantifies system performance; $\mathbb{F}_i(\cdot)$ captures antenna/array rotation-related constraints (e.g., limited rotation range, discrete antenna rotation, minimum antenna spacing, and maximum antenna rotation speed); $\mathbb{G}_i(\cdot)$ captures resource limitations; and $\mathbb{Q}_i(\cdot)$ represents coupled constraints involving both rotation and resource allocation.

\subsubsection{Utility Function}
The utility function $\mathbb{U}(\bm{\Theta},\mathcal{S})$ in \eqref{eq:framework-A} reflects the performance objective of the RA-enabled system. In communication systems, it may represent the achievable rate, secrecy rate, coverage probability, or outage performance~\cite{Wu2025Modeling,Zheng2025Rotatable,Dai2025RotatableSecure,Pan2025Rotatable}. In sensing systems, typical metrics include the Cramér--Rao bound (CRB), detection probability, and estimation accuracy~\cite{Xiong2025Efficient,Zhou2025Rotatable}. System-level metrics such as energy efficiency, spectral efficiency, or latency can also be incorporated depending on the network architecture and service requirements~\cite{Peng2025Rotatable,Tan2026Rotatable,Zhang2026Polarization}. For ISAC systems, $\mathbb{U}(\cdot)$ may be designed to strike a balance between communication throughput and sensing accuracy~\cite{Wu2025Rotatable,Zhang2026Rotatable}. In intelligent computing networks, it may further include computation latency, computing efficiency, and task completion rate, thereby capturing the interplay between communication and computation resources~\cite{Wang2026Rotatable,Wang2026RotatableAntennaAssisted}.

\subsubsection{Antenna/Array Rotation Constraint}
Regardless of whether mechanical or electronic rotation is used, the feasible rotation range is physically limited. To ensure that each RA or the array platform operates within the respective hardware constraints, the rotation angles must satisfy
\begin{subequations}\label{deqn_ex1c}
    \begin{align}
	    &\hspace{-0.4cm}[\bm{\theta}_{\mathrm{lower},n}]_i \leq [\bm{\theta}_n]_i \leq [\bm{\theta}_{\mathrm{upper},n}]_i,\; \forall n\in \mathcal{N},\; i\in \{1,2,3\},\hspace{-0.3cm}\label{deqn_ex1c1}\\
        &\hspace{-0.4cm}[\bm{\theta}_{\mathrm{lower,array}}]_i \leq [\bm{\theta}_{\mathrm{array}}]_i \leq [\bm{\theta}_{\mathrm{upper,array}}]_i,\; i\in \{1,2,3\},\hspace{-0.3cm}\label{deqn_ex1c2}
    \end{align}
\end{subequations}
where $\bm{\theta}_{\mathrm{lower},n}$ (or $\bm{\theta}_{\mathrm{lower,array}}$) and $\bm{\theta}_{\mathrm{upper},n}$ (or $\bm{\theta}_{\mathrm{upper,array}}$) denote the lower and upper bounds of the rotation angles of the RA~$n$/array, respectively.

In compact arrays, large antenna rotations may increase mutual coupling and distort radiation patterns due to spatial overlap. To avoid excessive boresight deviation and mitigate coupling effects, an additional constraint is imposed \cite{Zheng2026Rotatable,Wu2025Modeling}:
\begin{align}
	\label{deqn_ex2c}
	0 \leq \arccos{\left(\vec{\mathbf{f}}_{\perp,n}^T \mathbf{e}_1\right)} \leq \theta_{\max},\; \forall n\in \mathcal{N},
\end{align}
where $\theta_{\max} \in [0,\frac{\pi}{2}]$ specifies the maximum allowable deviation of each RA’s boresight from its initial nominal direction (i.e., the positive $x$-axis).

\subsubsection{Other Constraints}
Constraints $\mathbb{G}_i(\mathcal{S})\ge0$ represent system resource limitations, such as maximum transmit power, available bandwidth, beamforming constraints, and maximum transmission duration. These constraints characterize the physical and operational limits of the RA system and ensure that resource usage remains within practical and regulatory bounds.
In addition, constraints $\mathbb{Q}_i\left(\bm{\Theta},\mathcal{S}\right) \geq 0$ capture the joint impact of antenna orientation and resource allocation. Examples include minimum signal-to-interference-plus-noise ratio (SINR) requirements, minimum received signal power, maximum tolerable sensing error, and outage probability constraints. These constraints reflect the quality-of-service (QoS) and reliability requirements associated with dynamic antenna rotations.


{\color{black}Compared with fixed-antenna systems, RA systems can significantly enhance performance by reconfiguring directional gain patterns and concentrating radiation energy toward desired directions. However, antenna/array rotation also introduces additional implementation cost, energy consumption, control latency, and computational complexity. Practical RA optimization must therefore account for hardware imperfections (e.g., mechanical inaccuracies, RF distortions, mutual coupling, and calibration errors), imperfect channel state information (CSI) caused by estimation errors and channel variations, finite rotation speed, limited angular resolution, and rotation energy consumption~\cite{Zheng2025Rotatable}.
These practical factors may cause model mismatch between the ideal RA response used for optimization and the actual hardware response, which can degrade coherent combining, reduce the intended array gain, and lead to unintended interference leakage toward undesired users or sensing directions. Consequently, beamforming and RA orientation optimization based solely on ideal responses may become suboptimal under practical hardware impairments. This motivates robust and hardware-aware optimization frameworks that ensure reliable operation under real-world constraints. In particular, frameworks for robust beamforming and RA orientation optimization can be developed by modeling coupling- and calibration-induced errors as bounded or stochastic channel uncertainties \cite{Vorobyov2003Robust, ZhengRAImperfectCSI2026}.

To further improve the feasibility of real-time RA control, the optimization framework can be implemented in a two-timescale manner. On the long timescale, RA orientations can be optimized offline according to representative environment states, user/target distributions, and communication/sensing requirements, and the resulting RA orientations can be stored in a database or codebook. On the short timescale, the RA controller selects or interpolates suitable RA orientations based on observed environmental information and service demands, while beamforming and other resource-allocation variables are updated using instantaneous CSI. In this way, real-time RA control is transformed from repeatedly solving high-dimensional non-convex optimization problems into a lower-complexity environment-to-orientation mapping problem. Moreover, discrete orientation codebooks, learning-assisted orientation prediction, and hardware-aware robust designs can be incorporated to account for finite angular resolution, actuator delay, CSI uncertainty, rotation errors, and rotation energy consumption.}


\section{Antenna/Array Rotation Optimization}\label{sec:optimization}
Determining the optimal antenna orientations/boresights is essential for fully exploiting the performance gains offered by RA systems.
Since wireless channels depend on antenna/array orientation in a highly nonlinear manner, the resulting optimization problems are often challenging and require carefully designed solution strategies. In this section, we introduce efficient optimization methods for antenna/array rotation in various wireless system settings and demonstrate their performance advantages over conventional fixed-antenna architectures.
{\color{black}To characterize the fundamental performance bounds of RA systems, the optimization in this section is performed under several common implementation assumptions. First, the required CSI or geometry-related channel parameters are assumed to be available at the RA controller. In practice, they can be obtained through the channel estimation/acquisition methods provided in Section~\ref{sec:estimation}. Second, the RA orientations are continuously adjusted within their allowable rotation regions subject to mechanical or electronic rotation constraints. Third, if the rotationally symmetric cosine radiation pattern is adopted, self-rotation around the boresight axis does not affect the directional gain, and thus the RA orientation can be simplified to 2D boresight rotation characterized by the pointing vector.
Unless otherwise specified, the channel and optimization models in this section assume ideal or well-calibrated RA responses. Under this assumption, the adopted directional gain patterns, array responses, and orientation-dependent channels can reasonably approximate practical RA systems, and the obtained results mainly serve as representative performance benchmarks or theoretical bounds.}

\subsection{RA-Enabled MISO/SIMO System}
In MISO/SIMO systems, deploying an RA array at the BS enables significant array-gain enhancement by jointly adjusting the orientations/boresights of all antennas. The resulting spatial DoFs allow the RA array to concentrate radiated energy toward the intended user, thereby reducing energy leakage and improving transmission efficiency.
{\color{black}We note that the generic RA model in Section~\ref{sec:fundamentals} is not restricted to a specific antenna type or array geometry, since it is fully characterized by the antenna positions, antenna orientations/boresights, and the directional gain pattern $G(\epsilon,\varphi)$.
In contrast, to obtain tractable closed-form insights and illustrate the fundamental array-gain scaling behavior, in this subsection, we adopt a representative implementation model: a single-user MISO system ($K=1$) with a ULA-based RA transmitter as well as free-space and narrowband propagation. The RAs are modeled as compact patch-like directional antennas whose dominant mainlobe is approximated by the cosine pattern in \eqref{deqn_ex1b}, while the user is assumed to employ an isotropic antenna. 
This representative model is chosen because ULA structures are standard and practically relevant for BS deployments, patch-like directional antennas are compact and widely used in array implementations, and the cosine pattern provides an analytically tractable approximation for antennas with a dominant mainlobe and negligible sidelobes. Based on this setting, we investigate a tractable example to reveal how near-field geometry and limited rotation range affect RA orientation design.}
Owing to uplink-downlink duality, the analytical framework can be directly extended to the SIMO case with receive beamforming.

Under the cosine pattern model, the directional gain depends solely on the projection between the pointing vector and the user direction, as shown in~\eqref{deqn_ex6b}. Hence, a 2D rotation model suffices to characterize RA orientation, and optimizing the rotation angles $\{\bm{\theta}_n\}$ is equivalent to optimizing the pointing vectors $\{\vec{\mathbf{f}}_{\perp,n}\}$, thereby mitigating the coupling among the rotation angles.
Dropping the user index in the near-field LoS channel model \eqref{deqn_ex8b} for simplicity, the signal-to-noise ratio (SNR) maximization problem is formulated as follows:
\begin{subequations}\label{eq:MISO}
	\begin{alignat}{2}
		\hspace{-0.3cm}\text{(P-MISO):} \; \max_{\mathbf{w}_{\mathrm{t}},\{\vec{\mathbf{f}}_{\perp,n}\}} \quad & \gamma = \frac{1}{\sigma^2} \left|\mathbf{w}_{\mathrm{t}}^{T} \mathbf{h}_{\mathrm{LoS}}\left(\mathbf{F}\right)\right|^2 & \label{eq:MISO-A}\\
		\mbox{s.t.} \quad 
		& 0 \leq \arccos{(\vec{\mathbf{f}}_{\perp,n}^T \mathbf{e}_1)} \leq \theta_{\max},\; \forall n,\hspace{-0.1cm} \label{eq:MISO-B}\\
        & \|\vec{\mathbf{f}}_{\perp,n}\| = 1,\; \forall n, \label{eq:MISO-C}\\
        & \|\mathbf{w}_{\mathrm{t}}\|^2 \leq P, \label{eq:MISO-D}
	\end{alignat}
\end{subequations}
where $\mathbf{w}_{\mathrm{t}}\in \mathbb{C}^{N\times 1}$ is the transmit beamforming vector, and $P$ and $\sigma^2$ denote the maximum transmit power and the noise power, respectively. In addition, constraint \eqref{eq:MISO-C} ensures that $\vec{\mathbf{f}}_{\perp,n}$ is a unit-norm vector, and constraint \eqref{eq:MISO-D} guarantees that the transmit power does not exceed its budget.

For any given RA pointing vectors $\{\vec{\mathbf{f}}_{\perp,n}\}$, the optimal transmit beamformer for problem (P-MISO) can be obtained via maximum-ratio transmission (MRT), i.e., $\mathbf{w}_{\mathrm{MRT}} = \sqrt{P}\frac{\mathbf{h}_{\mathrm{LoS}}^{\ast}\left(\mathbf{F}\right)}{\|\mathbf{h}_{\mathrm{LoS}}\left(\mathbf{F}\right)\|}$. Thus, substituting $\mathbf{w}_{\mathrm{MRT}}$ into \eqref{eq:MISO-A} yields
\begin{align}
	\label{deqn_ex0d}
	\hspace{-0.25cm}\gamma = \frac{P}{\sigma^2} \| \mathbf{h}_{\mathrm{LoS}}\left(\mathbf{F}\right)\|^2 = \frac{P \beta_0 G_{\max}^{\mathrm{cos}}}{\sigma^2} \sum_{n=1}^{N}{\frac{1}{d_{n}^{2}}{\left[\vec{\mathbf{q}}_{\mathrm{U},n}^T \vec{\mathbf{f}}_{\perp,n}\right]_{+}^{2\rho}}}.\hspace{-0.2cm}
\end{align}
{\color{black}It can be observed from \eqref{deqn_ex0d} that the orientation of each RA affects the received SNR through the projection of its pointing vector $\vec{\mathbf{f}}_{\perp,n}$ onto the user direction $\vec{\mathbf{q}}_{\mathrm{U},n}$. Therefore, the RA orientation design can be interpreted as a per-antenna projection maximization problem. Maximizing this projection under the limited rotation range in \eqref{eq:MISO-B} yields the following optimal pointing vector~\cite{Wu2025Modeling,Zheng2026Rotatable}:}
\begin{align}
	\label{deqn_ex1d}
	\vec{\mathbf{f}}_{\perp,n}^{\star} = \left[\cos{\theta_{\mathrm{z},n}^{\star}},\;\sin{\theta_{\mathrm{z},n}^{\star}}\cos{\theta_{\mathrm{a},n}^{\star}},\;\sin{\theta_{\mathrm{z},n}^{\star}}\sin{\theta_{\mathrm{a},n}^{\star}} \right]^T,
\end{align}
where $\theta_{\mathrm{z},n}^{\star}$ and $\theta_{\mathrm{a},n}^{\star}$ are given by
\begin{subequations}\label{deqn_ex2d}
	\begin{align}
		\theta_{\mathrm{z},n}^{\star} & = \min 	\left \{\arccos\left(\vec{\mathbf{q}}_{\mathrm{U},n}^T \mathbf{e}_1\right),\theta_{\max} \right \}, \label{deqn_ex2d1}\\
		\theta_{\mathrm{a},n}^{\star} & = 	\operatorname{arctan2}\left(\vec{\mathbf{q}}_{\mathrm{U},n}^T \mathbf{e}_3, \vec{\mathbf{q}}_{\mathrm{U},n}^T \mathbf{e}_2 \right), \label{deqn_ex2d2}
	\end{align}
\end{subequations}
respectively. This indicates that each RA should align its boresight with the user direction as closely as allowed by the rotation constraints to maximize its directional gain. 

\begin{figure}[!t]\centering
	\includegraphics[width=3.4in]{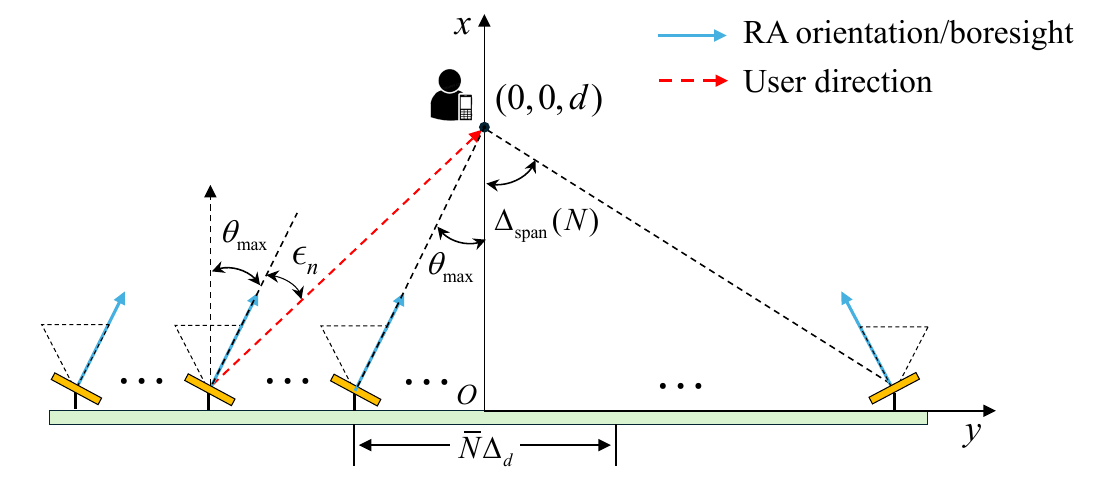}\vspace{-0.25cm}
	\caption{Illustration of the geometric relationship between the user and RAs in the considered RA-enabled MISO system.}
	\label{fig_span_angle}
\end{figure}
For a ULA-based MISO system with antenna directivity factor $\rho=\frac{1}{2}$ and a user located directly in front of the array as shown in Fig.~\ref{fig_span_angle}, the analyses of \cite{Zheng2026Rotatable} show that the maximum SNR based on the optimal RA pointing vector in~\eqref{deqn_ex1d} admits the following closed-form expression:
\begin{align}
	\label{deqn_ex3d}
	\hspace{-0.2cm}{\gamma} =
	\begin{cases}
		{\frac{2\zeta P}{\pi^2 \sigma^2} \triangle_{\mathrm{span}}(N),}&\hspace{-0.25cm}{N \leq \bar{N}} \\
		{ \frac{2\zeta P}{\pi^2 \sigma^2} \left[\theta_{\max} + \sin\left(\triangle_{\mathrm{span}}(N) - \theta_{\max}\right) \right],}&\hspace{-0.25cm}{N > \bar{N},}
	\end{cases}\hspace{-0.2cm}
\end{align}
where $\zeta \triangleq \frac{\Delta_d}{d} \ll 1$ with $d$ being the distance between the center of the ULA and the user, $\triangle_{\mathrm{span}}(N) \triangleq \arctan\left(\frac{N\zeta}{2}\right)$ denotes the user span angle, defined as the angle between the two line segments connecting the user to the center and to one end of the ULA as illustrated in Fig.~\ref{fig_span_angle}, and $\bar{N} \triangleq 2\left \lfloor  \frac{\tan \theta_{\max}}{\zeta}\right \rfloor + 1$ is the maximum number of antennas whose boresights can be aligned with the user direction.
The maximum SNR in \eqref{deqn_ex3d} scales with the number of RAs $N$ according to the span angle $\triangle_{\mathrm{span}}(N)$ and is fundamentally limited by the allowable rotation range.


{\color{black}{\bf SNR Scaling with $N$:} The closed-form result in \eqref{deqn_ex3d} reveals two distinct scaling regimes. When $N\le\bar{N}$ (i.e., $\Delta_{\mathrm{span}}(N)\le\theta_{\max}$), all RAs can steer their boresights toward the user within the allowable rotation range. Using the approximation $\arctan(x)\approx\frac{\pi}{4}x$ for $|x|\le 1$~\cite{Rajan2006Efficient}, the SNR in the first case of \eqref{deqn_ex3d} becomes $\gamma \approx \frac{P\zeta^2}{4\pi \sigma^2}N$. In this regime, the received SNR increases approximately linearly with $N$.
When $N>\bar{N}$ (i.e., $\Delta_{\mathrm{span}}(N) > \theta_{\max}$), the span angle exceeds the feasible rotation range for part of the array, and the additional RAs cannot be perfectly aligned with the user.
It can be verified that $f(x) \triangleq \sin\left(\arctan\left(x\right) - \theta_{\mathrm{max}}\right)$ is a concave increasing function, and $\lim_{x\to \infty} f'(x) = 0$. Consequently, the SNR growth rate gradually diminishes as $N$ increases. In particular, letting $N\to\infty$, the second case of \eqref{deqn_ex3d} yields the following asymptotic SNR:
\begin{align}
	\label{deqn_ex4d}
	\lim_{N \to \infty} \gamma = \frac{2\zeta P}{\pi^2 \sigma^2} \left(\theta_{\max} + \cos\theta_{\max}\right),
\end{align}
which indicates that a larger allowable rotation range leads to a higher asymptotic SNR. Thus, the limited rotation range introduces a practical saturation effect: simply increasing the number of antennas cannot fully compensate for insufficient rotation capability. As shown in~\cite{Zheng2026Rotatable}, a similar SNR scaling law also holds for the general UPA case.}

\begin{figure}[!t]\centering
	\includegraphics[width=3in]{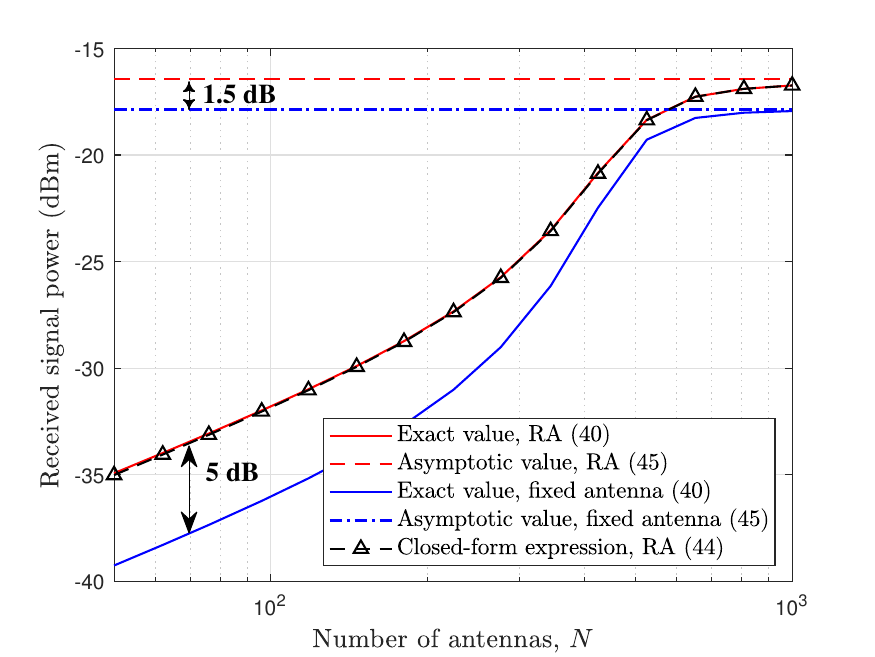}
	\caption{Received signal power versus the number of antennas $N$ for the RA-enabled MISO system.}
	\label{fig_single_ULA}
\end{figure}
To evaluate the array-gain improvement achieved by RAs, we consider a single-user MISO system with a ULA transmitter and a user located at distance $d = 15$~m from the array center.
The system operates at 2.4~GHz ($\lambda=0.125$~m), with noise power $\sigma^2=-80$~dBm, antenna spacing $\Delta_d=\lambda/2$, transmit power $P=10$~dBm, and maximum deviation angle $\theta_{\max}=\pi/6$.
For these settings, Fig.~\ref{fig_single_ULA} shows the received signal power versus the number of antennas $N$ for both RA-enabled and fixed-antenna MISO systems.
{\color{black}For a small to moderate value of $N$, both systems exhibit approximately linear power scaling. As $N$ increases further, the received powers of both systems gradually approach their asymptotic values, consistent with the SNR scaling law discussed above.
Throughout the entire range, the RA system achieves higher received power because each RA can independently steer its boresight toward the user, thereby improving the effective directional gain. The performance advantage is particularly pronounced (up to 5~dB) when $N \le 100$, indicating that RA can achieve a higher SNR with fewer antennas (an important benefit when the array size is constrained).
Even as $N \to \infty$, the RA system maintains a non-negligible asymptotic gain (nearly 1.5~dB) over the fixed-antenna system, demonstrating that the inherent benefits of RA do not vanish with increasing array size.}

\begin{figure}[!t]\centering
	\includegraphics[width=3in]{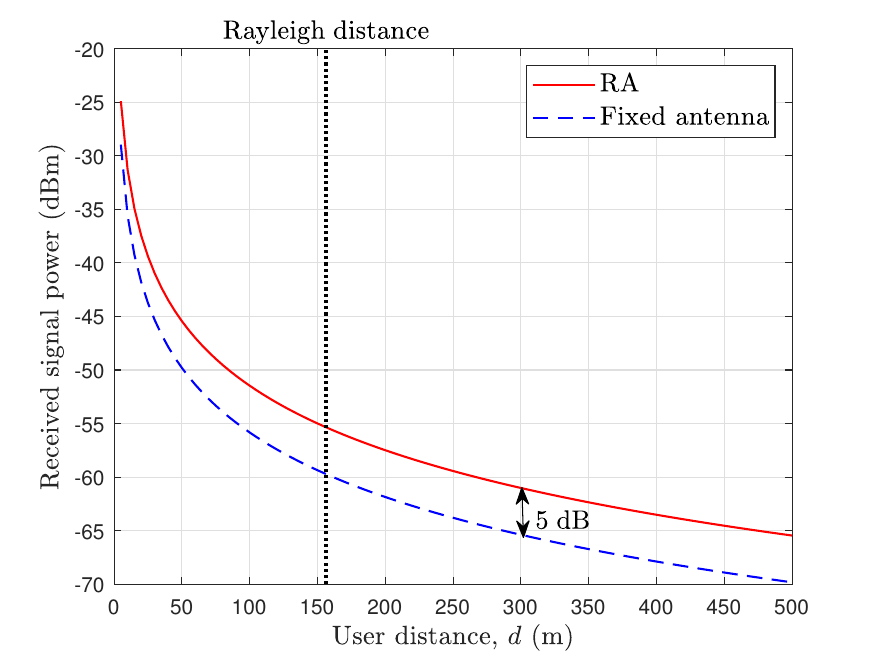}
	{\color{black}\caption{Received signal power versus user distance $d$ for RA-enabled and fixed-antenna MISO systems with $N = 50$ antennas.}
	\label{fig_single_ULA_distance}}
\end{figure}
{\color{black}To further examine the impact of user distance $d$ and the transition between the near-field and far-field regimes, Fig.~\ref{fig_single_ULA_distance} shows the received signal power versus $d$ for a user located at an azimuth angle offset of $\frac{5\pi}{12}$ (in radians) from the array boresight. As expected, the received power of both systems decreases with distance due to more severe path loss, while the RA system consistently achieves higher power.
In the near field, different antennas observe slightly different user directions because of the non-negligible angular spread across the array. RA orientation control can exploit this antenna-dependent angular variation to enhance the directional gain. As the user moves farther away, the angular spread diminishes, and the optimized RA orientations gradually converge to a similar direction. Nevertheless, since the user is not located directly in front of the array, the fixed-antenna system suffers from off-boresight gain loss, which becomes more uniform and pronounced as $d$ increases. In contrast, RAs can continue to align their boresights with the user direction, maintaining a clear performance advantage of about 5~dB even beyond the Rayleigh distance (at $d = 156$~m).}


{\color{black}We note that the closed-form SNR expressions in \eqref{deqn_ex3d}--\eqref{deqn_ex4d} and the numerical results in Figs.~\ref{fig_single_ULA} and \ref{fig_single_ULA_distance} are specific to the representative ULA-based model with cosine-pattern directional antennas. For other antenna types or array geometries, the exact scaling behavior may differ depending on the underlying radiation pattern, antenna placement, and propagation conditions. Nevertheless, the key physical insight remains general: RA improves the effective channel gain by reducing the angular mismatch between the antenna boresight and the desired propagation direction.}
{\color{black}Beyond single-user scenarios, RA can also be leveraged to control spatial interference by steering nulls toward multiple undesired directions~\cite{Feng2026Rotatable,Wen2025Rotatable}. As shown in~\cite{Wen2025Rotatable}, RA enables an orthogonality condition for both isotropic and directional antenna patterns, allowing the beam gain in a desired direction to be maximized while simultaneously nulling several other directions. This capability arises from RA’s ability to adjust the spatial correlation among steering vectors and jointly shape both the mainlobe and sidelobe behavior through orientation control.}

\subsection{RA-Enabled MIMO System}
Compared with MISO/SIMO systems where antenna/array rotation is applied only at one side of the link, RA-enabled MIMO architectures allow joint orientation/boresight adjustment at both the transmitter and receiver.
The resulting additional spatial DoFs enable dynamic beam alignment with dominant propagation paths, mitigating angular mismatch caused by user mobility or environmental variations.
By jointly optimizing the transmit- and receive-side RA orientations/boresights, the system can simultaneously concentrate radiated energy toward desired directions and enhance the received signal power at the target user. As a result, the overall channel condition improves, inter-path correlation is reduced, and the MIMO channel matrix becomes higher-rank and better conditioned. This transceiver-side spatial reconfiguration strengthens the spatial multiplexing capability and ultimately increases the achievable channel capacity.

To illustrate these advantages, we consider an RA-enabled MIMO communication system, where the transmitter and receiver are equipped with $N_{\mathrm{t}}$ and $N_{\mathrm{r}}$ RAs, respectively.
Let $\mathbf{H}(\mathbf{F}_{\mathrm{t}},\mathbf{F}_{\mathrm{r}})\in \mathbb{C}^{N_{\mathrm{r}}\times N_{\mathrm{t}}}$ denote the MIMO channel matrix, where $\mathbf{F}_{\mathrm{t}}\in \mathbb{R}^{6\times N_{\mathrm{t}}}$ and $\mathbf{F}_{\mathrm{r}}\in \mathbb{R}^{6\times N_{\mathrm{r}}}$ are the orientation matrices of the transmit and receive RAs, respectively. In particular, the $(n_{\mathrm{r}},n_{\mathrm{t}})$-th entry of $\mathbf{H}(\mathbf{F}_{\mathrm{t}},\mathbf{F}_{\mathrm{r}})$ corresponds to the channel coefficient between the $n_{\mathrm{t}}$-th transmit RA and the $n_{\mathrm{r}}$-th receive RA, where $n_{\mathrm{t}}\in \mathcal{N}_{\mathrm{t}}\triangleq \{1,2,\dots,N_{\mathrm{t}}\}$ and $n_{\mathrm{r}}\in \mathcal{N}_{\mathrm{r}}\triangleq \{1,2,\dots,N_{\mathrm{r}}\}$. The channel coefficient of each transmit-receive RA pair can be modeled using the LoS, multipath, wideband, and polarization-aware channel models in \eqref{deqn_ex7b}, \eqref{deqn_ex17b}, \eqref{deqn_ex19b}, and \eqref{deqn_ex24b}, respectively.

Given channel matrix $\mathbf{H}(\mathbf{F}_{\mathrm{t}},\mathbf{F}_{\mathrm{r}})$ and transmit covariance matrix $\mathbf{S}\in \mathbb{C}^{N_{\mathrm{t}}\times N_{\mathrm{t}}}$ with $\mathbf{S}\succeq \mathbf{0}$, the MIMO channel capacity is given by
\begin{align}
    \label{deqn_ex1h}
    &C_{\mathrm{MIMO}}\left(\mathbf{F}_{\mathrm{t}},\mathbf{F}_{\mathrm{r}},\mathbf{S}\right)\nonumber\\
    = &\log_2 \det \left(\mathbf{I}_{N_{\mathrm{r}}} + \frac{1}{\sigma^2} \mathbf{H}(\mathbf{F}_{\mathrm{t}},\mathbf{F}_{\mathrm{r}}) \mathbf{S} \mathbf{H}(\mathbf{F}_{\mathrm{t}},\mathbf{F}_{\mathrm{r}})^H\right).
\end{align}

Based on the cosine pattern model in \eqref{deqn_ex1b}, we assume that the initial boresights of the transmit and receive RAs are aligned with the positive and negative $x$-axis, respectively. The corresponding MIMO channel capacity maximization problem jointly optimizes the transmit and receive RA orientation matrices $\mathbf{F}_{\mathrm{t}}$ and $\mathbf{F}_{\mathrm{r}}$, as well as the transmit covariance matrix $\mathbf{S}$. Then, the resulting problem is formulated as
\begin{subequations}\label{eq:MIMO}
    \begin{alignat}{2}
		\hspace{-0.4cm} \text{(P-MIMO):}& \max_{\mathbf{F}_{\mathrm{t}},\mathbf{F}_{\mathrm{r}},\mathbf{S}\succeq \mathbf{0}} \; C_{\mathrm{MIMO}}\left(\mathbf{F}_{\mathrm{t}},\mathbf{F}_{\mathrm{r}},\mathbf{S}\right) & \label{eq:MIMO-A}\\
		\mbox{s.t.} \quad 
		& 0\leq \arccos \hspace{-0.05cm}\left((\vec{\mathbf{f}}_{\perp,n_{\mathrm{t}}}^{\mathrm{(t)}})^T \mathbf{e}_1\right)\leq \theta_{\max}, \forall n_{\mathrm{t}}\hspace{-0.05cm} \in \mathcal{N}_{\mathrm{t}},\hspace{-0.2cm} \label{eq:MIMO-B}\\
        & 0\leq \arccos \hspace{-0.05cm} \left(-(\vec{\mathbf{f}}_{\perp,n_{\mathrm{r}}}^{\mathrm{(r)}})^T \mathbf{e}_1\right)\leq \theta_{\max}, \forall  n_{\mathrm{r}}\hspace{-0.05cm}\in \mathcal{N}_{\mathrm{r}},\hspace{-0.2cm} \label{eq:MIMO-C}\\
		& \|\vec{\mathbf{f}}_{\perp,n_{\mathrm{t}}}^{\mathrm{(t)}}\| = 1,\; \forall n_{\mathrm{t}}\in \mathcal{N}_{\mathrm{t}}, \label{eq:MIMO-D}\\
        & \|\vec{\mathbf{f}}_{\perp,n_{\mathrm{r}}}^{\mathrm{(r)}}\| = 1,\; \forall n_{\mathrm{r}}\in \mathcal{N}_{\mathrm{r}}, \label{eq:MIMO-E}\\
        & \operatorname{Tr}(\mathbf{S})\leq P,\label{eq:MIMO-F}
   \end{alignat}
\end{subequations}
where constraint \eqref{eq:MIMO-F} ensures that the transmit power does not exceed its maximum value $P$.

{\color{black}Problem (P-MIMO) is highly non-convex due to the strong coupling between the transmit covariance matrix and the orientation-dependent MIMO channel matrix. The transmit covariance $\mathbf{S}$ determines power allocation across spatial eigenmodes, while the transmit and receive orientation matrices $\mathbf{F}_{\mathrm{t}}$ and $\mathbf{F}_{\mathrm{r}}$ reshape the channel matrix and modify its singular values and eigenstructure. Moreover, the rotational-range constraints in \eqref{eq:MIMO-B}--\eqref{eq:MIMO-C} and the unit-norm constraints in \eqref{eq:MIMO-D}--\eqref{eq:MIMO-E} create a non-convex spherical-cap feasible set, further complicating the joint optimization.
To address this difficulty, the alternating optimization (AO) algorithm in~\cite{Zheng2026JointTransceiver} can be employed. For fixed $\mathbf{F}_{\mathrm{t}}$ and $\mathbf{F}_{\mathrm{r}}$, the transmit covariance matrix $\mathbf{S}$ is optimally updated via eigenmode transmission and water-filling. With $\mathbf{S}$ fixed, the RA orientation matrices are optimized by alternately updating the transmit-side and receive-side pointing vectors. Each update decomposes the high-dimensional orientation problem into a sequence of single-RA subproblems, which can be efficiently solved over the spherical-cap domain using a Riemannian Frank-Wolfe method. Iteratively updating $\mathbf{S}$, $\mathbf{F}_{\mathrm{r}}$, and $\mathbf{F}_{\mathrm{t}}$ yields a high-quality locally optimal solution with manageable computational complexity.}


\begin{figure}[!t]\centering
	\includegraphics[width=3in]{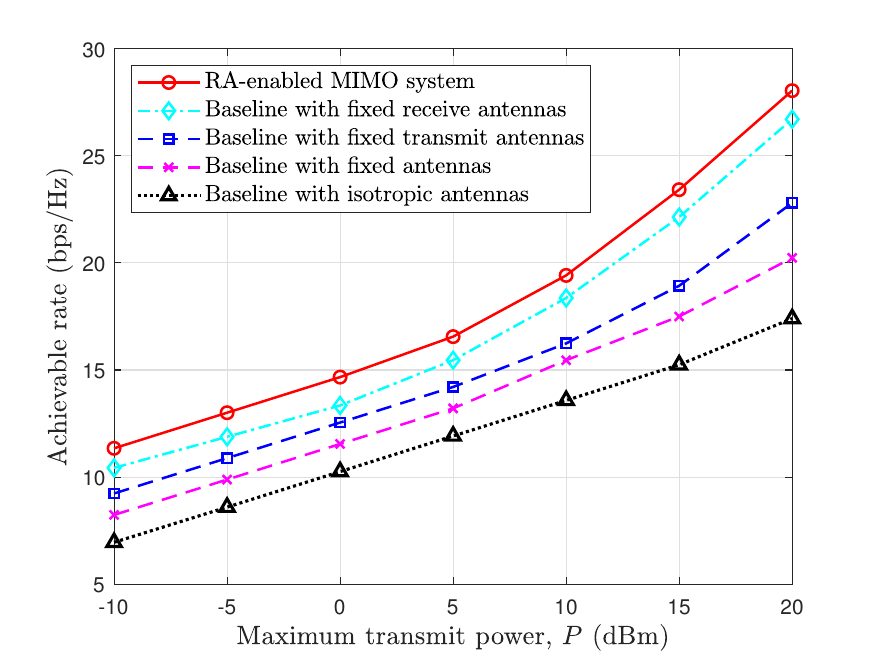}
	\caption{\color{black}Achievable rate versus maximum transmit power $P$ for the RA-enabled MIMO system.}
	\label{fig_MIMO_capacity}
\end{figure}

{\color{black}To demonstrate the enhanced spatial multiplexing capability of RA-enabled MIMO, we consider a system with $N_{\mathrm{t}}=4\times 4$ transmit RAs and $N_{\mathrm{r}}=4\times 4$ receive RAs. In addition to fixed-antenna and isotropic-antenna baselines, two one-sided benchmarks are included: one optimizes only the transmit-side RA orientations, and the other optimizes only the receive-side orientations.
Fig.~\ref{fig_MIMO_capacity} shows the achievable rate versus the maximum transmit power $P$. The proposed RA-enabled MIMO system consistently achieves the highest rate across the entire power range, demonstrating that joint transceiver orientation optimization can effectively reshape the MIMO channel, enhance the effective channel gain, and improve spatial multiplexing capability.
Furthermore, the proposed scheme outperforms both one-sided benchmarks, indicating that optimizing the RA orientations at only one side cannot fully exploit the orientation-domain DoFs. For the considered setup, the baseline with fixed receive antennas performs better than that with fixed transmit antennas, suggesting that transmit-side orientation control plays a more dominant role, while receive-side adaptation remains essential for maximizing end-to-end MIMO capacity. For a given target rate, the RA-enabled MIMO system also requires lower transmit power than conventional fixed-antenna MIMO systems.}

\subsection{RA-Enabled Multi-User System}
The enhanced array and multiplexing gains enabled by the RA architecture can also be leveraged in multi-user communication systems~\cite{Zhu2026Rotatable,Dai2026RotatableAntenna,Li2026Energy,Jiang2026Rotatable}.
In single-user scenarios, all RAs can orient their boresights toward the same user to maximize the directional gain.
In contrast, in multi-user settings, different RAs can steer their boresights toward different user directions, thereby enhancing directional gains for multiple users simultaneously.
Moreover, in multipath environments, RA orientations/boresights can be adapted to the spatial distribution of the users and scatterers to balance multiple propagation paths and maximize the effective channel gain.

To illustrate this capability, we consider an RA-enabled multi-user uplink system operating in a multipath channel, where $K$ users (each equipped with a single isotropic antenna) simultaneously transmit to a BS equipped with a UPA comprising $N$ RAs. Using the multipath channel model in \eqref{deqn_ex18b}, the SINR for decoding the signal of user~$k$ at the BS is given by
\begin{align}
	\label{deqn_ex1e}
	\gamma_k = \frac{P_k |\mathbf{w}_{\mathrm{r},k}^{T} \mathbf{h}_{\mathrm{MP},k}(\mathbf{F})|^2}{\sum_{j\ne k}{P_j|\mathbf{w}_{\mathrm{r},k}^{T} \mathbf{h}_{\mathrm{MP},j}(\mathbf{F})|^2}+\sigma^2},
\end{align}
where $P_k$ is the transmit power of user~$k$, and $\mathbf{w}_{\mathrm{r},k}\in \mathbb{C}^{N\times 1}$ is the linear receive beamforming vector for user~$k$ with $\|\mathbf{w}_{\mathrm{r},k}\| = 1$.

Adopting the cosine pattern model in \eqref{deqn_ex1b}, we formulate the following max-min SINR problem by jointly optimizing the receive beamforming matrix $\mathbf{W}_{\mathrm{r}}\triangleq \left[\mathbf{w}_{\mathrm{r},1},\mathbf{w}_{\mathrm{r},2},\dots,\mathbf{w}_{\mathrm{r},K}\right] \in \mathbb{C}^{N\times K}$ and the RA pointing vectors $\{\vec{\mathbf{f}}_{\perp,n}\}$:
\begin{subequations}\label{eq:multiuser}
	\begin{alignat}{2}
		\hspace{-0.25cm}\text{(P-MU):} \; \max_{\mathbf{W}_{\mathrm{r}},\{\vec{\mathbf{f}}_{\perp,n}\}} \quad & \min_{k}\; \gamma_k & \label{eq:multiuser-A}\\
		\mbox{s.t.} \quad 
		& 0\leq \arccos (\vec{\mathbf{f}}_{\perp,n}^T \mathbf{e}_1)\leq \theta_{\max},\; \forall n,\hspace{-0.2cm} \label{eq:multiuser-B}\\
		& \|\vec{\mathbf{f}}_{\perp,n}\| = 1,\; \forall n, \label{eq:multiuser-C}\\
		& \|\mathbf{w}_{\mathrm{r},k}\| = 1,\; \forall k.\label{eq:multiuser-D}
	\end{alignat}
\end{subequations}
Since the optimization variables are tightly coupled and the objective in \eqref{eq:multiuser-A} is highly non-convex, it is difficult to obtain the optimal solution of problem (P-MU) directly.
To address this issue, an AO algorithm can be developed to alternately optimize the receive beamforming and the RA pointing vectors in an iterative manner. Specifically, for fixed RA pointing vectors, linear receivers such as zero-forcing (ZF) and minimum mean-square error (MMSE) beamforming can be used to enhance the SINR of each user. On the other hand, for fixed ZF/MMSE beamforming, the RA pointing-vector subproblem remains non-convex.
By relaxing \eqref{eq:multiuser-C} to $\|\vec{\mathbf{f}}_{\perp,n}\|\le1$, the successive convex approximation (SCA) technique can be applied to obtain a locally optimal solution iteratively.
Furthermore, for the special case of $\rho = 1$ in \eqref{deqn_ex6b} and \eqref{deqn_ex16b}, problem (P-MU) can be reformulated as a convex semidefinite program (SDP), which can be solved efficiently without iteration, as shown in~\cite{Zheng2026Rotatable}.

\begin{figure}[!t]\centering
	\includegraphics[width=3in]{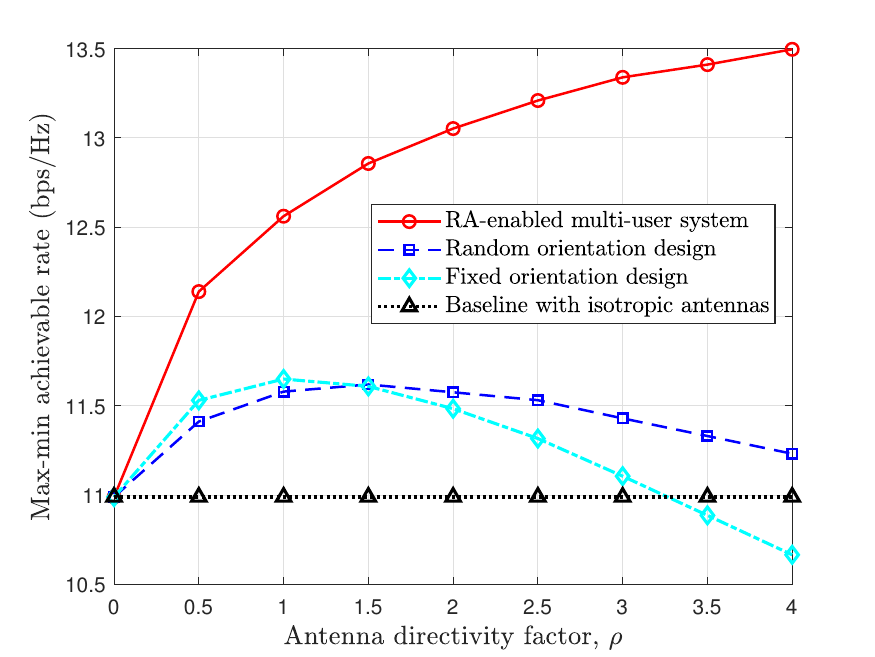}
	\caption{Max-min achievable rates of different systems versus the antenna directivity factor $\rho$.}
	\label{fig_multi_pattern}
\end{figure}
To evaluate the performance gain of RAs in multi-user communication under multipath propagation, Fig.~\ref{fig_multi_pattern} depicts the max-min achievable rate
\begin{align}
	\label{deqn_ex2e}
	C_{\mathrm{MU}} = \min\limits_{k} \log_2(1+\gamma_k) = \log_2(1 + \min\limits_{k}\gamma_k),
\end{align}
for different schemes versus the antenna directivity factor $\rho$.
Specifically, the BS is equipped with a square UPA comprising $N=4\times4$ RAs, serving $K=4$ users uniformly distributed in four distinct directions in front of the BS. User distances are drawn uniformly from the interval $[30,50]$~m, and $Q=8$ scatterer clusters are randomly placed around the users. The maximum deviation angle is $\theta_{\max}=\pi/6$, and the transmit power of each user is $P_k=10$~dBm.

The results in Fig.~\ref{fig_multi_pattern} show that the max-min achievable rate of the RA system increases with $\rho$, since a larger directivity factor yields higher boresight gain and a narrower mainlobe.
This enables the RA array to more effectively achieve directional gains across multiple user directions, thereby providing a higher max-min achievable rate.
In contrast, for fixed-antenna systems, the max-min rate decreases when $\rho\ge1$, since a narrower mainlobe reduces the directional gain for users located away from the array’s main pointing direction.
Additionally, although the random orientation design can disperse the radiation power in several directions, it is significantly inferior to the RA system since it fails to strategically allocate antenna orientations to balance directional gains across multipath channels and fairly improve the communication performance of all users. 
These results highlight the importance of RA architectures for improving multi-user performance, especially when antennas exhibit strong directivity and narrow mainlobes.

\begin{figure}[!t]\centering
	\includegraphics[width=2.25in]{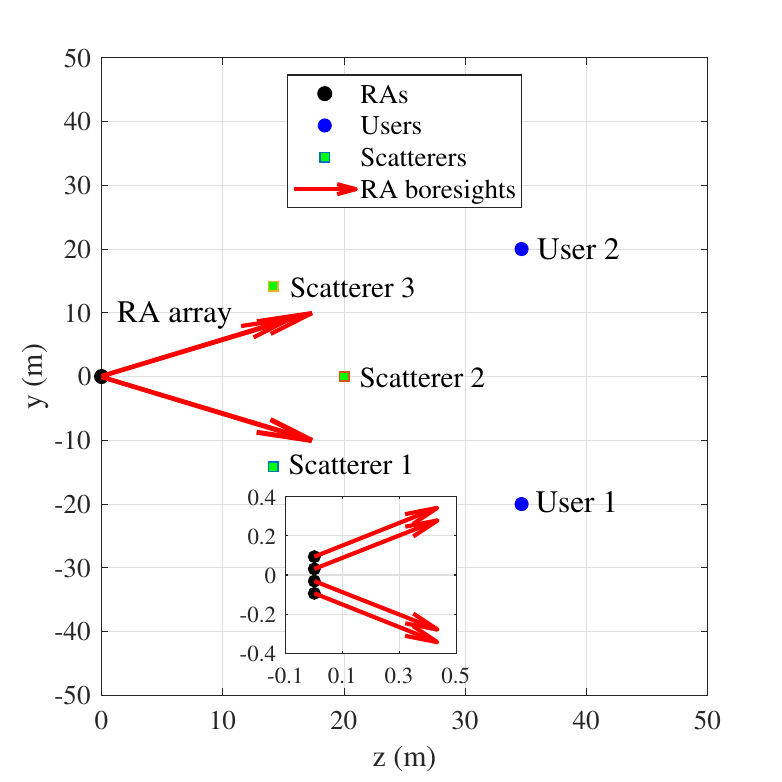}
	{\color{black}\caption{Illustration of optimized RA boresights in an RA-enabled multi-user system.}
	\label{fig_RA_pointing}}
\end{figure}
{\color{black}To further illustrate how RA orientation optimization exploits the propagation geometry, Fig.~\ref{fig_RA_pointing} visualizes the optimized RA boresight directions in a representative multi-user scenario. For clarity, a simplified $4\times 1$ ULA with $K=2$ users and $Q=3$ scatterer clusters is considered, and the optimized RA boresights are plotted in the $y$-$z$ plane.
As shown in Fig.~\ref{fig_RA_pointing}, the optimized RA boresights naturally divide into two groups, each aligned with a dominant user-specific propagation direction. One group points toward User~1, where Scatterer~1 lies close to the corresponding path, while the other group points toward User~2, where Scatterer~3 is located. Scatterer~2, positioned between the two users, is not selected as a primary boresight direction. This demonstrates that RA orientation optimization prioritizes dominant user-specific paths rather than indiscriminately pointing toward all users or scatterers. Consequently, the RA array can flexibly allocate antenna orientations to different dominant paths, providing additional spatial DoFs for multi-user transmission.}

\subsection{RA-Enabled Wideband System}

By dynamically adjusting antenna orientations/boresights to capture dominant propagation paths while attenuating delayed or weaker multipath components arriving from other directions, RAs provide additional spatial DoFs that can be exploited to enhance overall system performance across multiple frequency bands in wideband systems. This spatial selectivity helps mitigate frequency-selective fading and yields a more favorable effective channel response across subcarriers. In this subsection, we study an RA-enabled wideband OFDM system to demonstrate how RA architectures balance multiple propagation paths and enhance wideband communication performance.
{\color{black}The wideband system considered below is based on the space-frequency channel response in Section~\ref{Wideband_channel} and therefore naturally accommodates near-field wideband RA operation. For high-frequency large-aperture arrays operating within the Rayleigh distance, the propagation distance, delay, phase, and directional gain may vary across both RAs and subcarriers. As a result, the optimal RA orientations are determined by the aggregate wideband channel quality across subcarriers rather than by a single far-field geometric direction.
}

Extending the multi-user system to the wideband setting, we consider an OFDM channel with $L$ subcarriers as modeled in \eqref{deqn_ex21b}.
Let $a_{k,l}$ be a binary variable indicating whether subcarrier~$l$ with $l\in \mathcal{L}$ is assigned to user~$k$ ($a_{k,l}=1$) or not ($a_{k,l}=0$).
If $a_{k,l}=1$ and perfect synchronization (including symbol timing, frame alignment, and carrier-frequency-offset compensation) is assumed at the BS, the received signal from user~$k$ on subcarrier~$l$ is given by
\begin{align}
	\label{deqn_ex1f}
	\tilde{\mathbf{y}}_{k,l} = \mathbf{h}_{\mathrm{WB},k,l}(\mathbf{F})\sqrt{P_{k,l}} {s_{k,l}} + \mathbf{n}_l,
\end{align}
where $s_{k,l}$ and $P_{k,l}$ denote the transmitted symbol and power of user~$k$ on subcarrier~$l$, respectively, and $\mathbf{n}_l \sim \mathcal{N}_c(\bm{0},\tilde{\sigma}_l^2\bm{I}_N)$ is the additive white Gaussian noise (AWGN) vector of subcarrier~$l$ with variance $\tilde{\sigma}_l^2$.

Since each subcarrier is allocated to at most one user, there is no inter-user interference on any subcarrier. 
Thus, the BS can apply maximum-ratio combining (MRC), i.e., $\mathbf{w}_{\mathrm{r},k,l} = \frac{\mathbf{h}_{\mathrm{WB},k,l}^{\ast}(\mathbf{F})}{\|\mathbf{h}_{\mathrm{WB},k,l}(\mathbf{F})\|}$, to maximize the received SNR. The achievable rate of user~$k$ in bits per second per Hertz (bps/Hz) is then given by
\begin{align}
	\label{deqn_ex2f}
	\hspace{-0.25cm}\tilde{R}_k = \frac{1}{L + L_{\mathrm{CP}}}\sum\limits_{l = 1}^L {{a_{k,l}}{{\log }_2}\left( {1 + \frac{{{P_{k,l}}{{\|\mathbf{h}_{\mathrm{WB},k,l}(\mathbf{F}) \|}^2}}}{{\tilde{\sigma}_l^2}}} \right)},\hspace{-0.2cm}
\end{align}
where $L_{{\rm{CP}}}$ is the cyclic-prefix (CP) length.

We now formulate the sum-rate maximization problem by jointly optimizing the RA pointing vectors $\{\vec{\mathbf{f}}_{\perp,n}\}$ and the subcarrier allocation variables $\{a_{k,l}\}$.
\begin{subequations}\label{eq:wideband}
	\begin{alignat}{2}
		\hspace{-0.25cm}\text{(P-WB):} \; \max_{\{\vec{\mathbf{f}}_{\perp,n}\},\{a_{k,l}\}} \quad & \sum_{k=1}^{K} \tilde{R}_k & \label{eq:wideband-A}\\
		\mbox{s.t.} \quad 
		& 0\leq \arccos (\vec{\mathbf{f}}_{\perp,n}^T \mathbf{e}_1)\leq \theta_{\max},\; \forall n,\hspace{-0.2cm} \label{eq:wideband-B}\\
		& \|\vec{\mathbf{f}}_{\perp,n}\| = 1,\; \forall n, \label{eq:wideband-C}\\
        & \sum\limits_{k = 1}^K {{a_{k,l}}}  \le 1,\;\forall l, \label{eq:wideband-D}\\
        & \sum\limits_{l = 1}^L {{a_{k,l}}} {P_{k,l}} \le {P_k},\;\forall k, \label{eq:wideband-E}\\
        & a_{k,l}\in\{0,1\},\; \forall k,l, \label{eq:wideband-F}
	\end{alignat}
\end{subequations}
where constraint \eqref{eq:wideband-D} ensures that each subcarrier is allocated to at most one user to avoid inter-user interference, constraint \eqref{eq:wideband-E} guarantees that the total transmit power of user~$k$ across all subcarriers does not exceed its maximum value $P_k$, and constraint \eqref{eq:wideband-F} ensures that $a_{k,l}$ is a binary variable.

Problem (P-WB) is non-convex due to the coupling between the RA orientation and subcarrier allocation.
To tackle this difficulty, we can alternately optimize the subcarrier allocation $\{a_{k,l}\}$ and the RA pointing vectors $\{\vec{\mathbf{f}}_{\perp,n}\}$ in an iterative manner. On the one hand, for fixed RA pointing vectors $\{\vec{\mathbf{f}}_{\perp,n}\}$, the subproblem of optimizing $\{a_{k,l}\}$ is a binary integer program that can be solved using standard optimization tools. On the other hand, for fixed $\{a_{k,l}\}$, the subproblem of optimizing $\{\vec{\mathbf{f}}_{\perp,n}\}$ remains highly non-convex. By approximating the objective in \eqref{eq:wideband-A} as a concave function with respect to $\{\vec{\mathbf{f}}_{\perp,n}\}$ and relaxing the equality constraint \eqref{eq:wideband-C}, the SCA technique can be applied to solve this subproblem efficiently.

\begin{figure}[!t]\centering
	\includegraphics[width=3in]{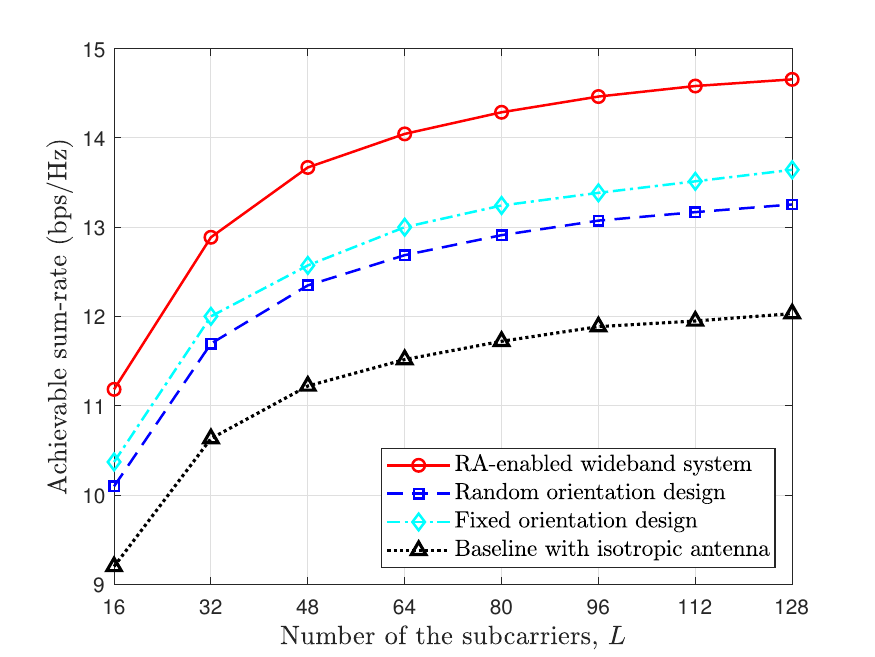}
	\caption{Achievable sum-rates of different schemes versus the number of subcarriers $L$.}
	\label{fig_polar_carrier}
\end{figure}
To validate the benefits of the RA architecture in wideband systems, we consider an OFDM system with carrier frequency $f_c=2.4$~GHz, bandwidth $B=40$~MHz, $L=64$ subcarriers, and CP length $L_{\mathrm{CP}}=6$.
Other system parameters follow those of the previous multi-user setup.
Fig.~\ref{fig_polar_carrier} shows the achievable sum-rate versus the number of subcarriers. It is observed that the sum-rate increases with $L$ for all schemes because the relative CP overhead decreases as the number of subcarriers grows, improving spectral efficiency. Additionally, for RA systems, a larger $L$ also provides greater flexibility in subcarrier allocation, which can be jointly optimized with antenna orientations.
The RA-enabled wideband system consistently achieves the highest sum-rate by jointly optimizing pointing vectors and subcarrier assignment. This performance gain stems from the ability of RAs to flexibly adjust their boresights to balance multiple propagation paths across subcarriers, thereby improving the overall wideband channel quality.

{\color{black}It is also worth noting that RA architectures offer an effective means to mitigate \emph{beam-squint} effects in wideband systems without relying on costly true-time-delay hardware. For example, the authors in \cite{Xie2025THz} studied a wideband beam coverage problem in which transmit beamforming and array rotation are jointly optimized to maximize the worst-case beam gain over a given angular region and frequency band. Their results show that beam-squint effects can be significantly alleviated by employing a rotatable ULA, since the severity of beam squint inherently depends on the angle of departure. By adjusting the array orientation, the RA architecture can partially compensate for frequency-dependent beam deviations and improve wideband beam coverage.}

\subsection{RA-Enabled ISAC System}
ISAC is widely regarded as a key enabling technology for 6G wireless networks, offering unified communication and sensing functionalities through shared hardware resources.
Leveraging the additional spatial DoFs introduced by antenna rotation, RA architectures can adaptively control their boresights according to ISAC performance requirements and the spatial distribution of communication users and sensing targets.
Specifically, by steering antenna orientations/boresights toward intended users or regions of interest, an RA array can significantly enhance communication capacity and link reliability, while also increasing the echo power for sensing, thereby improving detection accuracy and sensing range.
Furthermore, joint optimization of RA orientations/boresights and probing signals enables flexible trade-offs between communication coverage and sensing performance, improving both spectral and energy efficiency~\cite{Wu2025Rotatable,Wu2026Antenna}.

To examine the potential of RAs for ISAC applications, we consider an RA-enabled ISAC system in which a BS transmits both communication signals and dedicated probing signals to simultaneously serve $K$ downlink users and sense a potential target located within a region $\mathcal{A}$. The BS transmits and receives probing signals to estimate the target’s position. To avoid mutual interference between communication and sensing, we assume that they share the same BS architecture but use separate time or frequency resources.
Let $\mathbf{q}_{\mathrm{U},k}\in \mathbb{R}^{3\times 1}$ denote the position of user~$k$. By denoting the transmit beamforming vector for user~$k$ as $\mathbf{w}_{\mathrm{t},k}\in \mathbb{C}^{N\times 1}$, the achievable rate of user~$k$ can be expressed as
\begin{align}
	\label{deqn_ex1g}
	\hspace{-0.35cm}\bar{R}_k(\mathbf{F},\mathbf{W}_{\mathrm{t}}) \hspace{-0.05cm}=\hspace{-0.05cm} \log_2\hspace{-0.1cm} \left(1 + \frac{|\mathbf{h}^T(\mathbf{F},\mathbf{q}_{\mathrm{U},k}) \mathbf{w}_{\mathrm{t},k}|^2}{\sum_{j\ne k}{|\mathbf{h}^T(\mathbf{F},\mathbf{q}_{\mathrm{U},k}) \mathbf{w}_{\mathrm{t},j}|^2} \hspace{-0.05cm}+\hspace{-0.05cm} \sigma_k^2}\right)\hspace{-0.1cm},\hspace{-0.2cm}
\end{align}
where $\mathbf{h}(\mathbf{F},\mathbf{q})$ denotes the channel vector between the RA-based BS and any spatial point $\mathbf{q}\in \mathbb{R}^{3\times 1}$, which can be constructed using the channel models in Section~\ref{Channel_Model}, $\mathbf{W}_{\mathrm{t}}\triangleq \left[\mathbf{w}_{\mathrm{t},1},\mathbf{w}_{\mathrm{t},2},\dots,\mathbf{w}_{\mathrm{t},K}\right] \in \mathbb{C}^{N\times K}$ is the transmit beamforming matrix, and $\sigma_k^2$ is the noise power at user~$k$.

For sensing, we model the potential target as an unstructured point located at $\mathbf{q}_{\mathrm{T}}\in\mathcal{A}$. 
The round-trip channel matrix for the BS to receive the probing signal reflected by the target, denoted by $\bar{\mathbf{H}}(\mathbf{F},\mathbf{q}_{\mathrm{T}}) \in \mathbb{C}^{N\times N}$, is expressed as
\begin{align}
	\label{deqn_ex2g}
	\bar{\mathbf{H}}(\mathbf{F},\mathbf{q}_{\mathrm{T}}) = \sigma_{\mathrm{T}} \mathbf{h}(\mathbf{F},\mathbf{q}_{\mathrm{T}}) \mathbf{h}^T(\mathbf{F},\mathbf{q}_{\mathrm{T}}),\;\mathbf{q}_{\mathrm{T}}\in \mathcal{A},
\end{align}
where $\sigma_{\mathrm{T}}$ denotes the RCS of the target.
Let $T_{\mathrm{s}}$ denote the total number of symbols during one sensing period. Let $\mathbf{s}[t]\in \mathbb{C}^{N\times 1}$ represent the $t$-th transmit probing symbol sent by the BS, where $t\in \{1,2,\dots,T_{\mathrm{s}}\}.$
Using the echo power as a practical sensing metric, the received echo power at $\mathbf{q}_{\mathrm{T}}$ is given by
\begin{align}
	\label{deqn_ex3g}
	\hspace{-0.15cm}\bar{P}(\mathbf{F},\bar{\mathbf{S}},\mathbf{q}_{\mathrm{T}}) =& \mathbb{E}\left\{\|\bar{\mathbf{H}}(\mathbf{F},\mathbf{q}_{\mathrm{T}}) \mathbf{s}[t]\|^2\right\} \nonumber\\
	=& |\sigma_{\mathrm{T}}|^2 \|\mathbf{h}(\mathbf{F},\mathbf{q}_{\mathrm{T}})\|^2 \mathbf{h}^T(\mathbf{F},\mathbf{q}_{\mathrm{T}}) \bar{\mathbf{S}} \mathbf{h}^{\ast}(\mathbf{F},\mathbf{q}_{\mathrm{T}}),\hspace{-0.1cm}
\end{align}
where $\bar{\mathbf{S}}\triangleq \mathbb{E}\left[\mathbf{s}[t]\mathbf{s}^H[t] \right]$ is the probing-signal covariance matrix.
Additionally, the CRB, which is widely adopted as a theoretical bound for localization performance, can also serve as a metric to characterize the sensing performance in RA-enabled ISAC systems, as shown in \cite{Zhou2025Rotatable}. In general, a higher echo power received from the target leads to a lower CRB, and thus better sensing performance.

To achieve consistent regional sensing over $\mathcal{A}$ while guaranteeing communication performance, we jointly optimize the RA orientation matrix $\mathbf{F}$, the communication beamforming matrix $\mathbf{W}_{\mathrm{t}}$, and the probing-signal covariance matrix $\bar{\mathbf{S}}$ under the cosine pattern model in \eqref{deqn_ex1b}. The goal is to maximize the minimum received echo power over $\mathcal{A}$ while ensuring that each user meets a minimum communication-rate requirement.
The resulting optimization problem is formulated as
\begin{subequations}\label{eq:ISAC}
	\begin{alignat}{2}
		\hspace{-0.5cm}(\text{P-ISAC}): \max_{\mathbf{F},\mathbf{W}_{\mathrm{t}},\bar{\mathbf{S}}\succeq \mathbf{0}} \quad & \min_{\mathbf{q}_{\mathrm{T}}\in \mathcal{A}}{\bar{P}(\mathbf{F},\bar{\mathbf{S}},\mathbf{q}_{\mathrm{T}})} & \label{eq:ISAC-A}\\
		\mbox{s.t.} \quad
		& \bar{R}_k(\mathbf{F},\mathbf{W}_{\mathrm{t}}) \geq R_{\min},\; \forall k, \label{eq:ISAC-B}\\
		& \sum_{k=1}^{K}{\|\mathbf{w}_{\mathrm{t},k}\|^2} \leq P_{\mathrm{max,c}}, \label{eq:ISAC-C}\\
		& \operatorname{Tr}(\bar{\mathbf{S}}) \leq P_{\mathrm{max,s}}, \label{eq:ISAC-D}\\
		& 0 \leq \arccos(\vec{\mathbf{f}}_{\perp,n}^T \mathbf{e}_1) \leq \theta_{\max},\forall n,\hspace{-0.3cm} \label{eq:ISAC-E}\\
		& \|\vec{\mathbf{f}}_{\perp,n}\| = 1, \; \forall n, \label{eq:ISAC-F}
	\end{alignat}
\end{subequations}
where constraint \eqref{eq:ISAC-B} ensures that user~$k$ is served with its required minimum communication rate $R_{\min}$, and constraints \eqref{eq:ISAC-C} and \eqref{eq:ISAC-D} prevent the average transmit powers of the communication and sensing signals from exceeding their budgets $P_{\mathrm{max,c}}$ and $P_{\mathrm{max,s}}$, respectively.

Problem (P-ISAC) is difficult to solve due to the non-convexity of \eqref{eq:ISAC-A}, \eqref{eq:ISAC-B}, and \eqref{eq:ISAC-F}, as well as the semi-infinite nature of \eqref{eq:ISAC-A} since $\mathcal{A}$ is continuous.
To address this issue, we discretize $\mathcal{A}$ into $\tilde{M}$ sampling points $\{\mathbf{q}_{\mathrm{T},\tilde{m}}\}_{\tilde{m}=1}^{\tilde{M}}$, yielding the approximate problem:
\begin{subequations}\label{eq:ISACD}
	\begin{alignat}{2}
		(\text{P-ISAC-D}): \max_{\eta,\mathbf{F},\mathbf{W}_{\mathrm{t}},\bar{\mathbf{S}}\succeq \mathbf{0}} \quad & \eta & \label{eq:ISACD-A}\\
		\mbox{s.t.} \quad
		& \bar{P}(\mathbf{F},\bar{\mathbf{S}},\mathbf{q}_{\mathrm{T},\tilde{m}}) \geq \eta,\; \forall \tilde{m}, \label{eq:ISACD-B}\\
		& \eqref{eq:ISAC-B}-\eqref{eq:ISAC-F},\label{eq:ISACD-C}
	\end{alignat}
\end{subequations}
where $\eta$ denotes the minimum received echo signal power, and $\mathbf{q}_{\mathrm{T},\tilde{m}}$ with $\tilde{m}\in \mathcal{M}\triangleq \{1,2,\dots,\tilde{M}\}$ is the position of the $\tilde{m}$-th spatial sampling point.
Then, to address the non-convexity of problem (P-ISAC-D), we adopt a block coordinate descent (BCD) framework, which alternately optimizes $\{\mathbf{W}_{\mathrm{t}},\bar{\mathbf{S}}\}$ and $\mathbf{F}$ until convergence is achieved.
For a fixed RA orientation matrix $\mathbf{F}$, the subproblem of optimizing the transmit beamforming matrix $\mathbf{W}_{\mathrm{t}}$ and the probing-signal covariance matrix $\bar{\mathbf{S}}$ can be reformulated as an SDP and solved using the CVX solvers.
For a fixed transmit beamforming matrix $\mathbf{W}_{\mathrm{t}}$ and probing-signal covariance matrix $\bar{\mathbf{S}}$, it has been proved in \cite{Wu2026Antenna} that $\mathbf{h}^T(\mathbf{F},\mathbf{q}_{\mathrm{U},k}) \mathbf{w}_{\mathrm{t},k} \mathbf{w}_{\mathrm{t},k}^H \mathbf{h}^{\ast}(\mathbf{F},\mathbf{q}_{\mathrm{U},k})$ in \eqref{eq:ISAC-B} and $\mathbf{h}^T(\mathbf{F},\mathbf{q}_{\mathrm{T},\tilde{m}}) \bar{\mathbf{S}} \mathbf{h}^{\ast}(\mathbf{F},\mathbf{q}_{\mathrm{T},\tilde{m}})$ in \eqref{eq:ISACD-B} are convex with respect to the amplitude components of $\mathbf{h}(\mathbf{F},\mathbf{q}_{\mathrm{U},k})$ and $\mathbf{h}(\mathbf{F},\mathbf{q}_{\mathrm{T},\tilde{m}})$, respectively. Therefore, the subproblem of optimizing the RA orientation matrix $\mathbf{F}$ can be approximated as a convex problem and efficiently solved using the SCA technique.

\begin{figure}[!t]  \centering
	\includegraphics[width=3in]{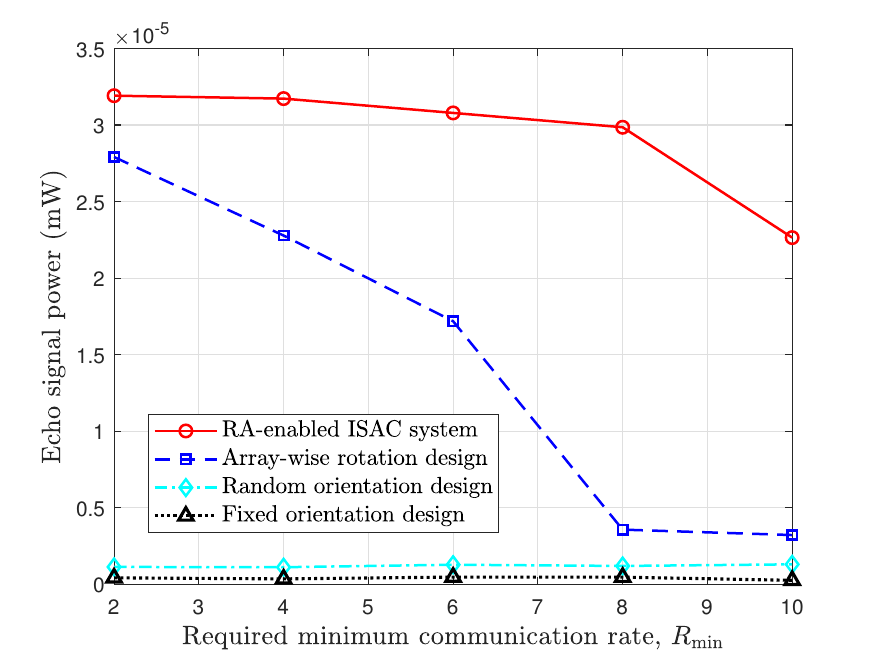} 
	\caption{Echo signal power of different systems versus the required minimum communication rate $R_{\min}$.}
	\label{fig_ISAC_rate}
\end{figure}
To illustrate the trade-offs between communication and sensing performance and to demonstrate the advantages of RAs in enhancing ISAC capability, we consider an ISAC system with $K = 3$ users and a circular horizontal sensing region centered at $[40\sin\frac{\pi}{3},40\cos\frac{\pi}{3},-10]^T$~m with a radius of $5$~m. Specifically, three users are uniformly distributed in three distinct directions in front of the BS at a distance of 50~m, and we uniformly select $\tilde{M} = 8$ spatial sampling points within this region. The BS employs a UPA with $N = 16$ RAs. The maximum average transmit powers allocated to communication and sensing are set to $P_{\mathrm{max,c}} = 30$~dBm and $P_{\mathrm{max,s}} = 30$~dBm, respectively.
Fig.~\ref{fig_ISAC_rate} shows the minimum received echo power versus the required minimum communication rate $R_{\min}$ for different schemes.
In the baseline system with array-wise rotation, the entire RA array shares a common orientation, which is optimized by solving a problem similar to (P-ISAC-D). 
It is observed that the proposed RA-enabled ISAC system consistently achieves a higher minimum echo power than all baseline schemes. This improvement stems from the ability of individual RAs to independently adjust their orientations/boresights, enabling the array to reconfigure its directional gain pattern in response to both the wireless environment and the ISAC performance requirements. As a result, the RA-enabled ISAC system significantly enlarges the achievable trade-off region between communication and sensing performance.
Although the array-wise rotation can also enhance the echo power compared with random or fixed-orientation designs, its ability to balance communication and sensing is fundamentally limited. This limitation is because individual antennas cannot be independently oriented, leading to suboptimal ISAC performance compared to the RA-enabled system. Moreover, the performance gap between the RA-enabled system and the array-wise rotation widens as $R_{\min}$ increases, highlighting the benefits of RA architectures in ISAC applications with stringent communication and sensing requirements.

{\color{black} \textit{Remark 1:} The optimization problems for the above representative RA systems are highly non-convex and are typically handled by AO/BCD and SCA methods, where different variable blocks (such as RA orientations, beamforming vectors, transmit power, subcarrier allocation, and other resource-allocation variables) are updated alternately. When generic convex solvers are used, the complexity of each convexified subproblem is polynomial in the number of optimization variables and constraints, while the overall complexity further depends on the number of iterations required for convergence.
As a result, these methods may be sensitive to initialization and may become computationally demanding for large-scale or rapidly time-varying systems.
To improve practical implementation, several enhancements can be incorporated, including dominant-direction-based initialization, warm-started SCA using previous time-slot solutions, and two-timescale RA control strategies. These techniques help reduce computational overhead and improve convergence behavior, thereby enabling more efficient real-time RA optimization.
}

\section{RA Channel Estimation/Acquisition} \label{sec:estimation}
Accurate CSI is essential for enabling precise control of antenna orientation/boresight and unlocking the full potential of RA systems.
Unlike other flexible antenna architectures, such as FAS/MA/6DMA that mainly enhance spatial DoFs through antenna repositioning, RA systems achieve spatial adaptability by reorienting the antenna boresight while keeping the antenna position unchanged.
Importantly, antenna rotation does not alter the underlying propagation geometry of the wireless environment, such as path loss, angles of arrival/departure (AoAs/AoDs), or delays. Instead, rotation only reorients the radiation pattern, whose shape is assumed known after antenna characterization/calibration.
Therefore, for any given antenna orientation, the radiation pattern can be treated as a known directional gain function (e.g., see \eqref{deqn_ex1b} and \eqref{deqn_ex2b}), so existing fixed-antenna channel estimation techniques remain applicable in RA systems after accounting for orientation-dependent gains.

\subsection{RA Orientation Scheduling for Channel Estimation}
\begin{figure}[!t]\centering
\includegraphics[width=3.5in]{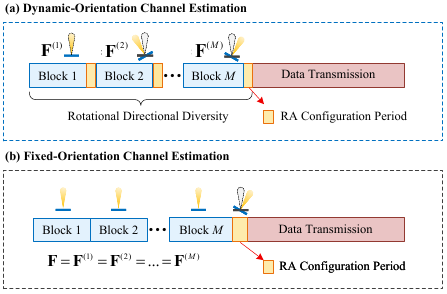}
	{\color{black}\caption{Two representative channel estimation strategies in RA systems: (a) Dynamic-orientation strategy; and (b) Fixed-orientation strategy.}
	\label{ChannelEst}}
\end{figure}

Depending on whether antenna rotation is exploited during the training phase, channel acquisition in RA systems can be broadly classified into two strategies, as illustrated in Fig. \ref{ChannelEst}.

\begin{itemize}
\item \textbf{Fixed-orientation channel estimation:} {\color{black}RA orientations remain fixed throughout the training interval. This yields a single-view observation and is equivalent to a conventional fixed-antenna system with a given radiation pattern. After channel acquisition, an additional RA configuration period is required to adjust the antenna orientations for data transmission. Thus, the main overhead arises from conventional pilot transmission and post-estimation RA configuration.} 
\item \textbf{Dynamic-orientation channel estimation:} {\color{black}RA orientations are periodically reconfigured across training blocks, with each block corresponding to a distinct antenna orientation. This provides multiple orientation-dependent pilot observations and introduces rotational directional diversity, which can improve parameter identifiability and estimation accuracy. However, it also incurs additional pilot overhead and orientation-switching cost.}
\end{itemize}
{\color{black} Overall, fixed-orientation estimation incurs lower orientation-control overhead but provides only a single observation view, whereas dynamic-orientation estimation exploits rotational directional diversity to improve estimation accuracy at the cost of increased RA configuration overhead. These two strategies therefore exhibit different overhead-accuracy trade-offs and can be selected based on system requirements and hardware capabilities.
}

Unless otherwise specified, we consider a general RA-enabled uplink communication system for ease of exposition, where a BS equipped with an $N$-element RA array serves $K$ single-antenna users in the presence of $Q$ scatterer clusters. 
Define the position of user $k$ as $ \mathbf{q}_{\mathrm{U},k}=[d_k \sin{\vartheta_k}\cos{\xi_k},d_k\sin{\vartheta_k}\sin{\xi_k},d_k\cos{\vartheta_k}]^T$, where $d_k$ denotes the distance between the BS's RA array and user~$k$, and $\vartheta_k$ and $\xi_k$ denote the zenith and azimuth angles, respectively. The location of scatterer cluster $q$, i.e., $ \mathbf{q}_{\mathrm{C},q}=[d_q \sin{\tilde{\vartheta}_q}\cos{\tilde{\xi}_q},d_q\sin{\tilde{\vartheta}_q}\sin{\tilde{\xi}_q},d_q\cos{\tilde{\vartheta}_q}]^T$, can be similarly defined.
Under geometric near-field propagation, the overall multipath channel between the BS and user $k$ in \eqref{deqn_ex18b} can be reformulated as
\begin{equation}
\begin{aligned}
    \mathbf{h}_k(\mathbf{F})
    &=\mathbf{h}_k^{\text{LoS}}(\mathbf{F})+\mathbf{h}_k^{\text{NLoS}}(\mathbf{F})\\
&=\tilde{\beta}_{k,0}\tilde{\mathbf{b}}_{k,0}(\mathbf{F},{\mathbf{q}}_{\mathrm{U},k}) + \sum_{q=1}^{Q}\tilde{\beta}_{k,q} \tilde{\mathbf{b}}_{k,q}(\mathbf{F},{\mathbf{q}}_{\mathrm {C}, q}), 
\end{aligned}
\label{channel_model}
\end{equation}
where $\tilde{\beta}_{k,0}=\frac{\sqrt{\beta_0}}{d_{k}}$ and $\tilde{\beta}_{k,q}=\frac{\sigma_q{\beta_0}}{{d}_{q}\bar{d}_{k,q}}$ denote the propagation coefficients of the LoS and the $q$-th NLoS components, respectively. 
The effective near-field array response vector $\tilde{\mathbf{b}}_{k,0}(\mathbf{F},{\mathbf{q}}_{\mathrm{U},k})$ for the LoS channel is defined as
\begin{align}\label{effective_array}
\tilde{\mathbf{b}}_{k,0}(\mathbf{F},{\mathbf{q}}_{\mathrm{U},k})= \text {diag}(\tilde{\mathbf{g}}_{\mathrm{U},k} (\mathbf{F}))^{\frac{1}{2}}\tilde{\mathbf{a}}_{k,0}(N,{\mathbf{q}}_{\mathrm{U},k}),
\end{align}
where $\tilde{\mathbf{g}}_{\mathrm{U},k} (\mathbf{F})=[{g}_k(\vec{\mathbf{f}}_1), \ldots,{g}_k(\vec{\mathbf{f}}_N)]^T$ denotes the directional gain vector, $ \tilde{\mathbf{a}}_{k,0}(N,{\mathbf{q}}_{\mathrm{U},k})= [\frac{d_{k}}{d_{1,k}}e^{-j\frac{2\pi}{\lambda}{d_{1,k}} },\ldots, \frac{d_{k}}{d_{N,k}}e^{-j\frac{2\pi}{\lambda}{d_{N,k}}}]^T$ denotes the near-field array response vector with $d_{n,k}$ being the distance between RA $n$ and user $k$. Then, the effective array response vector $\tilde{\mathbf{b}}_{k,q}$ for the NLoS channel can be similarly defined as in \eqref{effective_array}, i.e.,
\begin{align}\label{effective_array_scatter}
\tilde{\mathbf{b}}_{k,q}(\mathbf{F},{\mathbf{q}}_{\mathrm{C},q})= \text {diag}(\tilde{\mathbf{g}}_{\mathrm{C},q} (\mathbf{F}))^{\frac{1}{2}}\tilde{\mathbf{a}}_{k,q}(N,{\mathbf{q}}_{\mathrm{C},q}).
\end{align}
where $\tilde{\mathbf{g}}_{\mathrm{C},q} (\mathbf{F})=[\tilde{g}_q(\vec{\mathbf{f}}_1), \ldots,\tilde{g}_q(\vec{\mathbf{f}}_N)]^T$ denotes the directional gain vector with respect to scatterer cluster $q$, and $\tilde{\mathbf{a}}_{k,q}(N,{\mathbf{q}}_{\mathrm{C},q})$ denotes the array response vector with respect to scatterer cluster $q$.
Note that when the users and scatterers are located in the far-field region of the BS array, the distance variation across antennas becomes negligible in amplitude but approximately linear in phase. 
Thus, the directional gain vector $\tilde{\mathbf g}_{\mathrm{U},k}(\mathbf F) $ reduces to ${\mathbf g}_{k}(\mathbf F,\vec{\mathbf{q}}_{\mathrm{U},k}) $ as shown in \eqref{deqn_ex12b} or \eqref{deqn_ex13b}.
Moreover, the near-field array response vector $\tilde{\mathbf{a}}_{k,0}(N,{\mathbf{q}}_{\mathrm{U},k})$ reduces to a far-field steering vector $ \mathbf{a}_k(N,\vec{\mathbf{q}}_{\mathrm{U},k})$ in \eqref{deqn_ex9b} that only depends on the direction of user $k$, i.e., $ \vec{\mathbf{q}}_{\mathrm{U},k}= [\sin{\vartheta_k}\cos{\xi_k},\sin{\vartheta_k}\sin{\xi_k},\cos{\vartheta_k}]^T$.

During the uplink training phase, the users transmit mutually orthogonal pilots over $T_a$ time slots, based on which the BS estimates the key channel parameters for each user, e.g., propagation coefficients
$\mathcal{A}_k \triangleq\{\tilde{\beta}_{k,q}\}_{q=0}^{Q}$ and angular parameters
$\mathcal{B}_k  \triangleq\{ (\vartheta_{k},\xi_{k}),(\tilde{\vartheta}_{q},\tilde{\xi}_{q})\}_{q=1}^{Q}$, 
and then reconstructs the full RA channel $\{\mathbf{h}_k\}_{k=1}^{K}$.
Moreover, for user $k$, we define the channel parameters inherent in $\mathbf{h}_k$ as $\mathcal{C}_k= \{ \mathcal{A}_k,\mathcal{B}_k\}$.
In the following, we present two main estimation schemes for RA systems, with an emphasis on their main differences regarding the exploitation of the RA orientation during each channel training period. 
\subsubsection{Dynamic-Orientation Channel Estimation} 
In this scheme, the RA array dynamically adjusts antenna orientations/boresights across different pilot symbols to acquire multi-view observations. Let $x_{k}^{(t)}$ denote the pilot symbol transmitted by user $k$ in the $t$-th slot. Accordingly, the received signal at the BS is given by
\begin{align}
    \mathbf {y}^{(t)} = \sum_{k=1}^{K}\mathbf {h}_{k}\left(\mathbf {F}^{(t)}\right)x_{k}^{(t)} + \mathbf {n}^{(t)}, 
    \label{block_model}
\end{align}
with $t=1,\ldots,T_a$, where $\mathbf {F}^{(t)}$ is the RA orientation matrix in slot $t$, and $\mathbf {n}^{(t)}\sim \mathcal{N}_c(\mathbf{0},\sigma^2\mathbf{I}_N)$ denotes AWGN at the BS.
In principle, $\mathbf {F}^{(t)}$ can be updated at each slot to increase orientation diversity. 
However, frequent switching introduces control latency, calibration and feedback overhead, and higher complexity for multi-view data fusion. To reduce this overhead, a block-wise scanning strategy can be adopted, where RA orientations remain constant within each block and change only across blocks.  
Specifically, the $T_a$ slots are divided into $M$ blocks, each with $T_b=T_a/M$ consecutive slots.
Let $\mathbf {F}^{(m)}$ denote the RA orientation matrix in block $m$ with $m=1,\ldots,M$. Then, the received pilot signal at the BS during block $m$ is given by 
\begin{align}
    \mathbf {y}_{m}^{(t)} = \sum_{k=1}^{K}\mathbf {h}_{k}\left(\mathbf {F}^{(m)}\right)x_{m,k}^{(t)} + \mathbf {n}_{m}^{(t)}, 
    \label{block_model}
\end{align}
with $t=(m-1)T_b+1,\ldots, mT_b$ and $m=1,\ldots,M$, where $x_{m,k}^{(t)}$ denotes the pilot symbol transmitted by user $k$ in slot $t$ of block $m$, and $\mathbf {n}_{m}^{(t)}$ denotes the corresponding AWGN vector at the BS. 
By collecting measurements under different RA orientations, dynamic-orientation channel estimation enables joint processing of multi-view observations. 
Since key channel parameters remain invariant to antenna orientation, these observations can be coherently fused to improve estimation accuracy by exploiting the additional spatial DoFs offered by antenna rotation.
The performance gains, however, come at the cost of increased control overhead and signal processing complexity.
Efficient scanning strategies and low-complexity estimation algorithms are therefore essential to fully realize the benefits of dynamic-orientation channel acquisition.

\subsubsection{Fixed-Orientation Channel Estimation}
In this scheme, the RA array maintains fixed antenna orientations/boresights throughout the training period. Specifically, the orientation matrix $\mathbf F$ is preselected and kept unchanged over the $T_a$ pilot slots. Under this setting, the effective channels $\{\mathbf h_k(\mathbf F)\}_{k=1}^{K}$ remain constant during training, and thus the RA array operates equivalently to a conventional fixed-antenna array with a deterministic radiation pattern.
For notational simplicity, we drop the explicit dependence on $\mathbf F$ and define $\mathbf h_k \triangleq \mathbf h_k(\mathbf F)$. The received pilot signal at the BS is given by
\begin{align}
    \mathbf y^{(t)}= \sum_{k=1}^{K}\mathbf h_k x_k^{(t)} + \mathbf n^{(t)},
    \quad t=1,\ldots,T_a.
    \label{receive_fixed}
\end{align}
Since antenna orientations/boresights remain fixed during channel training, fixed-orientation channel estimation relies on a single-view observation of the propagation environment.
The main advantages of this approach are low implementation complexity, negligible orientation-control overhead, and full compatibility with existing channel estimation protocols and algorithms for conventional fixed-antenna systems. 
Nevertheless, the lack of orientation diversity limits sensing coverage and reduces estimation accuracy, particularly in rich multipath scenarios.  
\begin{figure}[!t]  \centering	\includegraphics[width=3in]{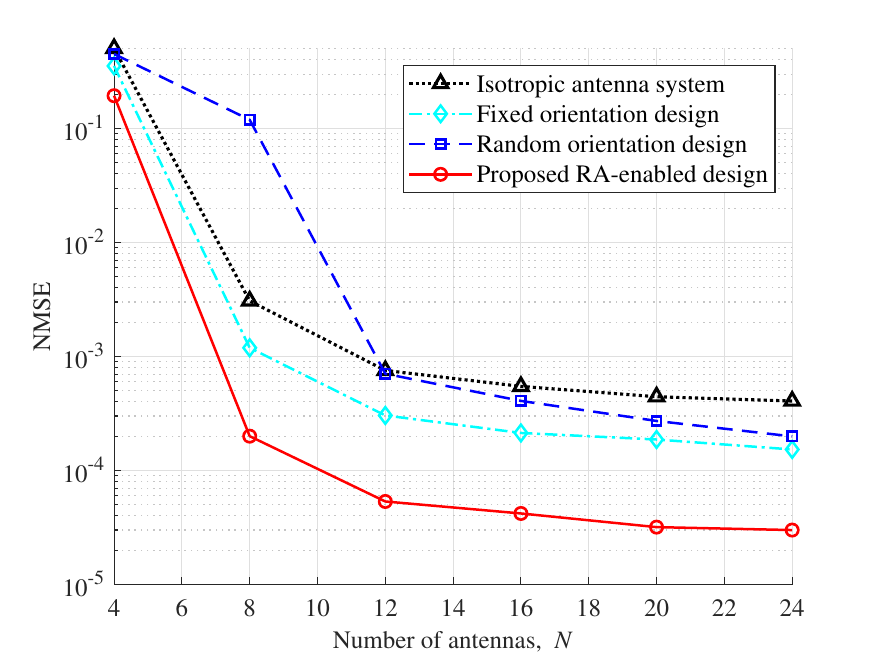} 
	\caption{NMSE versus the number of antennas $N$ for the RA-enabled uplink multi-user communication system.}
\label{NMSE_vs_antennas}
\end{figure}

From the above discussion, fixed-orientation estimation can be regarded as a direct application of conventional fixed-antenna channel estimation, requiring no fundamental modification to existing techniques. In contrast, dynamic-orientation channel estimation introduces observation diversity by varying antenna orientations during training.
{\color{black} Recent RA sensing studies have demonstrated the benefits of the dynamic-orientation mechanism. For example, the authors of \cite{Jiang2026Enhanced} employ sequential orientation scanning to mitigate signal attenuation and improve AoA sensing accuracy, while the authors of \cite{Ye2026Rotatable} leverage multi-configuration observations to provide additional spatial diversity and alleviate sparse-array ambiguity. These works show that dynamic orientation can enhance parameter estimation by improving the directional gain and enriching orientation-dependent observations. 
Similar benefits have also been reported for RA channel estimation~\cite{Xiong2025Efficient}. Fig.~\ref{NMSE_vs_antennas} illustrates the normalized mean square error (NMSE) versus the number of antennas $N$ for different schemes in an RA-enabled uplink multi-user communication system with $K=3$ users.}
It can be observed that the NMSE of all schemes decreases as $N$ increases, since more antennas provide higher spatial resolution and more informative observations for parameter estimation.
Moreover, the proposed RA-enabled design consistently outperforms the isotropic-antenna system, the fixed-orientation design, and the random-orientation design. This indicates that properly designed RA orientations can provide more informative multi-view observations and thereby improve channel estimation accuracy.

In the following, we focus on this dynamic-orientation framework and discuss how multi-view observations can be fused for channel estimation under different RA system setups and signal processing methods.


\subsection{Channel Estimation for Different RA System Setups}

Different RA system setups yield distinct observation structures, which directly affect the identifiability and resolvability of multipath components, as well as the training overhead and signal processing complexity. 
In this subsection, we discuss RA channel acquisition under two representative setups, i.e., single-RA and multi-RA array setups, with emphasis on their observation mechanisms and how multi-view measurements can be fused for channel estimation.

\subsubsection{Single-RA Setup}
In practical deployments, the single-RA setup is attractive for low-cost and resource-constrained platforms, such as APs and Internet-of-Things (IoT) devices, due to its lightweight hardware and compact structure.
Unlike traditional single-antenna transceivers with fixed radiation patterns that only observe the environment from a static viewpoint, a single-RA transceiver can acquire multi-view measurements under different orientations/boresights $\left\{ \vec{\mathbf{f}}^{(m)} \right\}_{m=1}^{M}$.
By sweeping the antenna boresight over time, the transceiver probes the environment from multiple angular perspectives, thereby introducing observation diversity even with a single antenna.

From an implementation perspective, channel estimation under the single-RA setup can be realized via either ``passive listening'' or ``active sensing'', depending on whether the RA node passively receives known pilot signals or actively transmits probing waveforms and processes echoes. 
Let $x_m^{(t)}$ denote the known transmitted signal in the time slot $t$ within the $m$-th orientation block: a pilot signal from an external transmitter (passive listening) or the probing waveform from the single-RA node (active sensing). 
In passive listening, the single-RA node receives pilot symbols broadcast by an external transmitter.
Within the $m$-th orientation block, the RA maintains a fixed orientation $\vec{\mathbf f}^{(m)}$ for $T_b$ slots, and the received signal in \eqref{block_model} reduces to a scalar form
\begin{align}
    y_m^{(t)}=h\left(\vec{\mathbf f}^{(m)}\right)x_m^{(t)}+n_m^{(t)},
    \label{singleRA_passive}
\end{align}
with $t=(m-1)T_b+1,\ldots,mT_b$ and $m=1,\ldots, M$.
By adjusting the antenna orientations/boresights over multiple time blocks, the receiver collects a sequence of observations, from which the channel parameters can be inferred by jointly processing the accumulated channel measurements.
In active sensing, the single-RA node transmits probing waveforms and processes echoes. 
Under a generic discrete-time baseband model with $\tilde{Q}$ resolvable reflection paths, the echo signal can be expressed as
\begin{align}
\hspace{-0.4cm}y_m^{(t)}=\sum_{\tilde{q}=1}^{\tilde{Q}}\beta_{\tilde{q}} g_{\mathrm{tx},\tilde{q}}\left(\vec{\mathbf f}^{(m)}\right) g_{\mathrm{rx},\tilde{q}}\left(\vec{\mathbf f}^{(m)}\right)e^{-j\frac{4\pi}{\lambda}d_{\tilde{q}}} x_m^{(t)}+n_m^{(t)},\hspace{-0.15cm}
\label{singleRA_active}
\end{align}
with $t=(m-1)T_b+1,\ldots,mT_b$ and $m=1,\ldots, M$, where $\beta_{\tilde{q}}$ and $d_{\tilde{q}}$ denote the round-trip propagation coefficient and propagation distance of the $\tilde{q}$-th path, respectively, and $g_{\mathrm{tx},\tilde{q}}(\cdot)$ and $g_{\mathrm{rx},\tilde{q}}(\cdot)$ represent the transmit/receive radiation coefficients.
For a monostatic single-RA transceiver, one may set 
$g_{\mathrm{tx},{\tilde{q}}}\left(\vec{\mathbf f}^{(m)}\right)=g_{\mathrm{rx},\tilde{q}}\left(\vec{\mathbf f}^{(m)}\right)= \sqrt{G\left(\epsilon(\vec{\mathbf f}^{(m)},\vec{\mathbf{q}}_{\tilde{q}}),\varphi(\vec{\mathbf f}^{(m)},\vec{\mathbf{q}}_{\tilde{q}}) \right)}
$,
with $\vec{\mathbf{q}}_{\tilde{q}}$ denoting the direction vector of the $\tilde{q}$-th path. 

Despite their different operating modes, both paradigms rely on accumulating multiple observations across time and antenna orientations. By stacking the received signal over $T_a$ time slots, the single-RA observation vector can be expressed as 
\begin{align}
\mathbf y \triangleq 
\left [ y_1^{(1)},\ldots,y_1^{(T_b)},\ldots,y_M^{((M-1)T_b+1)},\ldots,y_M^{(MT_b)}\right ] ^{T}\hspace{-0.15cm} \in \mathbb{C}^{T_a\times 1},
\label{singleRA_stacking}
\end{align}
with $T_a=MT_b$.
Even with a single RF chain, the single-RA system enables multi-perspective spatial observation via antenna rotation, reducing hardware cost while improving multipath resolvability.
Such a setup is simple and cost-effective, yet it often suffers from limited view diversity and constrained parameter resolvability under practical training constraints.
Specifically, since the spatial information is acquired sequentially via boresight sweeping, the number of orientations within one coherence block is bounded by switching latency and channel variation. 
In rich-scattering environments, weak paths may be obscured by dominant components unless sufficiently dense sampling is employed, which inevitably increases the pilot overhead and training latency.


\subsubsection{Multi-RA Array Setup}
In contrast to the single-RA setup that observes spatial information sequentially via boresight sweeping, the multi-RA array acquires spatial information over multiple antennas in parallel, producing an $N$-dimensional snapshot per training slot.
By concatenating snapshots under a controlled orientation schedule, one forms a spatio-temporal observation matrix that enables more reliable multipath resolvability and parameter identifiability than sequential single-RA acquisition.
As $N$ increases, the effective array aperture and spatial DoFs grow, thereby enhancing angular discrimination and the ability to resolve closely spaced multipath components. 



Consider a multi-RA array with $N$ directional antennas arranged in a general geometry.
At the $t$-th training slot, the received signal follows the RA channel model in \eqref{deqn_ex18b}, yielding an $N$-dimensional observation vector $\mathbf y^{(t)}\in\mathbb C^{N\times 1}$.
Collecting received signals over $T_a$ training slots gives the spatio-temporal observation matrix:
\begin{align}\label{multiview_obser}
    \mathbf{Y}\triangleq
    \left [ \mathbf{y}_1^{(1)},\ldots,\mathbf{y}_1^{(T_b)},\ldots,\mathbf{y}_M^{((M-1)T_b+1)},\ldots,\mathbf{y}_M^{(MT_b)}\right ]\hspace{-0.1cm} \in \mathbb{C}^{N \times T_a},
\end{align}
which jointly captures spatial array responses across antennas and temporal evolution induced by antenna rotation.

Obviously, as $N$ increases, the multi-RA setup generates high-dimensional multi-view observations, providing substantially more information for multipath resolution and parameter estimation. 
However, fully exploiting these gains requires carefully designed low-overhead estimation schemes.
On the one hand, the dimensionality of orientation configurations grows rapidly with $N$, making joint orientation design challenging under practical control and latency constraints.
On the other hand, high-dimensional observations demand advanced spatio-temporal signal processing to leverage geometric priors and correlations for channel extrapolation and parameter inference.  
Promising approaches include structured orientation control (e.g., hierarchical codebook scanning and adaptive boresight refinement) combined with spatio-temporal estimation algorithms such as subspace/parametric fitting \cite{Xiong2025Efficient,Schmidt1986Multiple}, sparse reconstruction \cite{Lee2016Channel,Davenport2013Signal,Nguyen2017Linear}, and tensor-based factorization \cite{Zhang2022Tensor} to exploit low-rank or sparse structures in $\mathbf{Y}$. Moreover, AI-based methods, such as neural networks \cite{Dong2019Deep}, can be employed to capture correlations across orientations and approximate the nonlinear mapping from training data to key channel parameters, enabling efficient channel estimation in complex environments. 



\subsection{Signal Processing Methods for RA Channel Estimation}
{\color{black}In this subsection, we present a structured overview of efficient RA channel estimation schemes based on several representative signal processing methods, including maximum-likelihood (ML), subspace-based, compressed sensing (CS), beam training, and machine learning approaches.
We elaborate on their fundamental principles for CSI acquisition and discuss how RA orientation control can be exploited to improve estimation accuracy under practical training and complexity constraints.} 
In the following, we consider a typical user $k$ and aim to estimate $\mathbf{h}_k$. For notational simplicity, the subscript $k$ is omitted whenever possible.
\subsubsection{ML-Based Estimation}
As demonstrated by the geometric channel model in \eqref{channel_model}, the RA channel can be parameterized by a set $\mathcal{C}_k$ that collects dominant path parameters, such as AoA pairs and propagation coefficients.
A key property of RA systems is that antenna rotation reconfigures the radiation pattern or array manifold but does not alter the underlying propagation geometry within a coherence interval.
Hence, a common practice is to estimate the sparse channel parameters and then reconstruct the CSI for arbitrary orientations based on these parameters.

Consider the received signal model in \eqref{block_model} for the dynamic-orientation case. By stacking the received signal vectors $ \{ \mathbf y^{(t)}\}_{t=1}^{T_a}$, the received signal in \eqref{multiview_obser} can be rewritten as follows:
\begin{align}\label{eq:ML_stacked_model}
    \mathbf y \triangleq \text{vec}(\mathbf{Y})= \tilde{\mathbf S} ({\mathbf{F}};{\boldsymbol\eta}) \boldsymbol\beta + \mathbf n,
\end{align}
where $\tilde{\mathbf S} ({\mathbf{F}};{\boldsymbol\eta}) \in \mathbb{C}^{NT_a\times (Q+1)}$ denotes the observation matrix depending on RA configuration $\mathbf F$ and pilot sequence $ \{ x^{(t)}\}_{t=1}^{T_a}$, $\boldsymbol\beta=[\tilde{\beta}_{k,0},\tilde{\beta}_{k,1},...,\tilde{\beta}_{k,Q}]^T $ denotes the propagation coefficient vector, $ {\boldsymbol{\eta}}=[\vartheta_k,\tilde{\vartheta}_1,...,\tilde{\vartheta}_Q,{\xi}_k,\tilde{\xi}_1,...,\tilde{\xi}_Q]^T$ collects angular parameters, and $\mathbf n \in \mathbb{C}^{NT_a\times 1}$ stacks the AWGN vectors. 
The ML estimator seeks $\boldsymbol\beta$ and ${\boldsymbol{\eta}}$ that maximize the likelihood function $p(\mathbf{y}|\boldsymbol\beta,\boldsymbol{\eta})$.

Based on the received signal in \eqref{eq:ML_stacked_model}, the
log-likelihood function is given by
\begin{align}\label{likelihood_function}
   \hspace{-0.25cm}\text{ln} (p(\mathbf{y}|\boldsymbol\beta,\boldsymbol{\eta} ))= -NT_a\text {ln}(\sigma^2 \pi)-\frac{1}{\sigma^2}\left\| \mathbf{y}-\tilde{\mathbf S}({\mathbf{F}};{\boldsymbol\eta})\boldsymbol{\beta}\right\|^2.\hspace{-0.15cm}
\end{align}
Thus, ML-based estimation is realized by solving the following optimization problem:
\begin{align}
    \{{\boldsymbol\eta}^{\star},{\boldsymbol{\beta}}^{\star} \}= \arg\min_{\boldsymbol\eta,\boldsymbol{\beta}} \big\|\mathbf y-\tilde{\mathbf S} ({\mathbf{F}};\boldsymbol\eta)\boldsymbol\beta\big\|^2.
    \label{ML_estimation}
\end{align}
When multi-view training data are sufficiently informative, ML estimation is statistically efficient and can approach the CRB.
However, \eqref{ML_estimation} is generally a high-dimensional non-convex problem, making joint optimization over $\{\boldsymbol\eta,\boldsymbol\beta\}$ computationally prohibitive.
This motivates efficient techniques such as alternating minimization (e.g., BCD over $\boldsymbol{\eta}$ and $\boldsymbol{\beta}$).

The performance of ML-based estimation is also influenced by the RA orientation schedule $\mathbf{F}$ in $\tilde{\mathbf S}(\mathbf F;\boldsymbol{\eta})$.
Repeated or highly similar orientations yield correlated measurements and an ill-conditioned sensing matrix, amplifying noise and causing parameter coupling, thereby requiring more pilots to achieve a desired estimation accuracy.
Therefore, a favorable orientation schedule should ensure diverse angular coverage and avoid redundant views within the feasible orientation region. 
For example, when RAs are constrained to steer within a spherical cap as in \eqref{deqn_ex2c}, naive uniform sampling in the angular domain leads to non-uniform coverage: grid points cluster near the pole and become sparse near the boundary. To achieve near-uniform sampling, spherical Fibonacci sampling can be employed to generate reference directions with improved angular uniformity.
{\color{black} 
The computational complexity of ML-based estimation mainly arises from iterative likelihood optimization. For a given angular parameter vector $\boldsymbol{\eta}$, the LS update of the propagation coefficient vector $\boldsymbol{\beta}$ incurs a complexity on the order of $\mathcal{O}\!\left(NT_a(Q+1)^2+(Q+1)^3\right)$.
If the angular search space contains $G_m$ grid points, the overall complexity scales as $\mathcal{O}\!\left(
G_m\left[NT_a(Q+1)^2+(Q+1)^3\right]
\right).$
Alternatively, when iterative optimization algorithms are employed, the complexity scales with the number of iterations $I_{\rm ML}$.
In general, ML-based estimation provides high estimation accuracy and can asymptotically approach the CRB, but at the expense of relatively high computational complexity.
}

\subsubsection{Subspace-Based Method}
In finite-scattering scenarios, the number of effective channel paths $(Q+1)$ is typically much smaller than the number of RAs $N$.
This finite set of dominant propagation paths induces a low-rank structure in the RA channel, which can be exploited by subspace-based estimation methods. Specifically, classical algorithms such as MUltiple SIgnal Classification (MUSIC) and Estimation of Signal Parameters via Rotational Invariance Techniques (ESPRIT) \cite{Schmidt1986Multiple} can be applied to estimate the angular parameters $\boldsymbol{\eta}$, followed by a least-squares (LS) estimation of the propagation coefficients $\boldsymbol{\beta}$~\cite{Xiong2025Efficient}.  
Based on the geometric channel model in \eqref{channel_model}, the RA channel vector can be rewritten as follows:
\begin{align}
    \mathbf{h}(\mathbf{F})= \mathbf{A}(\mathbf{F};\boldsymbol{\eta})\boldsymbol{\beta},
    \label{channel_array}
\end{align}
where $\mathbf{A}(\mathbf{F};\boldsymbol{\eta})=[\tilde{\mathbf{b}}_{k,0},\tilde{\mathbf{b}}_{k,1},\ldots,\tilde{\mathbf{b}}_{k,Q}] \in \mathbb{C}^{N\times (Q+1)}$ denotes the RA array manifold. 
By exploiting $T_b$ pilot slots within block $m$ under a given RA orientation $\mathbf{F}^{(m)}$, the sample covariance matrix of the received pilot signals is given by
\begin{align}
    \hat{\mathbf R}_m
    = \frac{1}{T_b}\sum_{t=1}^{T_b}\mathbf y_{m}^{(t)}\left(\mathbf y_{m}^{(t)}\right)^{H}.
\end{align}
Eigenvalue decomposition of $\hat{\mathbf R}_m$ yields the estimated signal-subspace and noise-subspace matrices, denoted by $\mathbf E_{\mathrm s,m} \in \mathbb{C}^{N\times (Q+1)}$ and $\mathbf E_{\mathrm n,m} \in \mathbb{C}^{N\times (N-Q-1)}$, respectively.  
The MUSIC algorithm can then be applied to construct the spatial spectrum
\begin{align}
	V_{m}(\vartheta,\xi) = \frac{1}{\tilde{\mathbf{b}}^H(\mathbf F^{(m)};\vartheta,\xi)\mathbf E_{\mathrm n,m}\mathbf E_{\mathrm n,m}^H \tilde{\mathbf{b}}(\mathbf F^{(m)};\vartheta,\xi)},
\end{align}
whose peaks reveal the angular parameters of the dominant paths in block $m$.
Since the physical AoAs/AoDs are common across blocks, information from different RA orientations can be combined by fusing their power spectra.
A simple yet effective approach is to average the block-wise spectra:
\begin{align}
	\bar{V}(\vartheta,\xi)= \frac{1}{M}\sum_{m=1}^{M} V_{m}(\vartheta,\xi).
\end{align}
The $(Q+1)$ largest peaks of $\bar{V}(\vartheta,\xi)$ are then selected as the final angular estimates $\hat{\boldsymbol{\eta}}$. 

Finally, by stacking all $T_a=MT_b$ pilot observations into a single vector $\mathbf y$ as in \eqref{eq:ML_stacked_model}, the LS estimate of the propagation coefficients is obtained as
\begin{align}
	\hat{\boldsymbol\beta}
	= \big(\tilde{\mathbf S}^{H}\tilde{\mathbf S}\big)^{-1}
	\tilde{\mathbf S}^{H}\mathbf y.
\end{align}
Substituting $\hat{\boldsymbol{\eta}}$ and $\hat{\boldsymbol{\beta}}$ into \eqref{channel_array} yields a parametric reconstruction of the RA channel.
{\color{black} If the angular search space contains $G_s$ grid points, the computational complexity of subspace-based estimation is dominated by three components:  
	1) signal covariance matrix construction with complexity $\mathcal{O}(MN^2T_b)$,  
	2) eigenvalue decomposition with complexity $\mathcal{O}(MN^3)$, and  
	3) spatial spectrum search with complexity $\mathcal{O}(MG_sN(N-Q-1))$.}

\subsubsection{CS-Based Estimation}
Due to the inherent sparsity of the finite-scattering channel in \eqref{channel_model} within the angular (and distance) domain, CS-based methods can be effectively applied for RA channel acquisition.
These methods aim to recover key channel parameters from a reduced number of pilot observations, thereby lowering training overhead and computational complexity. The core idea is to discretize the continuous parameter space into a finite grid and represent the channel as a sparse combination of dictionary atoms.
Let $\boldsymbol{\alpha}\in\mathbb C^{J\times 1}$ denote the sparse coefficient vector defined on a grid of size $J$.
By stacking pilot observations collected under multiple RA orientations, the received signal model can be formulated as a linear sparse model:
\begin{align}
\mathbf{y}=\mathbf{\Phi}({\mathbf{F}})\boldsymbol{\alpha}+\mathbf{n},
\end{align}
where $\mathbf{\Phi}({\mathbf{F}})\in \mathbb{C}^{NT_a\times J}$ is the sensing matrix constructed from the effective RA array manifold associated with orientation ${\mathbf{F}}$ and the known pilots.  
In this context, CS-based RA channel estimation reduces to a sparse recovery problem:
\begin{equation}
\begin{aligned}
    \underset{\boldsymbol{\alpha}}{\min} &\quad 
\|\boldsymbol{\alpha} \|_0\\
\text{s.t.} &\quad \|\mathbf{y}-\mathbf{\Phi}({\mathbf{F}})\boldsymbol{\alpha} \|^2\le \varsigma,
\end{aligned}
\end{equation}
where $\varsigma$ denotes the error tolerance, and $\|\cdot\|_0$ is the $\ell_0$-norm. While $\ell_0$-norm minimization is generally NP-hard, it can be efficiently approximated using practical sparse recovery algorithms. In particular, classical CS algorithms, such as orthogonal matching pursuit (OMP) \cite{Lee2016Channel} and compressive sampling matching pursuit (CoSaMP) \cite{Davenport2013Signal,Nguyen2017Linear}, can be employed to estimate the channel parameters corresponding to the columns of $\mathbf{\Phi}({\mathbf{F}})$ with non-zero coefficients in $\boldsymbol{\alpha}$. 

A distinctive feature of RA systems is that the sensing matrix $\boldsymbol{\Phi}(\mathbf{F})$ is orientation-dependent.
The RA orientation configuration reshapes the effective radiation patterns and directly impacts the mutual coherence and conditioning of $\mathbf{\Phi}$, both of which are critical to sparse recovery performance. 
Poorly designed orientation schedules may increase dictionary atom correlation and degrade recovery accuracy.
Therefore, it is essential to design orientation-aware sensing matrices whose columns jointly encode beamforming directions and antenna orientations.
{\color{black} The computational complexity of CS-based estimation mainly depends on the sparse recovery algorithm and the dictionary size. For example, for OMP with sparsity level $(Q+1)$ and dictionary size $J$, the dominant complexity arises from:  
	1) the correlation search over the dictionary, with complexity $\mathcal{O}((Q+1)NT_aJ)$, and  
	2) the propagation-coefficient updates, with complexity $\mathcal{O}(NT_a(Q+1)^2 + (Q+1)^3)$.  
	Thus, CS-based methods can significantly reduce the pilot overhead by exploiting channel sparsity, but their complexity grows with the dictionary size, sparsity level, and grid resolution.} 


\subsubsection{Beam Training-Based Estimation}
Beam training is a practical alternative for acquiring implicit CSI in sparse channels by searching over a finite set of candidate beam/orientation configurations. 
Unlike model-based estimation, which explicitly recovers the geometric parameters $\boldsymbol{\eta}$ and $\boldsymbol{\beta}$, beam training aims to select the best transmission mode (i.e., a codeword pair) that maximizes the received signal power or SNR. This makes it particularly appealing when low-complexity link alignment is the primary objective.
In RA systems, mode selection is jointly governed by the RA orientation configuration, which reshapes the radiation pattern, and the conventional transmit/receive beamforming weights.
Let $\mathcal{F}=\{\mathbf F^{(1)},\ldots,\mathbf F^{(|\mathcal F|)}\}$ denote the RA orientation codebook and $\mathcal{W}=\{\mathbf w^{(1)},\ldots,\mathbf w^{(|\mathcal W|)}\}$ denote a conventional beamforming/combining codebook. 
Then, a typical beam-training problem in an RA system can be formulated as
\begin{equation}
\begin{aligned}
\max_{\mathbf F,\mathbf w} &\quad
|\mathbf w^{T}\mathbf h(\mathbf F)|^2\\
  \text{s.t.} &\quad
\mathbf F\in\mathcal F,\mathbf w\in\mathcal W.
\end{aligned}
\label{BT_opt_}
\end{equation}
Accordingly, beam training is equivalent to selecting the codeword pair $(\mathbf F^{\star},\mathbf w^{\star})$ that yields the strongest effective channel gain. 
In addition, the codebooks $\mathcal{F}$ and $\mathcal{W}$ need to be designed to satisfy compactness and scalability. Specifically, the codebook size should be reduced as much as possible to reduce the beam training overhead. 
Different from fixed antenna configurations, RA introduces an additional controllable dimension, i.e., the boresight rotation, which changes the effective manifold and may alter the distinguishability between different beams.
Therefore, practical RA beam training critically depends on the orientation-aware codebook design and low-overhead mode selection strategies.
The RA orientation codebook $\mathcal F$ should provide sufficient coverage of the boresight domain and ensure good separability among codewords. 
To satisfy the coverage requirement, a baseline idea is to discretize the boresight domain into $(\vartheta,\xi)$ grids and map each grid point to an RA orientation $\mathbf F^{(i)}$. 
For near-field/wideband settings, $\mathbf F^{(i)}$ can further incorporate a focal-distance bin, yielding a multi-resolution codebook.
In addition, RA orientation induces varying radiation patterns across antennas that disrupt the orthogonality of conventional codebooks. 
Therefore, it is essential to develop radiation-pattern-aware codebooks where
each codeword jointly encodes a beamforming direction along
with specific antenna orientation configurations, thus improving codeword distinguishability. In this case, $\mathcal W$ can follow standard designs, e.g., discrete Fourier transform (DFT) codebooks~\cite{Suh2017Construction}, while $\mathcal F$ is tailored to RA features and radiation-pattern characteristics.

Given $\mathcal F$ and $\mathcal W$, beam training performs mode selection by sweeping a subset of candidate codeword pairs $\left(\mathbf{F}^{(i)},\mathbf{w}^{(j)}\right)$, measuring the received power metric $O(i,j)\approx |(\mathbf w^{(j)})^{T}\mathbf h(\mathbf F^{(i)})|^2$, and selecting the maximal value in \eqref{BT_opt_}. 
A straightforward approach is to conduct an exhaustive search over all possible codebooks. However, this may incur prohibitively high training overhead, especially when the codebook size $|\mathcal F||\mathcal W| $ is large. 
To reduce the training overhead, 
hierarchical (coarse-to-fine) scanning is commonly adopted: a coarse stage uses wide beams (and coarse orientation sectors) to localize a small candidate set, followed by refined probing with narrower beams within the selected sector to identify the optimal index $(i^{\star},j^{\star})$.
{\color{black} The computational complexity of beam training is mainly determined by the number of beam/orientation codeword pairs to be tested. For an exhaustive search, evaluating the received power metric $O(i,j)$ for all candidate pairs incurs a complexity on the order of $\mathcal{O}\!\left(N|\mathcal F||\mathcal W|\right)$. 
Hence, the complexity scales linearly with the number of antennas and the codebook sizes, while hierarchical beam training reduces complexity by progressively narrowing the candidate set.
}


\subsubsection{Learning-Based Channel Acquisition} 
Compared with traditional estimation schemes that rely on accurate mathematical models, learning-based methods provide a data-driven alternative capable of incorporating multi-view information to improve estimation accuracy.
The key idea is to learn a nonlinear mapping from the observed pilot data (input) to a compact representation of the channel (output), and then reconstruct the CSI for subsequent data transmission.
This paradigm is particularly attractive for RA systems because antenna orientations reshape the effective radiation pattern and thereby generate informative multi-perspective measurements.
Such multi-view diversity can be leveraged to reduce pilot overhead and enhance robustness against model mismatches, calibration errors, and hardware impairments.
{\color{black}In addition, orientation-dependent mutual coupling and calibration errors may lead to a mismatch between the assumed array responses and the actual received pilot observations. This mismatch can result in biased channel estimates, degraded angle estimation accuracy, and increased pilot overhead. Practical remedies include periodic recalibration, joint estimation of channel and calibration parameters, and learning-assisted estimators that exploit multi-orientation pilot measurements to compensate for model mismatch~\cite{Vieira2017Reciprocity, Luo2016Multiuser}.}


To facilitate standard learning architectures, we stack all multi-view observations in \eqref{multiview_obser} as $\mathbf y=\mathrm{vec}(\mathbf Y)\in\mathbb C^{NT_a\times 1}$ and convert them into a real-valued input:
\begin{align}\label{eq:input_learning}
    \tilde{\mathbf y} = [\Re\{\mathbf y\}^T,\Im\{\mathbf y\}^T]^T\in\mathbb R^{2NT_a\times 1}.
\end{align}
In finite-scattering environments, the RA channel can be well characterized using a compact set of key parameters (e.g., $\boldsymbol{\chi}=[\boldsymbol{\beta}^T,\boldsymbol{\eta}^T]^T \in \mathbb{R}^{3(Q+1)\times 1}$). Hence, instead of directly estimating the full CSI, a more efficient and common practice is to estimate these low-dimensional parameters and then reconstruct the channel. 
Accordingly, learning-based RA channel estimation constructs a nonlinear mapping between the input data (multi-view observations) and the output data (key channel parameters):
\begin{align}
\hat{\boldsymbol{\chi}} = f_{\boldsymbol{\xi}}\!\left(\tilde{\mathbf y};\{\mathbf F^{(m)}\}_{m=1}^{M} \right),
\label{eq:DL_mapping}
\end{align}
where $f_{\boldsymbol{\xi}}(\cdot):\mathbb{R}^{2NT_a\times1} \mapsto \mathbb{R}^{3(Q+1)\times1}$ denotes a learning model parameterized by $\boldsymbol{\xi}$.
The training data can be enriched by using different orientation configurations, thereby improving view diversity and facilitating feature learning.

Learning-based methods offer several advantages over model-based techniques in RA systems. Their data-driven structure provides robustness against measurement/model impairments, facilitates feature extraction and adaptation to environmental changes, and enables low-complexity inference compared with iterative optimization.
More importantly, RA rotations naturally provide multi-view pilot measurements, enabling the learning-based estimator to fuse complementary perspectives and recover a compact set of geometric channel parameters with reduced pilot overhead and improved tracking capability.
A variety of learning models, ranging from lightweight multilayer perceptrons to deep architectures such as convolutional neural networks (CNNs) and transformers, can be employed~\cite{MOHDNOOR2025Asurvey}. 
Depending on label availability and deployment constraints, one may adopt supervised learning using labels generated by model-based estimators or simulations, self-supervised/unsupervised learning via physics-consistent reconstruction losses, or federated learning to train a shared estimator across distributed devices with limited data sharing (e.g., \cite{Lim2020Federated,Jia2025AComprehensive}).
{\color{black} According to \eqref{eq:input_learning} and \eqref{eq:DL_mapping}, the network input and output dimensions scale with $2NT_a$ and $3(Q+1)$, respectively. The computational complexity of learning-based channel acquisition comprises the offline training complexity and online inference complexity.
For a fully connected neural network with $L_{\rm NN}$ layers and $D_l$ neurons in the $l$-th layer, the per-sample forward inference complexity is on the order of $\mathcal{O}\!\left(\sum_{l=1}^{L_{\rm NN}-1}D_lD_{l+1}\right)$. The offline training complexity further scales with the number of training samples, training epochs, and back-propagation operations. 
Once the model is trained, online channel acquisition only requires a forward pass, resulting in low inference latency compared with iterative model-based estimation methods. 
}

{\color{black} \textit{Remark 2:} In highly dynamic scenarios, the snapshot-based channel training model may become inaccurate because the channel can evolve during RA orientation adjustment. 
This issue becomes more critical when the RA orientation switching time is comparable to the channel coherence time. 
In such cases, the initially estimated channel parameters may become outdated before the RA orientations are configured, leading to channel aging, estimation mismatch, and degraded acquisition performance.
Nevertheless, RA systems may still preserve part of their directional gain for moderate-to-fast channel evolution, since dominant geometric parameters (e.g., AoAs/AoDs) often vary more slowly than the instantaneous small-scale fading coefficients. Thus, RA orientations configured based on previously estimated directions may remain approximately aligned with the dominant paths. To further mitigate the impact of severe mobility, predictive channel/beam tracking methods can be incorporated into the RA channel acquisition framework. 
Representative approaches include extended Kalman filter (EKF)-based tracking, geometric model-based prediction exploiting trajectory and location information, and learning-assisted prediction methods. 
Furthermore, the joint design of RA orientation adaptation and channel tracking constitutes a promising direction, where antenna orientations and channel parameters are continuously updated within each coherence interval.


}

\section{RA Configuration and Deployment}\label{sec:configuration}
From concept to implementation, RA configuration and deployment play a critical role in determining both system performance and hardware efficiency. 
While RA offers a compact and scalable solution to enhance spatial flexibility, its practical configuration and deployment remain challenging due to stringent hardware and control constraints. 
Key design aspects—such as boresight control methods, rotational range and granularity, array structure, and deployment strategy—directly affect control accuracy, latency, size, and cost, thereby impacting overall performance and integration complexity~\cite{Zheng2026Wireless}.
In the following, we discuss representative RA configurations/deployments, highlight their respective advantages and disadvantages, and provide guidelines for selecting suitable RA configurations/deployments in practice.

\subsection{Mechanical vs. Electronic Rotation}
From a hardware perspective, the RA boresight can be reconfigured through either mechanically driven or electronically driven mechanisms, as depicted in Fig.~\ref{fig:RA_implement}. Both approaches enable effective boresight control but exhibit different trade-offs in accuracy, agility, complexity, and cost.  

For mechanically driven rotation, each antenna/array is mounted on an actuation platform (e.g., a gimbal or servo-driven stage) to physically steer the boresight. A typical realization involves mounting the RA on a servo-motor-controlled platform, where platform rotation directly adjusts antenna orientation in 3D space \cite{Dai2025Rotatable,Dai2025ADemo}.
Alternatively, miniaturized electromechanical actuation based on MEMS can facilitate compact integration and simplify deployment~\cite{Baek2003AVband}.
Mechanical rotation generally provides a wide steering range and fine angular resolution, but at the cost of non-negligible actuation latency and potential mechanical wear.
For instance, MEMS-based actuators typically consume milliwatt-level power and exhibit response times from µs to ms.
In practice, mechanically driven methods are constrained by actuator complexity, maintenance requirements, and physical limitations such as friction and limited lifetime under repeated rotation.


For electronically driven rotation, the antenna remains physically fixed, while its boresight is reconfigured by electronically reshaping the excitation or impedance distribution, thereby emulating radiation pattern rotation without moving parts. 
Representative designs include
(i) reconfigurable feeding networks or multi-feed radiators, where switching among feed points steers the mainlobe \cite{Morshed2023Beam}; and (ii) parasitic-element-based designs with electronically tunable loads (e.g., varactor or PIN diodes), where induced currents and mutual coupling are adjusted to rotate the pattern \cite{Boyarsky2021Electronically}.
Since practical feed/load configurations are finite, the achievable steering directions are typically discrete and predefined.
To enable quasi-continuous adjustment, tunable materials and reconfigurable metasurfaces can be employed \cite{Lindle2016Efficient}, though they require sophisticated biasing and calibration and may face bandwidth or thermal stability challenges.
Overall, electronically driven rotation is more compatible with compact platforms and can achieve reconfiguration latencies ranging from ns to ms, depending on the specific implementation.
Nevertheless, because boresight rotation is realized via excitation/impedance reconfiguration, the synthesized patterns across electronic states are generally not exactly rotated replicas and may exhibit variations in mainlobe width, sidelobe structure, polarization, and phase center.

These two approaches are complementary: mechanical rotation provides wide-angle coarse steering, whereas electronic rotation enables low-latency fine adjustment. 
Hybrid RA architectures can integrate both, with mechanical rotation handling infrequent large-angle updates and electronic control compensating for residual pointing errors and supporting fast beam tracking.


\subsection{Continuous vs. Discrete Rotation}
Ideally, each antenna orientation/boresight can be adjusted continuously, allowing arbitrarily fine angular resolution.
The continuous feasible set of orientations is modeled as in \eqref{deqn_ex1c}.
In compact array deployments, however, large rotations may increase mutual coupling between adjacent antennas and induce pattern distortion. To account for such effects and restrict excessive deviation, an additional rotational constraint can be imposed as in \eqref{deqn_ex2c}.
This continuous orientation model provides theoretical performance limits of RA-enabled systems under practical constraints \cite{Zheng2026Rotatable,Wu2025Modeling}.  

While continuous adjustment is beneficial for optimizing communication and sensing performance, it is difficult to implement in practice because high-resolution rotation requires higher cost and more complex hardware design.
As a cost-effective alternative, antenna orientation can be implemented discretely, where each RA steers its boresight toward a finite set of candidate directions. 
Let $I_{\psi}$, $I_{\theta}$, and $I_{\phi}$ denote the numbers of quantization levels for rotation angles $\psi$, $\theta$, and $\phi$, respectively.
Then, the corresponding discrete angle sets are  
\begin{align}\label{discrete_rotation_general}
\mathcal{F}_{X}^{\prime}
=\left\{\bar{X}_{1},\bar{X}_{2},\ldots,\bar{X}_{I_X}\right\},
\end{align}
with $X \in \{\psi,\theta,\phi\}$, where $0\le \bar{X}_{1}<\cdots<\bar{X}_{I_X}< 2\pi $. 
The resulting discrete orientation codebook is the Cartesian product $\mathcal{F}^{\prime}= \mathcal{F}_\psi^{\prime}\times \mathcal{F}_\theta^{\prime}\times \mathcal{F}_\phi^{\prime}$, whose size scales as $|\mathcal{F}^{\prime}| = I_\psi I_\theta I_\phi$.
As a special case, uniform quantization discretizes each angle into evenly spaced grids.
For example, $\mathcal{F}_{\psi}^{\prime}
=\left\{\psi_{i}\right\}_{i=1}^{I_{\psi}}$ with $\psi_{i}=\psi_{1}+(i-1)\Delta_{\psi}$, where $\Delta_{\psi}$ is the quantization step size;
$\mathcal{F}_{\theta}^{\prime}$ and $\mathcal{F}_{\phi}^{\prime}$ can be defined similarly.
Finite-resolution control introduces a fundamental performance-complexity trade-off: increasing $\{I_\psi,I_\theta,I_\phi\}$ improves steering accuracy but enlarges the codebook size and feedback overhead, whereas fewer levels reduce cost at the expense of performance.
Moreover, discrete rotation complicates orientation optimization, since it introduces discrete variables that are generally harder to handle than continuous ones \cite{Peng2025Rotatable}.

\subsection{Sparse vs. Non-Sparse Array}
Considering practical constraints on cost, hardware complexity, and deployment space, the RA array structure needs to be carefully designed to balance sensing and communication performance across different wireless systems.
The array geometry determines the physical aperture and spatial sampling pattern, which in turn shapes the mainlobe width, sidelobe level, and grating-lobe behavior, ultimately impacting achievable performance.
Specifically, there are two representative RA array structures: non-sparse (compact) and sparse RA arrays.
A compact array typically employs approximately half-wavelength inter-antenna spacing, providing dense spatial sampling that enables stable beam steering over a broad angular range while suppressing grating lobes.
Such layouts also facilitate integrated packaging and calibration, making them attractive for platforms with limited size.
However, densely packing antennas increases hardware cost, energy consumption, and signal processing overhead.
Moreover, for a fixed number of antennas, compact arrays have a relatively limited physical aperture,
restricting spatial resolution and interference suppression.
In addition, compact deployments are also more susceptible to mutual coupling and near-field interactions, and RA boresight rotation may cause pattern overlap among adjacent antennas, further distorting the effective array response.


To overcome these limitations, sparse RA arrays have emerged as a promising architecture that enlarges the aperture without increasing the number of antennas. 
By relaxing the half-wavelength spacing constraint, sparse arrays allow greater inter-antenna separation.
Such sparse geometries yield sharper beams and finer spatial resolution, desirable for high-resolution sensing and improved user separability/interference suppression in ISAC systems.
Meanwhile, increased antenna spacing also alleviates mutual coupling and reduces pattern overlap during antenna rotation, thereby enhancing reconfiguration flexibility.
Furthermore, certain sparse geometries enable difference/sum co-arrays, creating enlarged virtual apertures and additional sensing DoFs, which are particularly beneficial for localization and multi-target resolution.
Nevertheless, sparse sampling raises the risk of elevated sidelobes and grating lobes, which may cause energy leakage and severe inter-user interference when users fall into overlapping grating-lobe regions.
In sensing, they may also induce angular ambiguities.
To address this issue, sparse RA arrays can jointly exploit aperture gain from antenna placement and radiation pattern reconfigurability from orientation/boresight control. Specifically, sparse placement sharpens the mainlobe via an enlarged aperture, while RA rotations reshape radiation patterns to regulate sidelobes/grating lobes and provide view diversity.
This joint design is particularly appealing for ISAC, as orientation diversity enhances identifiability and clutter robustness in sensing while enabling interference shaping and improving link reliability in communication without sacrificing the aperture advantage of sparse arrays.

\subsection{Distributed vs. Centralized Deployment}
While RA hardware and orientation control determine the local radiation behavior of each antenna/array, deployment across the network dictates global spatial diversity and coordination overhead.
The choice of deployment strategy for RAs should be tailored to system requirements, striking a balance between hardware cost and performance.  
At the network level, there are two main strategies in RA deployment:
\textbf{centralized deployment}, 
where antennas/arrays are co-located on a single platform to facilitate coordinated control and coherent processing \cite{Dai2025RotatableSecure,Zhou2025Rotatable,Wu2025Rotatable,Wang2026Rotatable}; and \textbf{distributed deployment}, 
where antennas/arrays are placed across multiple spatially separated nodes to enhance coverage and provide macro-diversity gains against shadowing and blockage~\cite{Pan2025Rotatable,Peng2026Cell}.

Deployment strategies of RAs significantly impact effective channel realizations and thus the fundamental performance limits of RA-enabled systems. 
Centralized deployment is generally favorable for coherent beamforming with tight synchronization, whereas distributed deployment offers location diversity but requires more stringent synchronization, CSI exchange, and calibration. 
Implementation considerations include operational cost, user/target distribution, space constraints, and propagation environment.
For instance, in rich-scattering or sparsely populated environments, distributed deployment improves coverage via macro-diversity and mitigates shadowing/blockage.
By contrast, aerial or spaceborne platforms, often constrained by size, weight, and power (SWaP), favor centralized compact arrays with small footprints and electronically driven or MEMS-actuated rotation for practical integration.
Moreover, these platforms are also subject to mechanical vibrations, pointing jitter, and thermal variations, which can induce pointing errors and distort the effective array response, thereby degrading performance.
To alleviate these effects, RA deployments should incorporate ruggedized hardware, robust packaging, and calibration-aware control.



\section{RA Prototypes and Related Products}\label{sec:prototypes}
This section provides an overview of existing prototypes that employ different implementation strategies for RA-enabled wireless sensing and communication, as well as their experimental results validating the performance gains practically achievable through antenna rotation. Furthermore, we present several representative commercial products whose design concepts align with the core principles of RA, thereby demonstrating the feasibility and potential of RA-like solutions in practical deployments.
\subsection{Single RA Prototypes}

\begin{figure*}[t]
	\vspace{-0.2cm}
	\center
	\includegraphics[width=6.2in,height=2.4in]{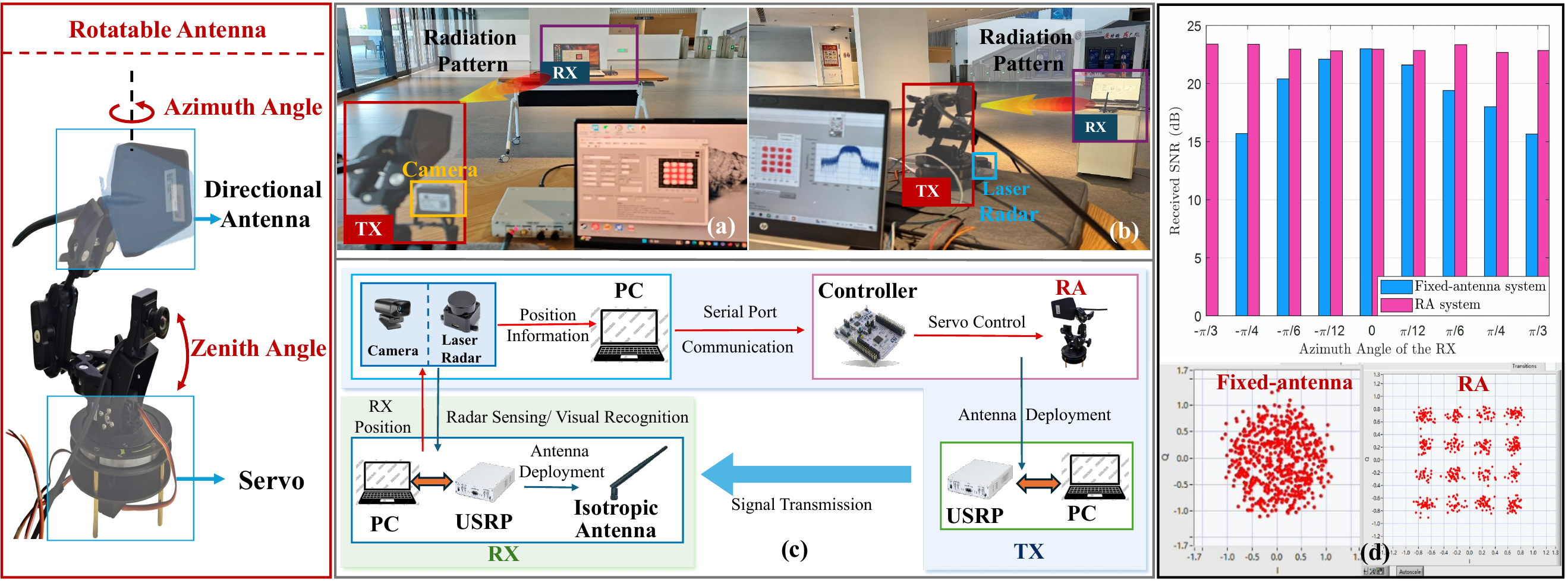}
	\vspace{-0.1cm}
	\caption{Prototypes of sensing-assisted RA. (a) Visual recognition~\cite{Dai2025Rotatable}.  (b) Radar sensing\cite{Dai2025ADemo}. (c) Prototype architecture. (d) Received SNR versus the azimuth angle of the receiver and constellation diagrams.}\vspace{-0.1cm}
	\label{RA_prototype_element}  
	\vspace{-0.2cm}
\end{figure*}

To validate the feasibility of single-RA implementations, Fig.~\ref{RA_prototype_element} presents two representative sensing-assisted RA prototypes and corresponding experimental results. In these prototypes, the core concept is to detect the spatial position of the receiver and convert it into the corresponding azimuth and elevation angles of the RA, enabling the servo system to precisely rotate the RA boresight direction toward the target. To acquire target/receiver position information, two representative sensing modalities are visual recognition and radar sensing. In the visual recognition-assisted RA prototype shown in Fig.~\ref{RA_prototype_element}(a)~\cite{Dai2025Rotatable}, a camera and a personal computer (PC) jointly form a vision module that detects and tracks the receiver as a visual target. Moreover, the ``You Only Look Once'' (YOLO) network~\cite{Redmon2016You} is employed for real-time target detection, while the ``DeepSORT'' algorithm~\cite{Du2023Strongsort} is adopted to track the movement of the receiver over time. Therefore, the spatial position of the receiver is captured within the image and serves as guidance for subsequent antenna boresight adjustment. Meanwhile, the radar sensing-assisted RA prototype illustrated in Fig.~\ref{RA_prototype_element}(b) acquires the receiver's spatial position information via radar detection. Specifically, a laser radar module mounted at the transmitter scans the surrounding environment and estimates the location of the receiver, using a time-of-flight (ToF) approach. Despite the different sensing modalities, both implementations rely on a common control process. Through geometric transformation, the acquired position information is converted into target azimuth and elevation angles, which are processed by a microcontroller (e.g., STM32) to drive a servo motor via a proportional-integral-derivative (PID) algorithm. This closed-loop mechanism dynamically steers the antenna to compensate for receiver mobility, thereby sustaining precise boresight alignment.

In the experiment, the sensing-assisted RA implementations operate at a carrier frequency of 5.8 GHz and employ 16-ary quadrature amplitude modulation (16-QAM), with a transmit power of 10 dBm and a data rate of 2 Mbps. As shown in Fig.~\ref{RA_prototype_element}(d), both sensing-assisted RA implementations achieve a notable performance gain compared with the conventional fixed-antenna system. In particular, an SNR improvement of approximately 7 dB is observed, accompanied by a significantly clearer 16-QAM constellation diagram. These experimental results confirm that accurate boresight adjustment enabled by environmental sensing effectively enhances link quality, validating the practicality of sensing-assisted RA systems.

\subsection{RA Array Prototype}
\begin{figure*}[t]
	\vspace{-0.1cm}
	\center
	\includegraphics[width=6.2in,height=2.1in]{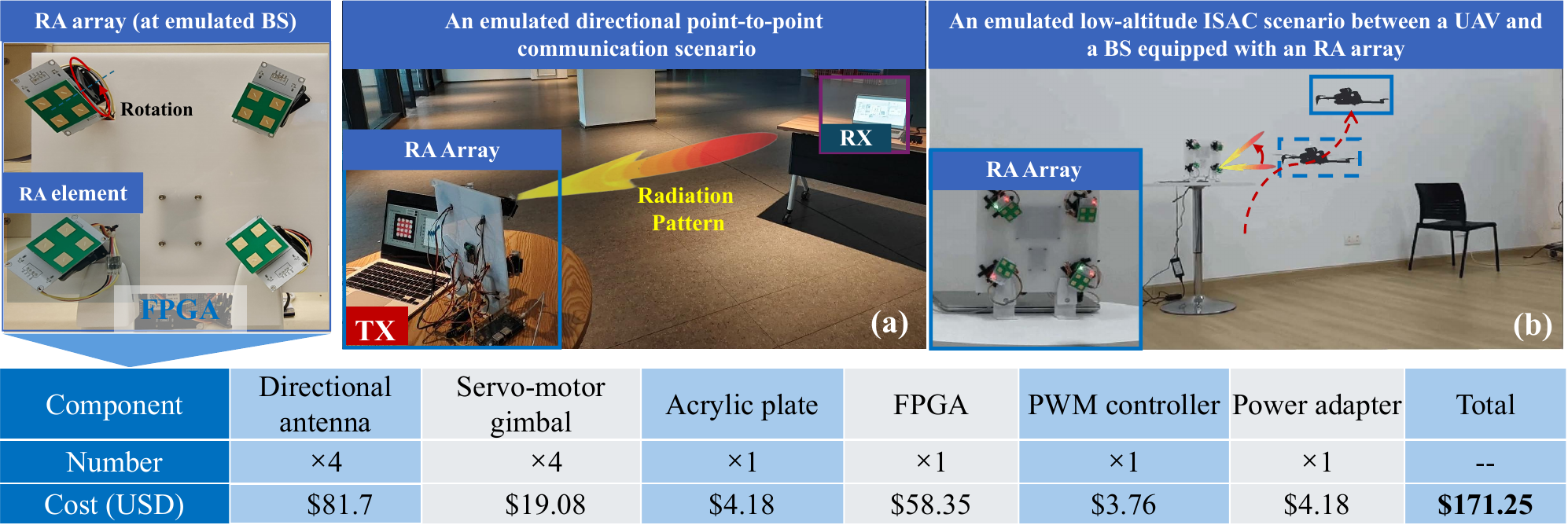}
	\vspace{-0.1cm}
	\caption{Prototype of RA array in different scenarios. }\vspace{-0.1cm}
	\label{RA_prototype_array}  
	\vspace{-0.0cm}
\end{figure*}
While the aforementioned prototypes validate the efficacy of the single RA, practical BSs typically employ antenna arrays to achieve high directional gain and spatial multiplexing. To demonstrate the feasibility of RA arrays in real-world deployments, Fig.~\ref{RA_prototype_array} illustrates two representative application scenarios. As shown in Fig.~\ref{RA_prototype_array}(a), the RA array enables high-gain point-to-point directional communication in indoor environments (such as AR/VR services) by dynamically adjusting the boresight direction of each antenna~\cite{Xiong2026Intelligent}. By accurately perceiving and tracking the target user’s location, the system can orient the RAs toward the desired direction, thereby concentrating signal power and supporting high-rate, low-latency transmission. In such scenarios, the RA array effectively exploits spatial DoFs to improve link reliability and throughput, thereby realizing perception-assisted directional communication.

Furthermore, the RA array can also be effectively applied in low-altitude ISAC scenarios as illustrated in Fig.~\ref{RA_prototype_array}(b). A short demonstration video is available on YouTube at \href{1}{https://www.youtube.com/watch?v=L1aMVaGj5rw}. In this experiment, the terminal emulator is deployed on a UAV platform to emulate the low-altitude scenario. Moreover, the RA array operates as a BS, supporting real-time sensing and communication with the UAV. In particular, the RA array can accurately estimate the AoA of incoming signals and track the UAV's trajectory in real time. Furthermore, the system maintains boresight alignment even during UAV maneuvers such as climbs, turns, and hovers. Consequently, the communication link remains stable, and the sensing results accurately track the UAV's motion. This dynamic tracking capability validates the robustness of the RA  array in managing low-altitude mobile targets and underscores its strong potential as a foundational infrastructure for future 3D network coverage.
Notably, the RA array prototype demonstrates remarkable cost-efficiency. As summarized in Fig.~\ref{RA_prototype_array}, the total hardware cost is only $171.25$ USD, including directional antennas, servo gimbals, field-programmable gate array (FPGA), and control modules. This competitive cost structure underscores the practicality and scalability of RA architectures, offering a cost-effective solution for dense low-altitude ISAC deployments.

{\color{black}It is worth noting that the above prototypes mainly serve as proof-of-concept validations of RA-enabled boresight adjustment. Although they demonstrate the practical gains brought by antenna rotation, their performance is still constrained by hardware limitations. For mechanically driven RA prototypes, the response time is limited by motor actuation and physical rotation. Moreover, the achievable angular resolution and control accuracy depend on the actuation hardware and control mechanisms, which may introduce boresight misalignment. As a result, the continuous boresight adjustment assumed in theoretical models is difficult to fully realize in practice, and most RA implementations rely on discrete, finite-resolution control. Therefore, for a balanced evaluation of practical RA prototypes, the achieved performance gains, actuation response time, angular resolution, and control accuracy have to be jointly considered.}

\subsection{Related Commercial Products}
\begin{figure*}[t]
	\vspace{-0.0cm}
	\center
	\includegraphics[width=5.8in]{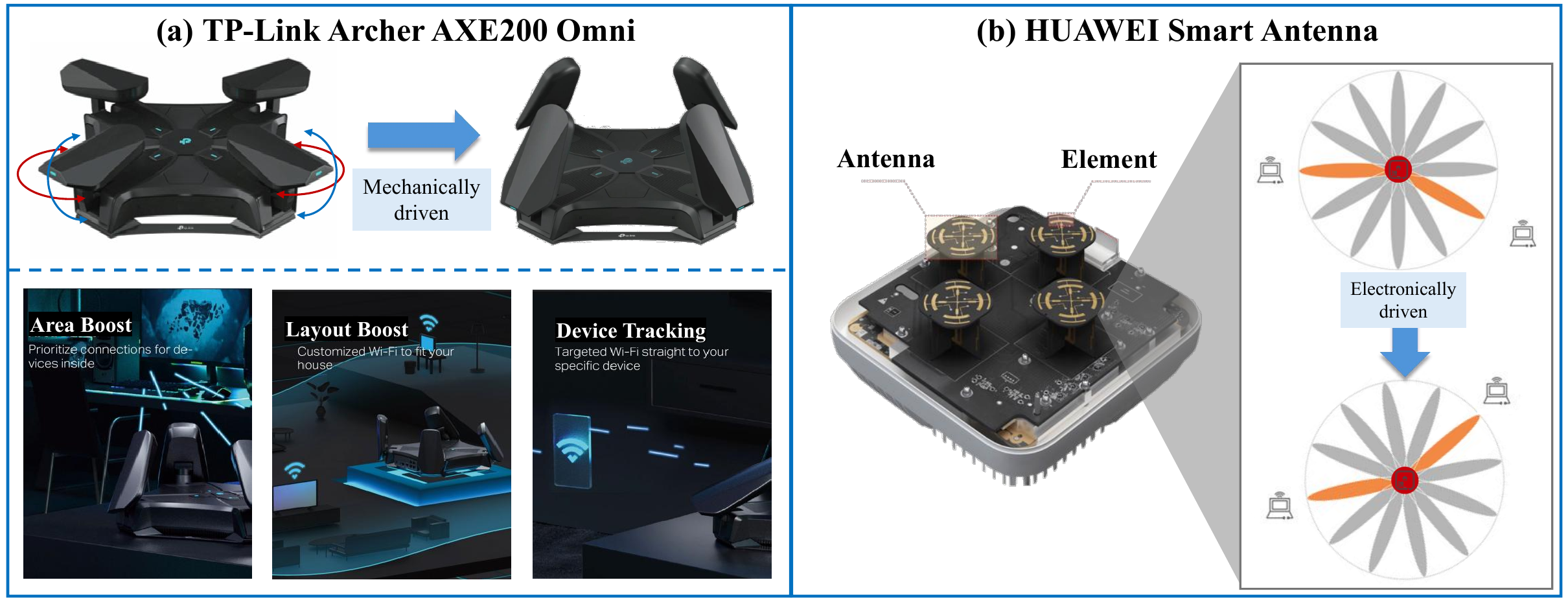}
	\vspace{-0.1cm}
	\caption{Representative commercial products. (a) TP-Link’s Archer AXE200 Omni is a Wi-Fi 6E router equipped with four motorized antennas~\cite{TP-Link2025Archer,TP-Link2025AXE11000}. (b) Huawei’s smart antennas  adopt similar principles by dynamically switching among multiple radiating elements~\cite{Huawei2025What,Huawei2025Wi-Fi}.}\vspace{-0.1cm}
	\label{Commercial_product}  
	\vspace{-0.2cm}
\end{figure*}
Beyond these academic prototypes, the concept of dynamic antenna orientation/boresight has also been adopted in recent commercial and industrial products, further validating its practical value. For instance, TP-Link has launched the Archer AXE200 Omni, which is a Wi-Fi 6E router equipped with smart mechanically driven antennas, as shown in Fig.~\ref{Commercial_product}(a)~\cite{TP-Link2025Archer,TP-Link2025AXE11000}.
Specifically, this device employs motorized mechanisms to dynamically reorient its antennas according to the spatial distribution and movement of connected devices. 
By tracking user locations and adapting to the layout of the environment, the router mechanically steers its antennas to enhance coverage and to provide signal enhancement toward specific devices. Such capability enables improved coverage uniformity as well as high-rate directional links for mobile users in indoor scenarios.

From the perspective of electronically adjusting the antenna boresight, Huawei has pioneered ``Smart Antenna" solutions, which have been widely deployed in the AirEngine Wi-Fi 6 APs and 5G BSs, as illustrated in Fig.~\ref{Commercial_product}(b)~\cite{Huawei2025What,Huawei2025Wi-Fi}. Unlike standard omnidirectional solutions, these systems typically employ a smart multi-antenna array managed by an intelligent switching algorithm. This architecture enables selective activation of different antennas to dynamically adjust the effective radiation direction of the array without mechanical movement. Such electronically driven boresight control enables fast adaptation to dynamic user mobility and channel variations.


In summary, the prototypes and commercial products reviewed in this section collectively demonstrate the transition of RA technology from theoretical modeling to practical implementation. The diverse realization strategies reviewed above highlight the practical value of flexible antenna orientation/boresight control; however, they constitute merely a small subset of the potential methods, suggesting considerable room for developing new hardware realization schemes. Looking ahead, antenna orientation/boresight control strategies are expected to evolve beyond single-modality sensing by incorporating multi-modal sensor fusion. Furthermore, the incorporation of machine learning-based trajectory prediction may help mitigate the latency incurred by practical antenna boresight adjustment and actuation. Finally, efforts should be directed toward developing cost-effective system architectures and establishing general and standardized frameworks to facilitate the practical implementation and widespread commercialization of this technology.

\section{Extensions and Future Directions}\label{sec:extensions}
In this section, we discuss several representative extensions of RA-aided wireless systems and outline open problems that are critical for broadening their applications. 

\subsection{Low-Altitude ISAC}
 Low-altitude ISAC has emerged as a key enabler for LAE, demanding both high performance and deployment flexibility to support real-time environmental perception and reliable connectivity~\cite{Song2025Overview}. 
 However, conventional low-altitude networks built upon fixed-sector antenna infrastructures are primarily designed for 2D ground coverage, with downtilt configurations optimized for terrestrial users. As a result, such architectures are inherently limited in providing effective 3D spatial coverage and reliable tracking for dynamic aerial users operating at varying altitudes and trajectories.
 To overcome these limitations, the RA architecture provides a flexible hardware solution that enhances spatial adaptability. By dynamically reconfiguring the boresight direction, RA arrays can effectively bridge coverage gaps, improve target tracking, and meet the stringent requirements of LAE. {\color{black}Recent studies have demonstrated these benefits in low-altitude sensing. For example, the authors in~\cite{Jiang2025Rotatable} exploited RA arrays for low-altitude direction sensing and showed that boresight rotation can mitigate the sensing degradation caused by misalignment between the target direction and the array boresight. Extending this idea, the authors in~\cite{Jiang2026Enhanced} further reduced the rotation and computation overhead by proposing a pre-rotation initialization and an iterative greedy local search method to improve target direction estimation.} Within the ISAC paradigm, RAs can be integrated into diverse deployment architectures to simultaneously boost sensing precision and communication quality~\cite{Wu2025Rotatable,Li2025Rotatable,Zhang2026Rotatable,Wang2025Rotatable,Wang2026RotatableAntenna}. A typical deployment is the ground-based RA architecture, where RAs are installed on terrestrial BSs to enhance coverage flexibility~\cite{Zheng2025Low,Li2026Cooperative}. This setup has the advantage of a reliable power supply and strong processing capabilities, allowing the system to perform persistent environmental scanning and high-precision beamforming. However, it also suffers from limited visibility in dense urban environments, where static obstacles frequently block LoS links to low-altitude targets. 
 {\color{black}To address such visibility limitations, an alternative architecture is the UAV-mounted RA system, which combines aerial mobility with agile boresight control. The authors in~\cite{Chen2025Rotatable} proposed a dual-level channel reconfiguration framework in which RAs are deployed on UAVs, and UAV trajectory and antenna orientation are jointly optimized to control both the large-scale path loss and the correlation between sensing and communication channels. This configuration enables UAV-mounted RAs to dynamically establish LoS links, illuminate blind spots, and support real-time sensing during motion.}
 

{\color{black}Despite the enhanced spatial adaptability enabled by RA, realizing reliable low-altitude ISAC in practice remains challenging due to the high mobility of aerial platforms and the need for coordinated ground-air deployment. Existing studies have demonstrated the benefits of RA for low-altitude direction sensing and channel reconfiguration enabled by UAV-mounted RAs, but they mainly investigate these functions separately. How to coordinate ground-based RA infrastructures and UAV-mounted RA systems to provide continuous 3D coverage, reliable target tracking, and robust communication support in dynamic low-altitude environments remains largely unexplored.}
Moreover, UAVs typically operate at high speeds with rapidly varying trajectories, imposing stringent requirements on real-time target tracking and low-latency antenna orientation adjustment. To maintain reliable high-rate links and accurate sensing under such mobility, RA systems need to support agile boresight control with fast actuation and high-precision alignment.
Furthermore, robust and continuous connectivity is also critical for ensuring safe UAV operations, including real-time monitoring and regulatory supervision. Communication disruptions may lead to loss of control or degraded situational awareness, raising safety concerns in dynamic low-altitude airspace. Addressing these challenges is essential for realizing secure, reliable, and resilient low-altitude ISAC systems.

\subsection{Cognitive Radio (CR) Systems}
CR systems enable secondary users (SUs) to access spectrum without causing harmful interference to primary users (PUs), addressing the critical issue of spectrum scarcity~\cite{Zhang2010Dynamic}. 
Due to the dynamic and spatially heterogeneous distribution of spectrum resources, reliable spectrum sensing is essential to identify spectrum holes in both spatial and spectral domains~\cite{Saruthirathanaworakun2012Opportunistic}. However, traditional CR systems provide limited spatial adaptability, which makes it difficult to fully exploit such spatial spectrum opportunities. 
Integrating RA technology into CR systems introduces enhanced spatial flexibility via adaptive boresight control, enabling more flexible and environment-aware spectrum sensing and access. By performing directional spatial scanning at the SU receiver, an RA-enabled CR system can observe the radio environment from multiple angular perspectives and collect fine-grained spectrum-occupancy information.
Once spatial spectrum holes are identified, the SU transceiver can support opportunistic transmission by rotating its antenna boresight toward available directions, thereby enhancing link efficiency while limiting unintended interference to nearby PUs. {\color{black}When spectrum resources are densely occupied, RA-assisted CR systems can also support interference-constrained spectrum sharing~\cite{Tan2026Rotatable,Peng2025Rotatable}.
In particular, joint optimization of transmit beamforming and antenna orientation enables the SU to concentrate radiation power toward intended receivers while spatially suppressing interference toward active PUs. This capability enhances spectrum utilization in interference-limited environments and provides additional spatial DoFs for coexistence.}

Despite these advantages, several open challenges remain in realizing the full potential of RA-aided CR systems. {\color{black}Existing studies have primarily focused on interference-constrained spectrum sharing, while RA-assisted spectrum sensing and opportunistic spectrum access remain relatively underexplored. One critical issue is the practical uncertainty and delay in obtaining accurate PU location information, which may hinder precise boresight alignment and degrade sensing performance. 
Moreover, persistent spectrum monitoring is indispensable during dynamic spectrum access. In opportunistic access scenarios, once the spectrum is reoccupied by a PU, the RA system must rapidly reconfigure its antenna orientations toward alternative available directions or transition to a spectrum-sharing mode with appropriate boresight control. This ensures that the SU maintains reliable connectivity while keeping interference to PUs below regulatory thresholds. Furthermore, UAV mobility, user dynamics, and fast channel variations introduce additional challenges for RA-assisted CR systems. Real-time boresight adjustment must be performed with low latency and high precision, while robust control mechanisms are needed to handle PU uncertainty, sensing errors, and rapid environmental changes. Coordinating RA-enabled CR functions across multi-antenna, multi-user, or distributed SU networks under imperfect CSI also remains an open research direction.
Overall, RA-aided CR systems present a promising pathway toward intelligent spectrum sensing and access. Future efforts should emphasize the joint design of adaptive access strategies and dynamic boresight control, while enhancing robustness to PU uncertainty and ensuring reliable coexistence in highly dynamic spectrum environments.}

\subsection{Physical Layer Security (PLS)}
The growing demand for secure wireless communications in complex and dynamic environments has spurred increasing interest in PLS techniques~\cite{Dai2025RotatableSecure}. By leveraging the inherent randomness and spatial characteristics of wireless channels, PLS provides a complementary approach to conventional cryptographic methods, particularly in latency-sensitive or infrastructure-limited scenarios. 
However, conventional PLS techniques typically rely on fixed-antenna architectures with limited spatial control, which restricts their ability to simultaneously enhance the legitimate link and suppress eavesdroppers.
With the integration of RA technology into wireless networks, flexible antenna orientation/boresight control introduces additional spatial DoFs for secure transmission~\cite{Dai2025RotatableSecure,Jiang2025Average,Zheng2026Joint,Wang2026Sensing,Dai2026Rotatable,Li2025Sensing}. {\color{black}In RA-enabled secure communication, antenna orientations can be adaptively adjusted to concentrate signal energy toward the legitimate receiver while suppressing radiation toward suspicious or vulnerable regions. For example, the authors in~\cite{Dai2025RotatableSecure} investigated an RA-enabled secure communication system, where confidential information is transmitted from an RA-based AP to a legitimate user in the presence of multiple eavesdroppers. By jointly optimizing transmit beamforming and RA rotation angles, the RA-enabled design strengthens the legitimate link and suppresses information leakage, thereby improving the achievable secrecy rate. Extending beyond instantaneous secrecy-rate optimization, the authors in~\cite{Jiang2025Average} studied average secrecy capacity maximization under practical scenarios with non-real-time RA adjustment. Their results show that RA orientations can be optimized based on statistical CSI, offering useful insights into secure RA deployment when frequent orientation updates or instantaneous CSI acquisition are difficult.}

{\color{black}However, the same spatial adaptability may also introduce new security risks if adversarial nodes are equipped with RA capabilities. An RA-enabled eavesdropper can dynamically align its boresight toward the legitimate transmitter to enhance interception capability, while an RA-assisted jammer can directionally inject interference into critical links with higher efficiency and lower detectability. Such orientation-aware attacks may significantly degrade secrecy performance and challenge existing countermeasures.
To address these emerging threats, advanced RA-aided PLS strategies should incorporate the joint design of antenna orientation, transmission, and interference management~\cite{Yang2026Rotatable}. For instance, an RA-equipped BS can proactively steer artificial noise, jamming signals, or deceptive waveforms toward suspicious spatial regions, thereby suppressing potential eavesdropping attempts while preserving reliable transmission toward legitimate users. 
In addition, adaptive boresight control combined with secure beamforming can rapidly form spatial nulls toward adversarial directions or reconfigure transmission paths upon threat detection.} Overall, RA-aided PLS extends secrecy enhancement from signal-domain processing to spatial-domain reconfiguration. With proper control and regulation, RA technology offers a promising and reconfigurable hardware-level solution for next-generation secure wireless systems.
\subsection{Cell-Free MIMO Networks}
Cell-free MIMO networks rely on a large number of distributed APs to cooperatively serve users in a cell-less manner, offering macro-diversity and eliminating inter-cell interference~\cite{Ammar2021User}. However, their performance hinges heavily on centralized coordination and dense fronthaul signaling, posing significant challenges to training overhead, scalability, and practical deployment.
Integrating RA into cell-free systems fundamentally alters this cooperation paradigm. Instead of relying solely on digital precoding to manage interference, each AP can reshape its effective channel through antenna rotation, strengthening desired propagation paths while attenuating unfavorable directions before baseband processing. 
Such orientation-induced channel reshaping reduces link imbalance across users and allows multi-AP cooperation to operate under more favorable channel conditions.
{\color{black}Recently, the authors in~\cite{Pan2025Rotatable} first incorporated RA technology into cell-free networks, where each single-RA AP adjusts its antenna boresight and performs AP-user pairing to enhance downlink performance.
Extending beyond such pairing-based designs, the authors in~\cite{Peng2026Cell} further investigated multi-RA array APs with joint optimization of antenna orientation and cooperative precoding, thereby substantially expanding the overall design space.}
Furthermore, RA naturally promotes spatially selective cooperation among APs. By concentrating radiation power and limiting coordination to the most relevant APs, RA systems reduce unnecessary cooperation and signaling overhead. This facilitates more fronthaul-efficient compression and control strategies that are crucial for scalable deployments, yielding a more favorable performance-complexity trade-off.

Nevertheless, RA-aided cell-free MIMO systems also introduce distinct challenges. 
Since orientation-dependent radiation gains are embedded into the effective channels, antenna rotation becomes tightly coupled with cooperative precoding and network-level coordination, making scalable joint optimization nontrivial. {\color{black}Moreover, AP-user association is no longer determined solely by proximity or large-scale fading, but is jointly influenced by antenna orientation control and discrete association decisions.
Furthermore, coordinated boresight control across a large number of APs remains difficult under limited fronthaul and imperfect CSI.
Future research should focus on scalable boresight control across distributed APs, joint AP-user association and antenna orientation optimization, and robust designs under imperfect CSI and limited control signaling. These directions are essential for enabling practical RA-assisted cell-free deployments in large-scale and dynamic network environments.}

\subsection{Simultaneous Wireless Information and Power Transfer (SWIPT)}
SWIPT has emerged as a promising solution for self-sustaining low-power devices in future IoT and wireless sensor networks\cite{Perera2017Simultaneous}. However, achieving an efficient balance between information decoding and energy harvesting remains a critical design challenge, especially in dynamic and rich multipath wireless environments~\cite{Pan2017Performance}. {\color{black}Although flexible antenna architectures have recently been explored for SWIPT, existing efforts have mainly focused on conventional beamforming-based or position-reconfigurable designs, while RA-aided SWIPT remains largely unexplored~\cite{Ding2015Application,Zhou2025Fluid}.} Integrating RA technology into SWIPT systems unlocks new opportunities for spatially adaptive transmission strategies. By adaptively adjusting the antenna boresight direction, RAs can dynamically align with dominant channel paths or high-energy regions, thereby improving the effective received power for energy harvesting. In parallel, directional boresight control enables selective signal reception to enhance the SINR for information decoding. This flexibility provides a new mechanism for managing the fundamental SWIPT trade-offs beyond traditional beamforming or power allocation techniques. For instance, under favorable channel conditions, the RA array can be steered toward energy-dense directions to maximize harvested power, whereas in interference-limited regimes, it can prioritize alignment with information-bearing signals to ensure reliable communication. Moreover, in an RA array, each antenna can be independently oriented to serve different objectives: some RAs may focus on maximizing SINR for data reception, while others align with regions of strong RF energy for harvesting. This spatial division of roles enables distributed optimization and functional decoupling, thereby enhancing the overall efficiency of SWIPT networks. 

In practice, the gains of RA-aided SWIPT critically depend on how accurately and how frequently antenna orientations are updated, since antenna rotation incurs non-negligible actuation energy and latency. Therefore, the net energy efficiency must account for both harvested energy and antenna rotation overhead, which can shift the optimal rate-energy operating point, particularly for low-power IoT devices. {\color{black}Beyond this overhead issue, RA-aided SWIPT introduces a joint antenna orientation and resource allocation problem across users with different service objectives. Although RA arrays enable spatial role division, dynamically coordinating antenna orientations to balance energy harvesting and information decoding remains inherently complex. The problem becomes even more challenging for co-located SWIPT users, where the RF power-maximizing direction generally does not coincide with the SINR-maximizing direction in the presence of interference. Future research may therefore focus on achieving a favorable trade-off between antenna rotation overhead and harvested energy, developing RA orientation allocation and scheduling strategies for SWIPT, and theoretically characterizing the achievable rate-energy region under practical rotation constraints.}


{\color{black}\subsection{Synergy of RA and IRS}
IRS has been widely investigated as a promising technology for reconfiguring wireless propagation environments through programmable reflections~\cite{Zheng2022ASurvey,Wu2021Intelligent,Liu2021Reconfigurable}. An IRS typically consists of a large number of passive or semi-passive reflecting elements whose reflection responses can be adjusted through controllable phase and/or amplitude coefficients. By creating controllable virtual propagation paths, IRS can enhance coverage, mitigate blockage, and improve energy efficiency. However, the performance of IRS-aided systems is highly sensitive to the geometric relationship among the transmitter, IRS, and receiver. In particular, when the IRS is weakly illuminated by the transmitter or operates under unfavorable incident/reflection angles, the cascaded channel may still suffer from severe path loss and limited reflection efficiency.

In this context, integrating RA into IRS-aided systems further enriches the spatial adaptability of the cascaded link. Instead of relying solely on IRS phase shifts to reshape the propagation environment, RA enables the transceiver to adjust its antenna orientations/boresights toward the IRS, intended users, or dominant propagation paths. Such transceiver-side orientation control improves the illumination and reception of IRS-related links before passive reflection, thereby strengthening the cascaded BS-IRS-user channel. Recently, the authors in~\cite{Zhang2026Joint} investigated multi-user uplink IRS-assisted communications with joint antenna rotation, receive beamforming, and IRS phase-shift optimization, showing that RA can better exploit controllable reflected paths. Another promising direction is to introduce the RA concept into IRS architectures, giving rise to ``rotatable IRS" designs. Unlike conventional IRSs with fixed panel orientations, a rotatable IRS introduces additional rotation DoFs by dynamically adjusting its panel orientation. In this way, the reflection coefficients and IRS orientation can be jointly optimized to improve incident/reflected link conditions~\cite{Zhou2025RotatableIRS,Peng2025RotatableIRS,Jiang2026LowAltitude}.  Building on these RA-assisted IRS and rotatable IRS designs, the authors in~\cite{Zhang2026MultiUser} further incorporated transceiver-side RA control into rotatable IRS systems, revealing that RA orientation, IRS orientation, and passive beamforming must be jointly coordinated to avoid inefficient illumination or reflection mismatch in the cascaded link.

Despite these promising opportunities, the synergy between RA and IRS remains largely unexplored. One key challenge lies in channel acquisition. Since the received pilot observations of the cascaded link are jointly affected by RA orientations, IRS phase shifts, and IRS orientation, conventional IRS channel estimation methods may become inefficient under joint RA-IRS reconfiguration. Low-overhead channel acquisition methods are therefore needed to exploit the geometric structure of RA-IRS links and avoid exhaustive training over all antenna and IRS configurations. 
Therefore, future research should focus on low-overhead channel acquisition exploiting the geometric structure of RA-IRS links, joint RA and IRS configuration design, IRS placement optimization, and practical prototype validation for RA-IRS integrated systems.
}

\subsection{Other Miscellaneous Topics}

In addition to the above directions, RA technology shows great potential in a wide range of emerging applications that require directional adaptability, spatial awareness, and low-cost hardware design. 
A unifying theme is that antenna boresight/orientation control provides a lightweight means of realizing directional alignment and spatial selectivity at the radio (or transducer) front end, which is particularly relevant to IMT-2030-oriented scenarios and highly dynamic, resource-constrained platforms~\cite{International2023Framework}.

Beyond conventional wireless communication, the capability of RA to realize low-complexity directional adaptation also makes it promising for a wide range of cross-disciplinary systems. For instance, in optical communication~\cite{Kaushal2016Optical}, RA enables dynamic alignment of highly directional links, mitigating degradation caused by mobility, drift, or mechanical offset. This is especially beneficial for aerial relays, space-ground free-space optical systems, and high-precision visible light communications. Similar benefits arise in acoustic communication~\cite{Stojanovic2009Underwater}, where rotatable acoustic transducers can adapt their boresight in rich multipath or reflective environments, such as underwater networks and smart audio systems, to improve signal reception and suppress interference without relying on large arrays.
The same directional flexibility is also valuable for sensing-related tasks. For example, in radar, imaging, and environmental sensing~\cite{Raymond2016Coexistence,Chen2022Millimeter,Jiang2025Rotatable}, RA supports flexible scanning and spatial selectivity for high-resolution target detection on mobile or resource-constrained platforms. Its agility enables cost-effective perception enhancement in UAVs, robots, and smart vehicles. Moreover, for localization and navigation~\cite{Jiang2010Localization,Ali2014ARotating}, RA supports geometry-aided positioning and angular separation in dense multi-user settings. Last but not least, RA can serve as a physical interface for AI-driven wireless systems. The antenna boresight direction can be optimized using reinforcement learning or predictive models, enabling autonomous and environment-aware boresight control with minimal signaling overhead. When integrated with edge intelligence, RA empowers real-time spatial decision-making, supporting self-organizing and adaptive network operations.

\section{Conclusions}\label{sec:conclusions}
By enabling flexible 3D antenna orientation/boresight rotation, RA introduces additional spatial DoFs that enhance wireless communication and sensing performance without requiring extra antenna resources or additional deployment space. Compared with conventional fixed-antenna architectures, RA offers a lightweight and practical means of spatial adaptation for future wireless networks. In particular, RA enables directional ``spotlight'' transmission toward desired users or targets and eye-like spatial scanning of the surrounding environment, thereby opening new possibilities for communication and sensing system designs. In this paper, we have provided a comprehensive tutorial on RA-empowered wireless networks. To this end, we have reviewed the historical development of RA-related technologies, established a unified framework for antenna/array rotation and channel modeling, and investigated rotation optimization in representative communication and ISAC scenarios. In addition, we have discussed RA channel estimation/acquisition, practical hardware architectures, deployment trade-offs, and recent prototypes that demonstrate the feasibility and potential of RA-enabled systems.
Overall, RA strikes a compelling balance between spatial flexibility, implementation complexity, and deployment cost, offering a promising pathway toward adaptive coverage, reliable transmission in interference-limited environments, and high-resolution sensing in emerging 6G systems. Looking ahead, further advances in the co-design of rotation optimization, channel estimation/acquisition, and practical RA implementation will continue to unlock the potential of RA for more agile, efficient, and intelligent wireless communication and sensing.
\ifCLASSOPTIONcaptionsoff
  \newpage
\fi

\bibliographystyle{IEEEtran}
\bibliography{RA_tutorial}

\end{document}